\documentclass[useAMS,usenatbib,fleqn]{mn2e}
\usepackage{graphicx}
\usepackage{comment}
\usepackage{subcaption}
\usepackage{caption}
\usepackage{amsmath}
\usepackage{amssymb}
\usepackage[normalem]{ulem}
\usepackage{color}
\newcommand{\htwo}{H$_2$}

\newcommand{\kms}{km\,s$^{-1}$}

\title[SPLASH: The Southern Parkes Large-Area Survey in Hydroxyl -- Data Description \& Release]{SPLASH: The Southern Parkes Large-Area Survey in Hydroxyl -- Data Description and Release}
\author[J. R. Dawson et al.]
{J. R. Dawson$^{1,2}$\thanks{Email: joanne.dawson@mq.edu.au}, P. A. Jones$^{1}$, C. Purcell$^{1,3}$, A. J. Walsh$^{4}$, S. L. Breen$^{5}$, \newauthor C. Brown$^{6}$, E. Carretti$^{7,2}$, M. R. Cunningham$^{8}$, J. M. Dickey$^{6}$, S. P. Ellingsen$^{6}$, \newauthor  S. J. Gibson$^{9}$, J. F. G\'omez$^{10}$, J. A. Green,$^{2}$, H. Imai$^{11}$, V. Krishnan$^{6,2}$, N. Lo$^{12}$, \newauthor V. Lowe$^{8}$, M. Marquarding$^{2}$, N. M. McClure-Griffiths$^{13}$ \\
$^{1}${Department of Physics and Astronomy and MQ Research Centre in Astronomy, Astrophysics and Astrophotonics, Macquarie}\\ {~University, NSW 2109, Australia}\\
$^{2}${Australia Telescope National Facility, CSIRO Space \& Astronomy, PO Box 76, Epping, NSW 1710, Australia}\\
$^{3}${Sydney Institute for Astronomy (SIfA), School of Physics, The University of Sydney, NSW 2006, Australia}\\
$^{4}${International Centre for Radio Astronomy Research, Curtin University, GPO Box U1987, Perth, WA 6845, Australia}\\
$^{5}${SKA Observatory, Jodrell Bank, Lower Withington,
Macclesfield, SK11 9FT, UK}\\
$^{6}${School of Natural Sciences, University of Tasmania, Private Bag 37, Hobart, TAS 7000, Australia}\\
$^{7}${INAF - Istituto di Radioastronomia, via P. Gobetti 101, I-40129 Bologna, Italy}\\
$^{8}${School of Physics, University of New South Wales, Sydney, NSW 2052, Australia}\\
$^{9}${Department of Physics and Astronomy, Western Kentucky University, Bowling Green, KY 42101, USA}\\
$^{10}${Instituto de Astrof\'{\i}sica de Andaluc\'{\i}a, CSIC, Glorieta de la Astronom\'{\i}a s/n, E-18008 Granada, Spain}\\
$^{11}${Amanogawa Galaxy Astronomy Research Center, Graduate School of Science and Engineering, Kagoshima University, } \\{~~1-21-35 Korimoto, Kagoshima 890-0065, Japan}\\
$^{12}${Departamento de Astronomia, Universidad de Chile, Camino El Observatorio 1515 Las Condes, Santiago, Chile}\\
$^{13}${Research School of Astronomy and Astrophysics, Australian National University, Cotter Road, Canberra, ACT 2611, Australia}
}

\begin{document}

\date{Accepted 2022 Month 00. Received 2022 Month 00; in original form 2021 Month 12}

\pagerange{\pageref{firstpage}--\pageref{lastpage}} \pubyear{2019}

\maketitle

\label{firstpage}

\begin{abstract}
We present the full data release for the Southern Parkes Large-Area Survey in Hydroxyl (SPLASH), a sensitive, unbiased single-dish survey of the Southern Galactic Plane in all four ground-state transitions of the OH radical at 1612, 1665, 1667 and 1720\,MHz. The survey covers the inner Galactic Plane, Central Molecular Zone and Galactic Centre over the range $|b|<$ 2$^{\circ}$,  332$^{\circ}$ $< l <$ 10$^{\circ}$, with a small extension between 2$^{\circ}$ $< b <$ 6$^{\circ}$, 358$^{\circ}$ $< l <$ 4$^{\circ}$. 
SPLASH is the most sensitive large-scale survey of OH to-date, reaching a characteristic root-mean-square sensitivity of $\sim15$ mK for an effective velocity resolution of $\sim0.9$\,\kms. The spectral line datacubes are optimised for the analysis of extended, quasi-thermal OH, but also contain numerous maser sources, which have been confirmed interferometrically and published elsewhere. We also present radio continuum images at 1612, 1666 and 1720\,MHz. Based on initial comparisons with $^{12}$CO(J=1--0), we find that OH rarely extends outside CO cloud boundaries in our data, but suggest that large variations in CO-to-OH brightness temperature ratios may reflect differences in the total gas column density traced by each. Column density estimation in the complex, continuum-bright Inner Galaxy is a challenge, and we demonstrate how failure to appropriately model sub-beam structure and the line-of-sight source distribution can lead to order-of-magnitude errors. Anomalous excitation of the 1612 and 1720\,MHz satellite lines is ubiquitous in the inner Galaxy, but is disabled by line overlap in and around the Central Molecular Zone. 
\end{abstract}


\section{Introduction}
\label{introduction}

The last decade has seen the resurgence of hydroxyl (OH) as a probe of the molecular interstellar medium (ISM). As the first radio molecular lines discovered in interstellar space \citep{weinreb63}, the 18-cm $\Lambda$-doubling transitions of ground-state OH were once widely used to study the distribution and properties of Galactic molecular clouds \citep[][]{robinson67, goss68, heiles69, knapp73, caswell74, sancisi74, mattila79a, turner79, turner82, wouterloot85}. 
The four transitions -- between the four sub-levels of the $^2\Pi_{3/2}$; J=$3/2$ OH ground state -- consist of two main lines at 1665.402 and 1667.359 MHz, and two weaker satellite lines at 1612.231 and 1720.530 MHz (relative strengths 1:5:9:1 in order of increasing frequency). All four lines exhibit strong maser emission, which traces a great variety of astrophysical phenomena, from star formation \citep[e.g.][]{caswell99}, to evolved stars \citep[e.g.][]{sevenster97}, to supernova shocks \citep[e.g.][]{green97}, to interstellar magnetic fields \citep[e.g.][]{reid90, fish03, green11}. Outside of compact, high-gain maser sites, the OH lines are generally weak, with brightness temperatures of no more than a few 100 mK in typical molecular cloud conditions. It is this class of emission and absorption that traces the bulk of the molecular ISM. Because the four lines are generally not in LTE (and may even be weakly masing, as we will discuss below), we do not refer to this emission and absorption as `thermal' OH. In order to draw the important distinction between the observed lines and a system whose levels are truly thermally populated, we will refer to this widespread weak and extended OH as `quasi-thermal' in this work. 

Despite its low brightness temperatures, OH has many advantages as a probe of the extended molecular ISM. Chief among these is that it can trace diffuse molecular gas that CO cannot. 
OH abundances remain relatively steady even in poorly shielded molecular regions 
\citep[e.g.][]{wolfire10,hollenbach12,nguyen18}, and OH 18-cm emission is observed to extend beyond CO-bright regions into diffuse cloud envelopes \citep{wannier93,barriault10,allen12,allen15,xu16}. As might be expected, the gas detected in OH but not CO appears to be warmer, more diffuse and lower column density material \citep{wannier93,li18,engelke20}. In some cases very faint OH emission even appears to be correlated with the distribution of H{\sc i} \citep[e.g.][]{allen12}, leading recently to the discovery of a thick disk of extremely diffuse molecular gas  ($n_{\mathrm{H2}}\sim5\times10^{-3}$ cm$^{-3}$) in the Outer Galaxy \citep{busch21}. Low-$A_V$, `CO-dark' \htwo\ may account for a significant fraction of the Milky Way's molecular gas mass, particularly in regions where the ambient density and pressure are relatively low \citep{reach94,grenier05,planck11,planck14,pineda13,langer14a,remy17,remy18}.  
OH provides a means of tracing this material on Galactic scales, along with 3D information that is difficult to obtain via measurements of the total proton column (e.g. from dust emission, reddening or gamma rays).

Quasi-thermal OH lines also provide a barometer of the physical conditions in the molecular ISM -- whether CO-dark or CO-bright. The ground state level populations are readily perturbed from their thermal ratios via small changes in physical conditions 
(see \citealt{elitzur92}), and all four ground-state OH transitions are usually anomalously excited 
\citep[e.g.][]{rieu76,crutcher79,turner82,dawson14,li18,engelke18}. 
The non-thermal line ratios, particularly in the satellite lines, can be modelled to constrain (at least) number density, column density, kinetic temperature and dust temperature \citep[e.g.][]{elitzur76,guibert78,vanlangevelde95}. Indeed, non-LTE modelling has been used to argue for elevated kinetic temperatures in CO-dark molecular gas \citep{ebisawa15}, and to demonstrate how commonly-seen excitation patterns in the OH satellite lines can be a marker of Galactic H{\sc ii} regions \citep{petzler20}. 

Finally, although the quasi-thermal OH lines are inherently weak in emission, 
the brightness of the radio sky at 18-cm means that they can often be observed strongly in absorption -- both against bright compact sources, and against the diffuse radio continuum of the inner Galaxy.  
Absorption observations can allow direct determination of line optical depth, and in some cases excitation temperature too \citep[e.g.][]{rieu76,liszt96,li18,engelke18}, providing direct information on the physical state of the gas, and removing a large source of uncertainty in the derivation of the molecular gas column. In the Galactic Plane the relative location of the continuum-emitting gas and the OH clouds along complex sightlines is a complicating factor (as we will discuss in detail in Section \ref{noh_is_hard}), but it can also a useful tool: for example, the relative strengths of OH absorption vs CO emission have been used to build 3D models of the gas in the the Central Molecular Zone (CMZ), by allowing components to be localised in front of or behind the bright Galactic Centre \citep{sawada04,yan2017}.

This paper presents the first full data release from SPLASH -- the Southern Parkes Large-Area Survey in Hydroxyl \citep{dawson14}. SPLASH is a sensitive, unbiased, fully-sampled, survey of the Southern Galactic Plane and Galactic Centre in the four ground-state transitions of OH and 1.6--1.7\,GHz radio continuum, using the Parkes 64-m telescope. The survey was designed to go deep enough to detect the widespread emission and absorption needed for studies of CO-dark \htwo, to simultaneously observe the full set of 4 lines needed for excitation modelling, and to identify numerous new OH maser candidates to unprecedented flux density limits. SPLASH maser sources have already been followed up at high resolution with the Australia Telescope Compact Array, resulting in the discovery of over 400 new ground-state OH maser sites. These are published in a separate set of catalogues (\citealt{qiao16,qiao18,qiao20}, see also \citealt{uno21}), and we do not focus further on them here. The single-dish data products presented in this data release are optimised for the analysis of extended, quasi-thermal OH (i.e. calibrated in brightness temperature units and gridded so that surface brightness is conserved).

This paper is organised as follows. We outline the observing programme in Section \ref{observations}, including the hardware setup and the observational strategy. Section \ref{reduction} describes in detail the data reduction process and measures of data quality, outlining key choices, assessing performance, quantifying uncertainties, and summarising the key characteristics of the final spectral line and continuum data products. While the bulk of this data release paper is focused on the data itself, we also take a first look at its use for science in Section \ref{discussion}. There we make an initial comparison with CO emission, discuss the critical importance of the line-of-sight and sub-beam source distribution in column density estimation, and present some initial statistics on the excitation of the satellite lines in the inner Galaxy. We conclude in Section \ref{conclusions}.

\begin{figure*}\includegraphics[width=\textwidth]{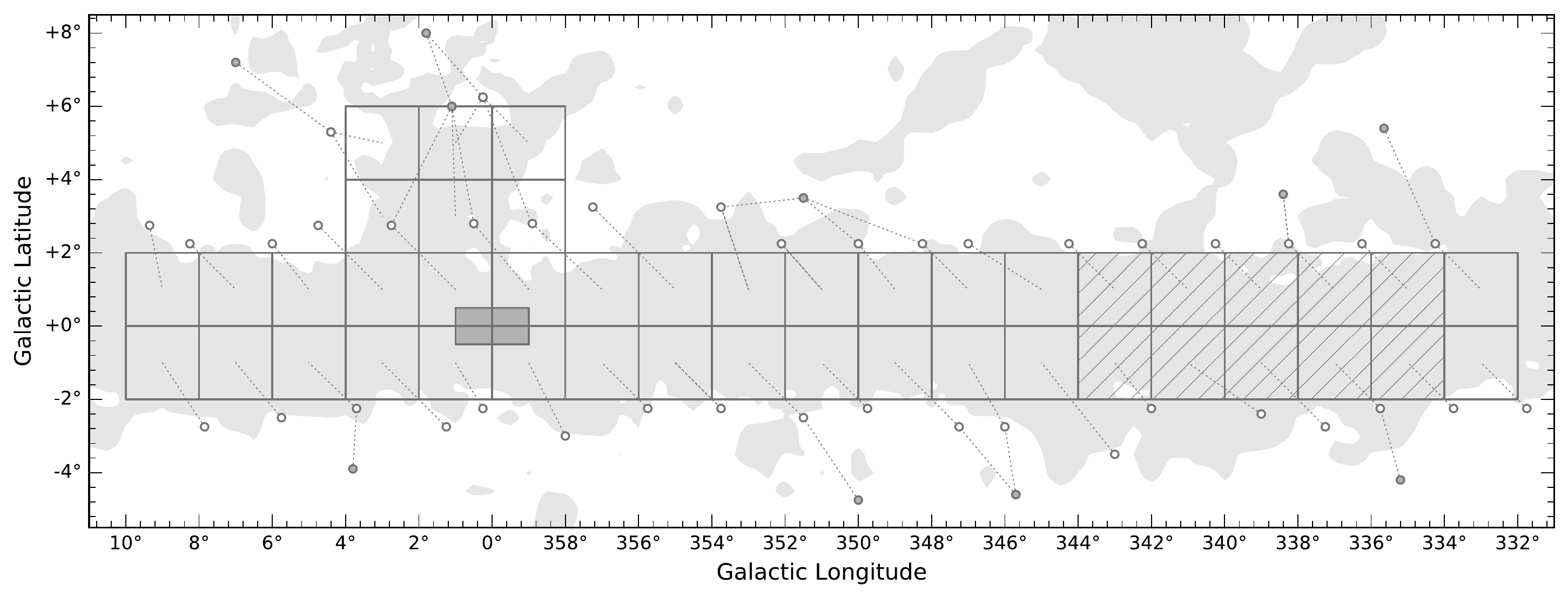}
\caption{SPLASH survey coverage, overlaid on the velocity integrated CO distribution from \citet{dame01} (grey shaded area). Each $2\times2$ degree square corresponds to a single SPLASH tile. The hatched zone indicates the Pilot Region \citep{dawson14}. Reference positions for each tile are shown as white-filled circles, with widths corresponding to the FWHM beam size at 1612~MHz ($12.6$ arcmin). Dotted lines join each reference position to its corresponding tile. Where deep observations were carried out to characterise low-level signal in the reference spectra (see text), the corresponding reference position for these observations is shown as a grey-filled circle. The Galactic Centre region over which observations were made with different attenuator settings spans $|l|<1^{\circ}$, $|b|<0.5^{\circ}$, and is shaded in dark grey. The CO contour level is 3 K\kms.}
\label{coverage_fig} 
\end{figure*}

\section{Observations}
\label{observations}

Observations were made between May 2012 and September 2014 with the Australia Telescope National Facility (ATNF) Parkes 64-m telescope (called `Murriyang' in Wiradjuri). The total time devoted to the survey was around 1800 hours, taken over 10 approximately evenly-spaced epochs.  

Data were taken in on-the-fly (OTF) mapping mode, in which spectral data is recorded continuously as the telescope is scanned across the sky. The scan rate was $34$ arcsec~s$^{-1}$, with data recorded to disk every 4\,s at intervals of $2.3$ arcmin. The spacing between scan rows was $4.2$ arcmin, which over-samples the beam both perpendicular and parallel to the scan direction (the FWHM at 1720~MHz is $12.2$ arcmin). The survey region was divided into $2\times2$ degree tiles, each of which was mapped a total of ten times to achieve target sensitivity. Repeat maps were scanned alternately in Galactic latitude and longitude to minimise scanning artefacts. Off-source reference spectra were taken every two scan rows, where the off-source position for each map was chosen to minimise the elevation difference between the reference position and the map throughout the course of observations. 

All reference positions were observed for a total integration time of 20 minutes prior to the main survey to ensure no emission or absorption was detected.  
However, upon preliminary reduction of the dataset it became clear that fifteen of the forty reference spectra contained emission or absorption at low levels in at least one of the four lines. To characterise this signal for later correction, affected positions were paired with new reference positions and re-observed in position-switching mode for a total on-source time of 100 minutes, achieving a $1\sigma$ sensitivity of $\sim$5 mK per 0.7\kms\ channel ($\sim$3 times better than the main survey). 
The new reference positions were chosen to be at least $0.5^\circ$ distant from CO detections \citep{dame01}, and were verified to be free of signal to within the $3\sigma$ detection limit of $\sim$15\,mK. Figure \ref{coverage_fig} shows a map of the survey coverage with all reference positions marked.  

The H-OH receiver provided two orthogonal linear polarisations, and spectral data were recorded by two digital filterbanks (DFB3 and DFB4). All four polarisation products were recorded, of which only XX and YY were retained for further processing. The dual inputs of the DFB3 were set to 1720 and 1666\,MHz, with the DFB4 hosting a singular input frequency of 1612\, MHz. The bandwidth for each of the three IFs was 8\,MHz, with 8192 spectral channels (note that the 1666\,MHz IF contains both of the main lines). This corresponds to a velocity coverage and channel width of $\sim1400$~km~s$^{-1}$ and 0.18~km~s$^{-1}$ respectively.

The ATNF standard calibrator source PKS B1934-638 was observed once per day in two orthogonal scans for intensity calibration, and the bright maser source G351.775-0.536 was observed daily as a systems check (since it shows emission in all four ground state transitions). The system temperature, $S_\mathrm{sys}$ (in flux density units), was derived by injecting a noise signal of known amplitude into the receiver feedhorn, and continuously monitored by a synchronous demodulator. 
Due to the limitations of the non-standard backend setup, and hardware-related instability in the 1666\,MHz system temperature measurements, $S_\mathrm{sys}$ for all three IFs was copied over from the 1720\,MHz band. Values were recorded once per 4\,s integration, and were typically between $25$--$30$ Jy, increasing to a maximum of $\sim 140$ Jy towards sight-lines with very bright continuum emission. Note that while on-source $S_\mathrm{sys}$ measurements are recorded, they are not directly used in the processing (see Equation \ref{eq:bandpass}). Slow changes in system response (e.g. due to elevation) are appropriately tracked via the off-source $S_\mathrm{sys}$ values. 

The high power levels towards the strong continuum emission in the Galactic Centre at Sgr A cause some level of saturation when the data are taken at standard attenuator settings. To address this, replacement data were taken over the Galactic Centre area ($|l|<1^{\circ}$, $|b|<0.5^{\circ}$, as shown in Figure \ref{coverage_fig}) at higher attenuation.

\section{Data Reduction}
\label{reduction}

The following subsections discuss the data reduction process in detail. 
The major challenge for SPLASH data is spectral baseline correction. The OH lines are weak (typically $<100$ mK); broad/blended spectral features may occupy a significant fraction of the spectral channels \citep[see e.g.][]{busch21} and may (particularly in the case of the satellite lines) `flip' between emission and absorption multiple times over the range of a feature \citep[see e.g.][]{petzler20}; and residual baseline structure may have broad, irregular bumps and humps or comparable widths and heights to real signal. The CMZ and Galactic Centre region are particularly problematic in this regard, due to the extreme velocity widths of the spectral features. These factors necessitated a careful, iterative approach to line-finding and baseline correction, which is described in more detail in Section \ref{baselining}, below.
Additional challenges include the mitigation of radio frequency interference (RFI), 
and the calibration and correction of the continuum data, which the survey was not originally optimised to produce (see Section \ref{cont_images}). 

The data reduction pipeline was primarily written in \textsc{Python}, incorporating a number of routines imported from the the python-based \textit{ATNF Spectral line Analysis Package} (\textsc{Asap})\footnote{http://svn.atnf.csiro.au/trac/asap}. The \textsc{Gridzilla} package \citep{barnes01} was used for gridding, and \textsc{Duchamp} \citep{whiting12} was used for intermediate-stage 3D source finding. Miscellaneous tasks from the Australia Telescope National Facility (ATNF) \textsc{Miriad} \citep{sault95} package were used in processing the gridded cubes. Unlike in the SPLASH pilot paper \citep{dawson14}, \textsc{Livedata}\footnote{https://www.atnf.csiro.au/computing/software/livedata/index.html} was used only for the bandpass calibration of standard calibrator data, and not for the main survey maps.  

\subsection{Flux-Scale Calibration and System Stability}
\label{intensity_cal}

The value of $S_\mathrm{sys}$ reported by the telescope tracks relative changes in system temperature, but does not provide accurate absolute flux-density calibration. Observations of the standard calibrator source PKS B1934-638 
were therefore used to calibrate the flux density scale. 
Calibrator observations were reduced in \textsc{Livedata}, with bandpass calibration performed using an off-source reference spectrum computed from the first and last 30 arcminutes of each 2-degree orthogonal cross-scan. The bandpass-calibrated 
flux density is given by: 
\begin{equation}
S^*=\frac{P_{\mathrm{ON}}-P_{\mathrm{OFF}}}{P_{\mathrm{OFF}}}~S_{\mathrm{sys,OFF}},
\label{eq:bandpass}
\end{equation}
where $S^*$ is the  
flux density of the source position (not yet absolutely calibrated), $P_{\mathrm{ON}}$ and $P_{\mathrm{OFF}}$ are the on-source and off-source power measured by the telescope, and $S_{\mathrm{sys,OFF}}$ is the system temperature (in flux-density units) at the off-source position. 

\begin{figure}\includegraphics[scale=0.4]{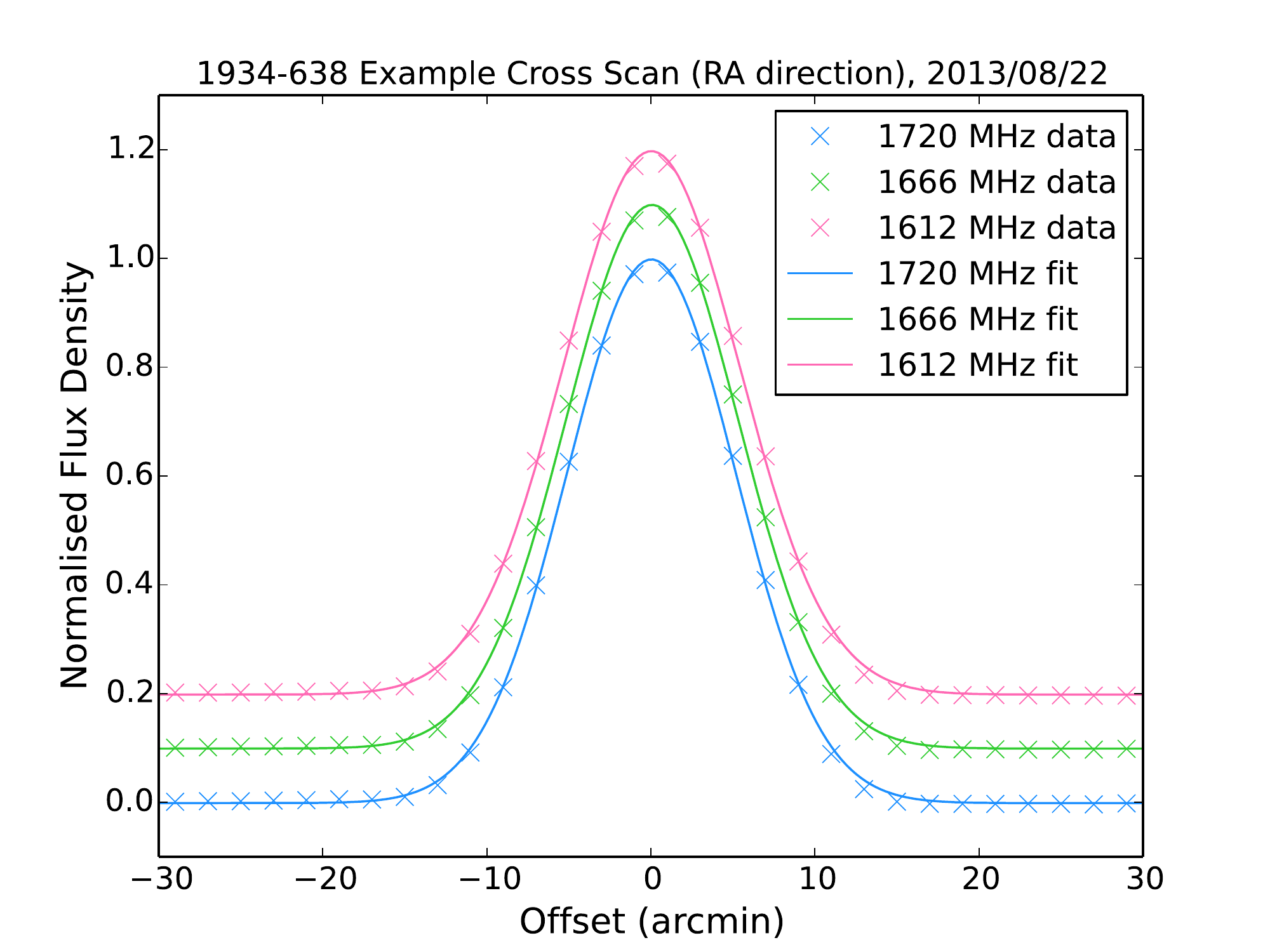}
\caption{Example of measured beam profiles and Gaussian fits obtained from cross scans of the standard calibrator source PKS 1934-638. Data are averaged over both polarisations and over the 8 MHz bandwidth. 1666 and 1612 MHz data are offset vertically for ease of viewing.}
\label{beamfits}
\end{figure}

\begin{figure}\includegraphics[scale=0.5]{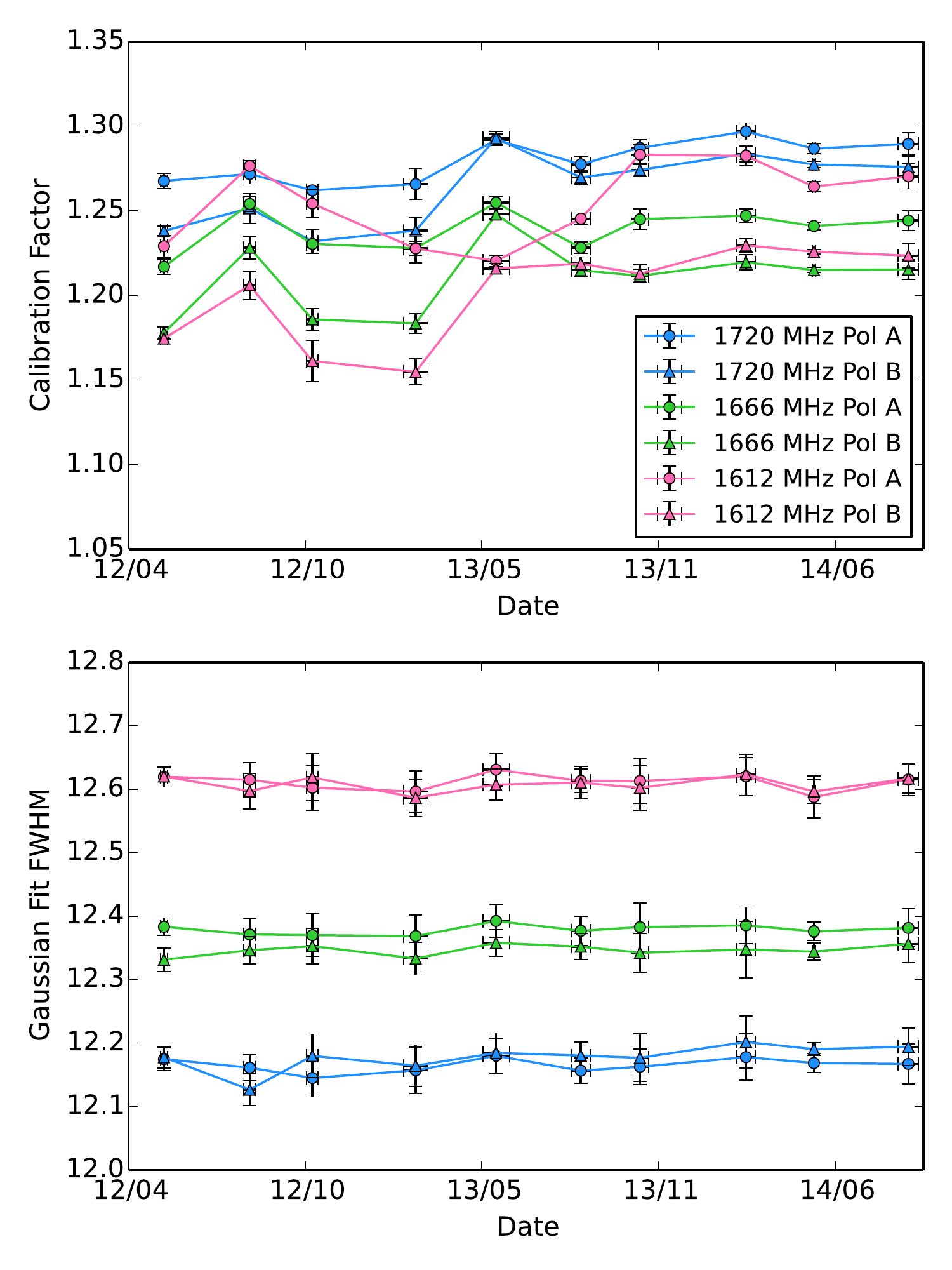}
\caption{Variation of calibration factors and beam FWHM measurements with time. The top panel shows the calibration factors applied to each IF/polarisation pair to map the raw data (in pseudo-Jy) onto the correct flux density scale, derived from Gaussian fits to the peak intensity of PKS B1934-638. 
Vertical error bars are the standard deviations of the sample of daily measurements from each epoch, and horizontal error bars show the length of each observing epoch. 
The bottom panel shows measured beam FWHM, derived from 1D Gaussian fits to the beam profile, taken in orthogonal directions and averaged for each IF/polarisation pair.}
\label{calibovertime}
\end{figure}

After flagging of bad channels at either end of the bandpass (removing $\sim10\%$ of channels in the 1666 and 1720 MHz bands and $\sim25\%$ in the 1612 MHz band), a single continuum flux density value for each integration was obtained from the mean of all remaining spectral channels within the 8 MHz bandpass. 1D Gaussian fits were then made to each scan direction and polarisation to obtain measured peak flux densities and beam sizes. Example data and fits are shown in Figure \ref{beamfits}. 

\begin{figure*}
\begin{center}
\includegraphics[scale=0.53]{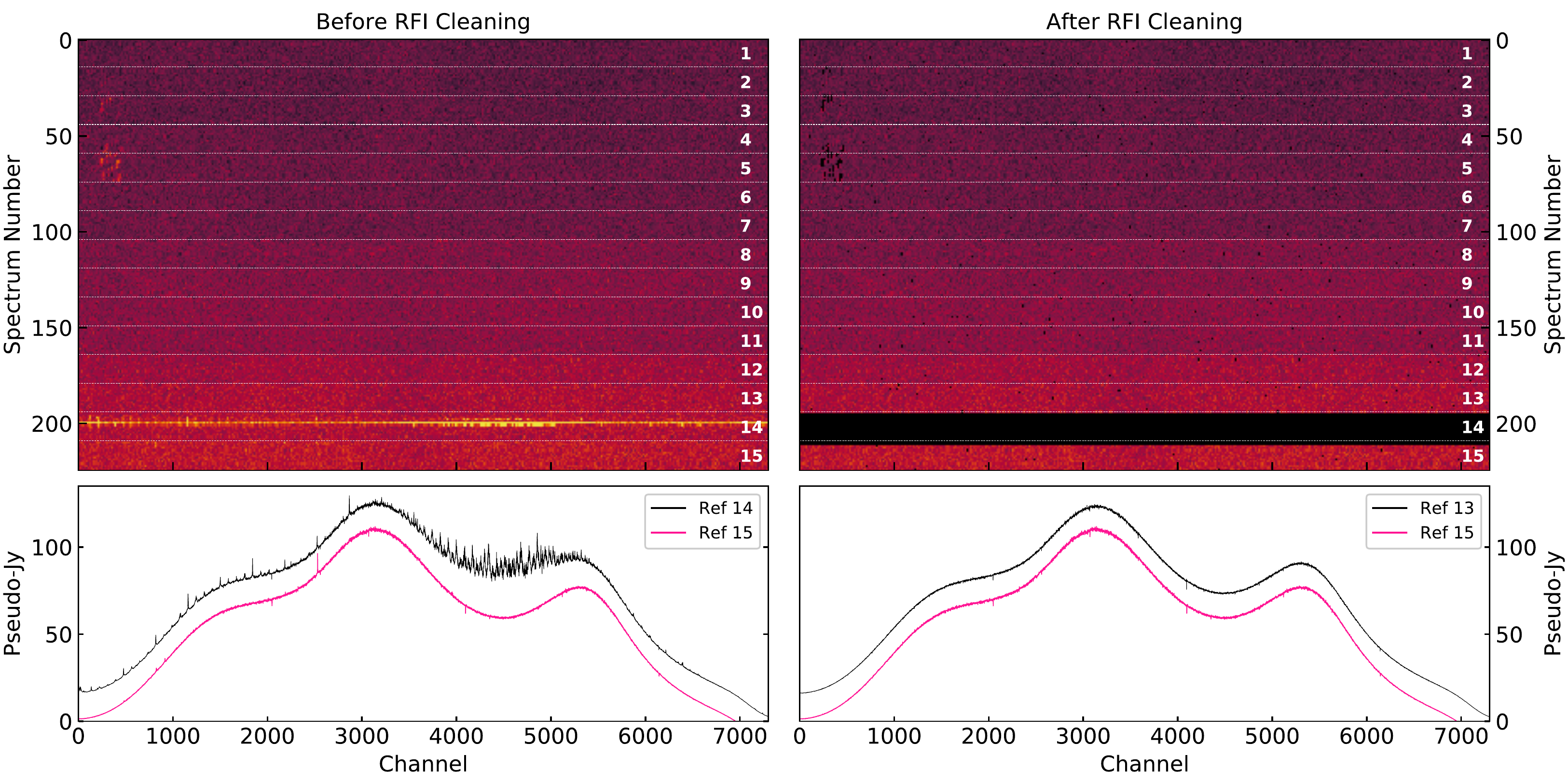}
\end{center}
\caption{Reference spectrum RFI cleaning. The top panels show waterfall plots of all reference integrations in a single OTF map file in the 1612 MHz band, where each row of the image represents a single spectral data dump with a 4s integration time. The colour scale runs from dark (low) to light (high), with black (right-hand panel only) indicating completely flagged channels. The waterfall plot spectra have been divided by a simple median spectrum of the entire set of reference observations for this map to temporarily remove the instrumental bandpass and make RFI easily visible. The white numbers and thin white dashed lines indicate the groups of integrations (total 60s integration time) comprising a single reference position observation -- 15 in total for a full $2\times2^{\circ}$ map file, separated by about 4 minutes in time. The horizonal axes show the channel number, truncated at 7400 to exclude low power channels at the edge of the 1612 MHz band. The bottom two panels show examples of the reference spectra as they would be applied to map data -- simple means of the full 60s of raw (non-bandpass-calibrated) data for each reference position observation (where the pink spectra are artifically offset for ease of viewing). It can be seen that RFI has severely affected the 14th reference observation in this file. The preceding reference that will replace it in processing is shown in its place in the right-hand plot.}
\label{fig_ref_rfi_flagging}
\end{figure*}

The calibration factors required to map the data onto the corrected flux-density scale are derived assuming true PKS B1934-638 flux densities of 14.34, 14.16 and 13.97 Jy at 1612, 1666 and 1720 MHz \citep{reynolds94}, and are shown in Figure \ref{calibovertime}.
Within epochs the system was extremely stable, with typical standard deviations in flux density of $< 1\%$ in each multi-day sample. Between epochs, we found minor but statistically significant variations in telescope response, necessitating the use of separate scaling factors. Note that each polarisation and frequency combination has a unique scaling factor.  

The SPLASH survey provides brightness temperature datacubes on a main beam temperature scale, $T_\mathrm{b}$. The conversion factor from a flux density scale is the main beam gain, $G_\mathrm{mb}$, such that $S=G_\mathrm{mb} T_\mathrm{b}$.
An idealised main beam gain is computed assuming an ideal circular Gaussian beam with FWHM obtained from our Gaussian fits (see the bottom panel of Figure \ref{calibovertime}), given by $G_\mathrm{mb} = 2k \Omega/ \lambda^2$, where $\Omega$ is the beam solid angle. We obtain values of 1.22, 1.25 and 1.29 Jy/K for 1612, 1666 and 1720 MHz, for measured beam FWHM of 12.6, 12.4 and 12.2 arcmin, respectively. 
These gains are $\sim8\%$ smaller than those listed in the Parkes telescope documentation at the time, where this difference arises from the slightly smaller beam sizes we measure here.

\subsection{Initial Removal of Problem Data}

A very small fraction of bad data was automatically identified and flagged prior to processing. The DFB3 and DFB4 digital filterbank correlators would occasionally fault, resulting in a small amount of spurious data being written to file before observations could be stopped and the problem corrected. Early epoch observations also occasionally suffered from a total loss of power from the receiver. Such problems are trivial to identify from exceptionally high or low flux densities recorded to the raw spectra, and all affected data was excluded prior to processing.

The DFB correlators also have some spurious bad channels; this was found to significantly affect every 1024th channel 
in the 8192 channel IF, with spurious data values persisting even after bandpass calibration. These channels were flagged in all data prior to processing. 

\subsection{Cleaning of Reference Spectrum RFI}

A single OTF map of a $2\times 2^\circ$ tile contains fifteen observations of the reference position, taken at 10-minute intervals. Each reference observation consists of fifteen four-second integrations which are averaged to form the reference spectrum for two map rows.  
RFI allowed to remain in any reference observation will therefore fold over into large portions of a map, and must be completely removed before the bandpass calibration step. While simple measures such as taking the median (rather than the mean) can effectively mitigate against some RFI, we prefer a simple mean for its superior noise characteristics. We therefore identify and remove RFI as follows.

A master reference for each complete OTF map file is first formed for each IF and polarisation from the median of all 225 ($15\times15$) integrations. Each individual integration is divided by this master reference to temporarily correct for the instrumental bandpass. The RMS noise is then computed for each of these spectra, along with a robust estimate of the spread in RMS values, as defined by the $S_n$ statistic of \citet{Rousseeuw93}: $S_n = 1.1926~\mathrm{med}_i\{\mathrm{med}_j|x_i-x_j|\}$. A second order polynomial is fit to the RMS noise as a function of time, which captures any slow drift (e.g. due to elevation change over the duration of a 2.5-hour map file). 
Any four-second integration whose spectral noise deviates by more than $3S_n$ from the fit line is flagged in its entirety. 

The above eliminates instances of strong RFI affecting a large fraction of channels in a spectrum. To remove isolated spikes that are localised in frequency, the $S_n$ statistic is computed again on a channel-by-channel basis for each remaining spectrum, and channels whose values deviate by more than $3S_n$ from the mean are flagged. If more than 50\% of channels in a given spectrum are bad, the entire spectrum (i.e. one four-second integration) is flagged. 

Finally, if more than 50\% of spectra in a given set of 15 integrations are bad, the entire reference observation is removed, and the pipeline defaults to its nearest neighbour in time when performing bandpass calibration. 

The fraction of reference position data removed in this process is 1.8\%, 1.1\% and 1.2\% for the 1612, 1666 and 1720\,MHz bands, respectively. 
Figure \ref{fig_ref_rfi_flagging} shows a waterfall plot of all reference spectra for a given map file before and after cleaning. Note that the temporarily bandpass calibrated reference spectra are not retained going forward; all flags are applied to the raw spectra.  

\subsection{Bandpass Calibration and Correction for Reference Position Emission/Absorption}

Each four-second integration in an OTF map is bandpass calibrated according to Equation \ref{eq:bandpass}, using the appropriate time-averaged, RFI-free reference observation (performed separately for each IF and polarisation). At this stage the flux density correction factors are applied and the corrected data converted to a main-beam brightness temperature scale (see Section \ref{intensity_cal}). The two polarisations are then averaged together.  

As described in Section \ref{observations}, and shown in Figure \ref{coverage_fig}, fifteen of the forty off-source reference positions were found to contain OH emission or absorption at a high enough level to contaminate the survey spectra, and were re-observed to enable the signal to be characterised. These position-switched reference position observations were bandpass calibrated, converted to LSRK velocity, and the appropriate brightness temperature calibration applied. Gaussian fits were performed, and
a lookup table of fit components made for all detected lines. 
These models are subtracted appropriately from each affected bandpass- and brightness-temperature-calibrated spectrum in the main dataset. 

\subsection{Spectral Baseline Correction}
\label{baselining}

\begin{figure*}
\begin{center}
\includegraphics[scale=0.55]{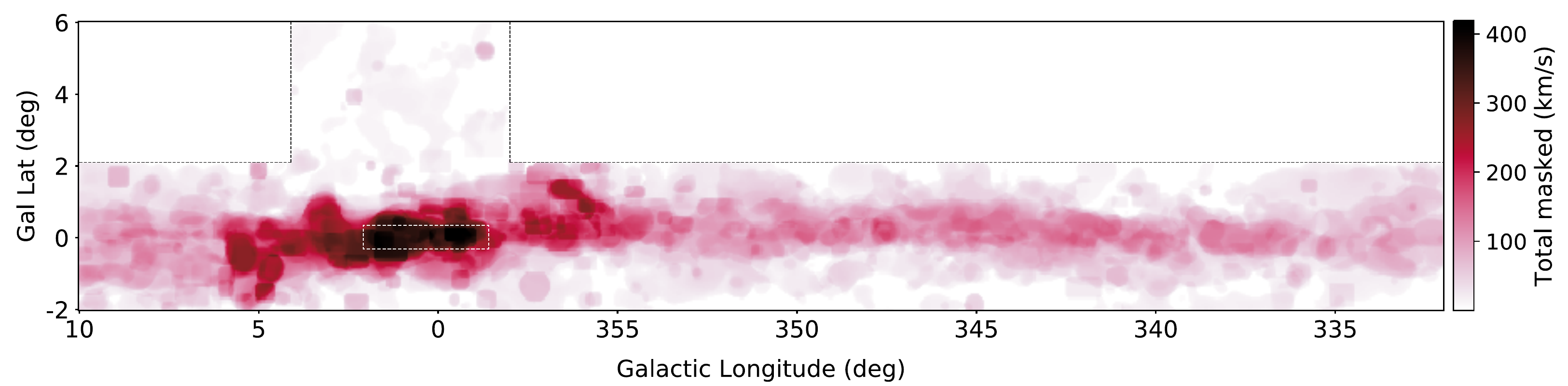}
\end{center}
\vspace{-0.2cm}
\caption{The 3D \textsc{Duchamp} mask used in baselining the 1720\,MHz line, summed in the velocity dimension. The colour scale indicates the total number of signal-bearing channels (in velocity units) masked out in the second-pass baselining step described in Section \ref{baselining_2}. For reference, 400\,\kms\ corresponds to $\sim30\%$ of the useable bandpass in this line. The masks are almost identical for the four lines, with the notable exception of some extra masked regions of maser emission at 1612\,MHz. Note that for the main lines (which both fall in the same IF), up to $\sim60\%$ of the useable bandpass may be masked. (See also Figure \ref{fig_baselining}.) The white dashed rectangle marks the approximate extent of the CMZ and GC. As discussed in the text (Section \ref{manual_correction}), no additional manual baseline corrections were attempted within this region.}
\label{fig_duchamp_map}
\end{figure*}

Correcting for residual baseline structure in the SPLASH dataset poses a significant challenge. After bandpass calibration, the spectral baselines contain residual structure on scales ranging from $\sim$200--8000 channels ($\sim$36--1440 \kms). In comparison, real signal may occupy $\sim$2--1500 contiguous channels ($\sim$0.36--270 \kms), where the larger end of this range arises from broad absorption in the Galactic Centre (GC) and Central Molecular Zone (CMZ). 
Furthermore, the OH lines can be very weak, with much of the emission/absorption at the limits of detectability. 
Clearly this presents some problems: distinguishing real signal from baseline structure can be difficult, yet the quality of the baseline correction we can achieve (and therefore the reliability of the final line profiles) depends strongly on our ability to do this well -- to identify and exclude real signal from any model solutions. There is also a fundamental limit to how well we can model baseline structure under very wide lines, where a baseline model will be relatively unconstrained. Baselining of the SPLASH dataset is therefore carried out in an iterative fashion, and in several stages, as described below. 

\subsubsection{First-pass Line Finding \& Rough Baselining}
\label{baselining_1}

Signal identification must ultimately be performed on a gridded cube, where 3D information is available, and the signal-to-noise ratio is maximised. 
This requires that we make a reasonable first-pass baseline correction for each spectrum prior to gridding. For this initial step, we use \textsc{Asap}'s native line-finder to identify the brighter lines in each four-second map integration, and perform a 5th order polynomial baseline correction over the full spectral channel range (with bad edge channels flagged). One difficulty is that the residual baseline structure is greatly amplified where continuum is strong, and in such locations can easily exceed typical OH line strengths elsewhere in the dataset. 
The pipeline therefore uses an iterative approach that begins with a sensitive search, but then tests for overflagging of baseline ripple and adjusts the line finder parameters until the flagging of overly large swathes of spectra is eliminated. The process is not perfect, but the relatively-low order polynomial ensures that the baseline solutions are not pathologically corrupted by the inclusion of some real signal (whether from line-wings and in completely-missed weak lines). Extreme RFI is also identified and removed at this stage by completely flagging any spectra for which the fit residuals exceed a threshold value. 
The resultant 
baseline solutions 
are sufficient to produce a cube on which 3D source-finding can be performed. These cubes are made according to the process described in Section \ref{gridding}, which includes outlier filtering to remove remaining RFI. 

\subsubsection{3D Source Finding and Master Mask Construction}

\begin{figure}
\begin{center}
\includegraphics[scale=0.45]{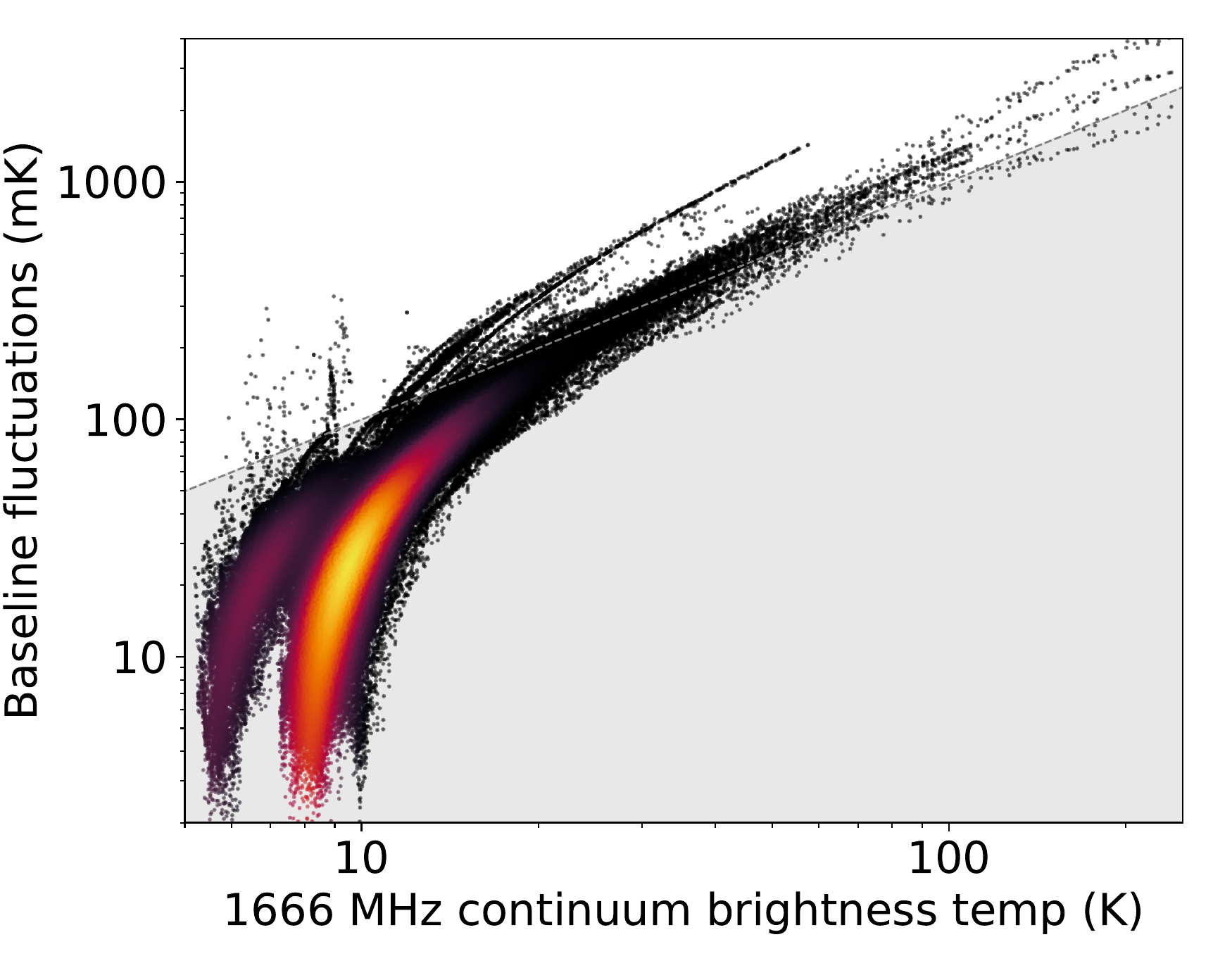}
\end{center}
\vspace{-0.3cm}
\caption{Amplitude of the residual baseline fluctuations subtracted during automatic baselining (see Section \ref{baselining_2}), plotted as a function of 1666\,MHz continuum brightness temperature. The y-axis shows half of the difference between the maximum and minimum points on the mean baseline fit at each gridded pixel, over the whole bandpass, excluding edge channels. The subtracted fluctuations for all three IFs are plotted together. The colour scale indicates the density of datapoints. The grey shaded region delimited by the grey dashed line shows the region where the baseline fluctuations are $<$1 per cent of the continuum brightness temperature; this contains 96 per cent of the data.}
\label{fig_auto_baseline_scatter}
\end{figure}

\begin{figure*}
\begin{center}
\includegraphics[scale=0.53]{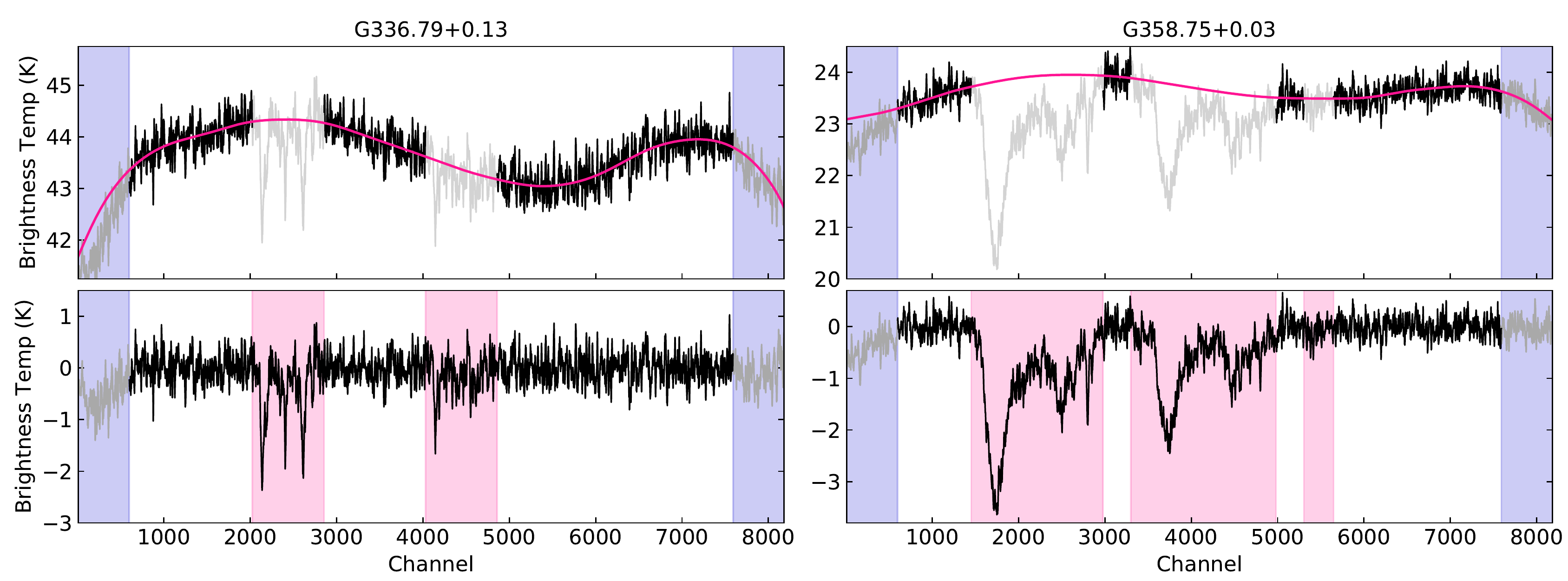}
\end{center}
\caption{Examples of baseline fitting and subtraction for single integrations of an OTF map. These plots show the full 8MHz bandwidth of the 1666 MHz band, which contains both the 1667.359 line (channels $\sim$1500--3000) and the 1665.402 MHz line (channels $\sim$3500--5000). The top panels show the bandpass- and flux-calibrated data prior to baselining, and the bottom panels show the baseline-subtracted data, both smoothed with a 10-channel boxcar kernel. The solid dark pink lines on the upper panels show the baseline models -- a combination of polynomial fits and low-pass filtering, as described in the text. Signal identified in the 3D datacubes by the \textit{Duchamp} source finder is shown as pink bands on the lower plots, and the equivalent spectral ranges greyed out on the upper plots; these ranges are excluded from the baseline model fitting. Both example positions shown here are continuum-bright, meaning that the OH absorption signal is strong enough to be seen even in these short integrations, and that residual baseline structure is fairly severe.} 
\label{fig_baselining}
\end{figure*}

We next use the \textsc{Duchamp} 3D source-finding package \citep{whiting12} to construct a 3D master mask of all emission and absorption from the roughly-baselined cubes. \textsc{Duchamp} searches for groups of connected voxels that lie above a defined threshold, and can produce a mask of these islands of emission. The user controls a large number of parameters such as minimum number of pixels or voxels, sub-regions over which to compute the spectral noise, an initial signal-to-noise (S/N) ratio required for detection, the S/N floor to which detected emission is grown to, and how neighbouring islands are merged. Since \textsc{Duchamp} does not perform well on cubes with varying noise levels, we create S/N cubes by dividing by a noise map computed from signal-free portions of the spectra. We also perform our own smoothing in both space and velocity before calling \textsc{Duchamp}. Since the algorithm only recognises emission, we invert the SPLASH cubes and repeat the process for absorption detections. 

A master $l$-$b$-$v$ mask for all four lines is constructed from a combination of the emission and absorption masks generated from the 1667, 1665 and 1720 MHz cubes, under the assumption that signal is present at some level in all lines if detected in any one. This is generally a good assumption for quasi-thermal OH, and while it can break down for compact, high-gain masers, 
the majority of main-line masers are associated with star formation, and are therefore coincident with quasi-thermal OH in any case. (Isolated 1720 MHz masers are rare enough that they are not a concern here.) 
The signal-finding parameters are chosen so that weak signal is recovered, at the cost of overflagging in places where the rough baseline solutions are poor (corrected as described below).
The 1612 MHz cube is not used in generating the master mask since it contains a large number of evolved star masers without counterparts in the other lines. A separate 1612 mask is generated and combined with the master mask for this frequency band. 

After removing all regions outside permitted Galactic velocities, the masks are inspected carefully by eye, and corrections made. Most commonly this involves the removal of residual baseline structure erroneously identified as signal -- easily identified as characteristic ``streaks'' in $l$-$v$ space coincident with locations of bright continuum -- but also includes some manual addition or removal of features, particularly in the CMZ and Galactic Centre, where the rough baseline solutions were poor. The most challenging cases to judge are where \textsc{Duchamp} identifies broad/weak signal only in the 1667 MHz line, that cannot be verified by comparison with the other three frequencies. In these cases we carefully inspect the non-baseline-corrected cubes, and also compare with the $^{12}$CO(J=1--0) data of \citet{dame01}. Generally a feature is retained if any \textit{one} of the following are true: (a) it can be discerned by eye in non-baselined spectra, (b) it appears to form part of the expected Galactic $l$-$b$-$v$ structure, (c) it is present in the $^{12}$CO line. A velocity-integrated image of the final corrected mask for the 1720\,MHz line is shown in Figure \ref{fig_duchamp_map}. 

\begin{figure*}
\begin{center}
\includegraphics[scale=0.48]{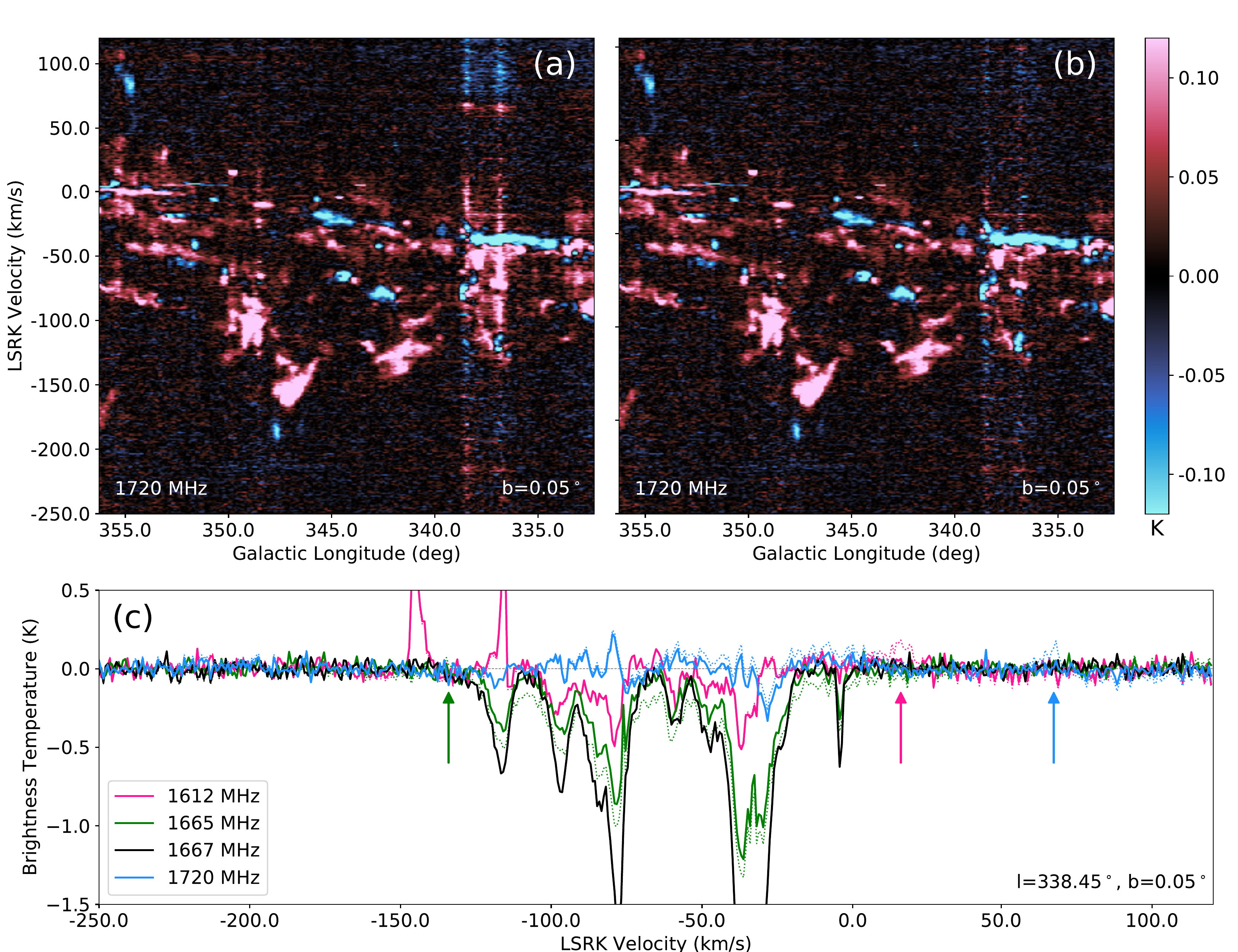}
\end{center}
\caption{Example longitude-velocity maps and spectra showing SPLASH data before and after the final manual baseline correction step described in Section \ref{manual_correction}. Panel (a) shows a single latitude slice at $b=0.05^{\circ}$ (with a pixel width $0.05^{\circ}$) of the 1720 MHz cube prior to correction. Significant striping can be seen around $l\approx336$--338.5$^{\circ}$, where the automatic baseline solutions have left some residual baseline structure towards a bright H{\sc ii} region complex. Narrow, line-like correlator artefacts can also be seen at $v\approx65$\kms. Panel (b) shows the same image post-correction. The noise level in the problem region remains elevated due to the higher system temperature associated with the bright continuum emission, but the residual baseline structure is greatly reduced. Minor striping has also been cleaned in some other regions of the image. Panel (c) shows a spectrum in all four transitions at one of the corrected positions. The dotted coloured lines show the pre-corrected data, and the solid coloured lines show the corrected spectra. The scale is chosen to highlight the baseline corrections. The coloured arrows indicate the location of the correlator artefacts in the three bands. The data displayed here is binned by a factor of 4 in the velocity axis, for a channel width of 0.7 \kms.}
\label{fig_before_after}
\end{figure*}

\subsubsection{Improved Baselining of the Pre-Gridded Data}
\label{baselining_2}

The $l$-$b$-$v$ masks of astrophysical emission and absorption generated in the previous step are used to generate spectral masks for each 4-second (bandpass-calibrated, polarisation-averaged) integration. The native \textsc{Asap} linefinder is now used (with appropriate parameters) to flag only RFI spikes that might otherwise corrupt the baseline solutions. 

We use a combination of polynomial fitting and low-pass filtering to produce baseline models. The polynomial fit is performed first, and captures the large-scale curvature, which is particularly important to represent well under broad spectral features. A seventh degree polynomial produces a visually good fit for most spectra (and higher orders produce negligible improvements in the fit residuals). But we switch to fifth or third order functions when many channels (3000--3800 and $>3800$ respectively) are masked and the model is unconstrained over a large fraction of the bandpass. 
These channel ranges correspond to $\sim$43--65\% and $\gtrsim$ 55--65\% of a full spectrum, excluding bad band edges. 
Such wide line masks are only found in the CMZ and Galactic Centre. 

The polynomial-subtracted spectrum is then low-pass filtered using a Gaussian smoothing kernel with $\sigma=200$ channels ($\sim36$\,\kms), to capture smaller scale structure. Where signal has been masked, we represent the data by a straight line between the median brightness temperatures at the edges of the flagged spectral data. The windows over which these medians are computed are set to 200 channels by default, and are allowed go as low as 50 channels in places where fewer than 200 unmasked channels are available. If fewer than 50 unmasked channels are present, neighbouring line masks are merged. The smoothing kernel is more responsive than the polynomial, and corrects most of the remaining baseline structure outside of masked regions. Where lines are masked, and the kernel can only respond to the information available at the mask edges, the solution tends towards the straight line drawn across the ``gap'' in the spectrum. While it cannot properly model the baseline shape underneath masked spectral lines, the smoothing step clearly improves the solutions in most cases, particularly in places where the polynomial curve has visibly over- or under-fit masked regions. 

Figure \ref{fig_auto_baseline_scatter} shows the characteristic amplitude of the baseline corrections performed at this stage as a function of 1666\,MHz continuum brightness temperature, defined as half the difference between the maximum and minimum points on the mean baseline fit (excluding edge channels). This is measured on a pixel-by-pixel basis from the gridded cubes of the baseline solutions, processed as described in Section \ref{cont_images}. The median amplitude of the corrections is $\sim40$\,mK in the 1612 and 1720\,MHz bands, and $\sim50$\,mK in the 1666\,MHz band. A strong correlation with the continuum brightness is evident above $\sim15$\,K. Figure \ref{fig_baselining} shows two examples of the baseline fits and line masks for two 4-second integrations towards different positions within the survey region.

\subsubsection{Final Baseline Correction \& Correlator Artefact Removal}
\label{manual_correction}

Once the baselined data are gridded into cubes (see section \ref{gridding}), a final manual correction step is performed towards positions where spectra show obvious residual baseline structure. This generally occurs in regions of bright continuum, and/or where a wide \textsc{Duchamp} mask has left insufficient information to constrain the fit, and manifests either as characteristic `streaks' in $l$-$v$ space, or as broad pedestals/troughs in one transition that are not mimicked in the others. (Although we note that for the 1667 MHz transition -- the strongest of the four -- a lack of counterparts in the other lines is not necessarily indicative of baseline problems, and is not treated as such; see e.g. \citealt{busch21}.) In these cases, line-free regions are masked more precisely by eye and an additional baseline solution generated by low-pass filtering with a Gaussian smoothing kernel of $\sigma=100$ channels ($\sim$18 \kms), following the method described above. Approximately 8 per cent of positions are corrected this way in at least one transition. The cubes of additional model corrections are smoothed with a Gaussian kernel of FWHM 3 pixels and subtracted directly from the survey datacubes.  We do not perform any additional corrections in the region of the Galactic Centre and CMZ ($-1.5 \lesssim l \lesssim +2.0^{\circ}, |b| \lesssim 0.3^{\circ}$), where the baseline solutions are too poorly constrained to justify further corrections. 

The residual baseline structure also includes spurious, line-like emission features, approximately Gaussian in shape, centred on channel 4096 in each of the three DFB bands. These are present throughout the data, but are significantly below the noise unless the background continuum is bright. We automatically fit and subtract one- or two-component Gaussian functions in all cases where the artefacts are well-separated from real OH signal. (The two-component fits are needed when data for a single map was taken in widely-spaced epochs such that channel 4096 had drifted in velocity). In cases where the spurious features are close to or blended with real signal, we utilise the fact that the shapes and strengths are consistent between the three bands (and the velocity differences between them are known exactly) to map the fit solutions from one band to another. 
We perform this correction process across the whole cube, whenever the velocity range of the artefacts is signal-free in at least one transition. This excludes the Galactic Centre, the CMZ and some sub-regions of broad emission to the positive side of $l$=0$^{\circ}$. The only uncorrected areas significantly affected are the CMZ and Galactic Centre (due to their bright continuum), towards which the baseline quality in any case is poor.

Figure \ref{fig_before_after} shows an example longitude-velocity slice in the 1720 MHz data before and after these final manual correction steps, together with example spectra in all four transitions.

\subsection{Gridding of the Spectral Datacubes}
\label{gridding}

The \textsc{Gridzilla} package \citep[described in][]{barnes01} was used to produce spectral datacubes and continuum images from the calibrated data. \textsc{Gridzilla} grids only in the spatial dimension, and expects the input spectra to be on a consistent frequency grid to within a certain tolerance. Prior to gridding, the spectra in each individual map file are shifted into the LSRK frame and resampled onto a fixed frequency grid in \textsc{Asap}. The grid is unique to each map file, with the exact frequency-to-channel registration and channel width determined by the Doppler factor for that position at the epoch of observation. \textsc{Gridzilla} then constructs a fiducial frequency grid that can accommodate as many of the input spectra as possible to within the specified tolerance, and performs the conversion to velocity using the radio definition of the Doppler formula, $v=c(f_0-f)/f_0$, where $f_0$ is the line rest frequency. For a velocity range of $\pm$300\,\kms, a tolerance of 0.6 channels ($\sim$0.1\,\kms) is sufficient to accommodate all input spectra. 

The final velocity resolution and velocity accuracy of the spectral cubes are affected by several factors. The initial resampling of each raw map file to a 
common frequency axis introduces some additional correlation between neighbouring channels, as well as a slight degradation in the frequency resolution. 
At the gridding step, \textsc{Gridzilla} must then align all input files to a fiducial grid, without resampling in the velocity dimension. (Note that this is a limitation of the \textsc{Gridzilla} package.) The fractional difference in channel width due to epoch and directional variations in the Doppler factor is very small -- $\lesssim0.01$\%. For a velocity range of $\pm$300\,\kms ($\sim\pm$1650 channels), this can result in a maximum misalignment of up to $\sim$0.15 channels by the edges of the bands. In addition, the frequency registration of each input file's central channel is essentially unconstrained, necessitating a tolerance of 0.5 channels to allow \textsc{Gridzilla} to shift and align each input spectrum. Together these effects determine the required tolerance of 0.6 channels, and degrade the frequency resolution of the gridded data by $\sim$1 raw channel ($\sim$0.18\kms). 
Finally, the assumption of a linear mapping of frequency to velocity under the radio definition of the Doppler formula becomes worse the further the line is shifted from the rest frequency. The maximum error at the the band edges, assuming a velocity range of $\pm300$\,\kms, is $\mp$0.14\,\kms, or approximately 0.8 channels. Given that quasi-thermal OH lines are rarely narrower than a few \kms\, these affects are unimportant to the intended science. For narrow maser lines, they could conceivably affect measured velocity profiles, but are not severe enough to affect our ability to cross-match sources between different datasets. 
In any case, dedicated interferometric follow-up observations of SPLASH maser sources are published separately in \citet{qiao16,qiao18,qiao20}. 

The pixels in each velocity plane are populated from a weighted mean of all nearby datapoints; a user-defined gridding kernel determines the weights 
as a function of distance from the target pixel. 
SPLASH uses a truncated Gaussian kernel with a FWHM of 10 arcminutes and cutoff diameter of 20 arcminutes, which provides a good balance between resolution and sensitivity, while minimising departures from Gaussianity in the final effective beam. A gridded pixel size of 3 arcminutes was chosen to oversample both the Parkes beam and the kernel. Prior to computing the weighted mean, outliers are filtered using \textsc{gridzilla}'s built-in outlier censoring. This uses the weighted median to compute the robust measure of spread, $S_n = 1.1926~\mathrm{med}_i\{\mathrm{med}_j|x_i-x_j|\}$ \citep{Rousseeuw93}, and rejects all values outside $3\times S_n$, achieving the robustness of the weighted median statistic, while retaining the linearity and efficiency of the simple mean. 
Since this step combines all data from the ten individual passes of each region, it filters out any RFI not caught and excluded in the baselining step, which manifests as statistical outliers at a single given epoch and map position. 

\begin{figure*}
\begin{center}
\includegraphics[scale=0.35]{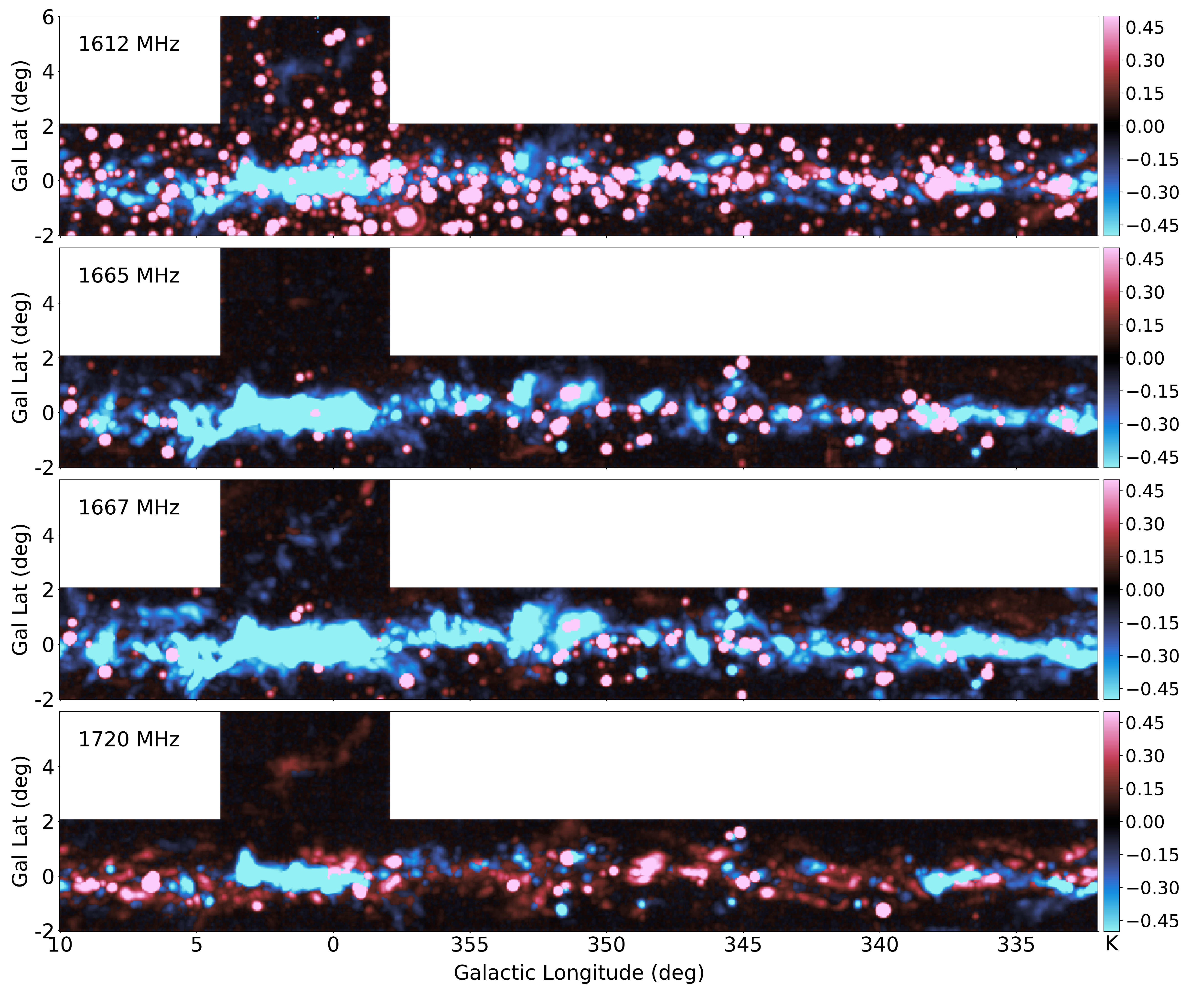}
\end{center}
\caption{Combined peak emission/absorption maps for all four transitions over the full SPLASH survey region. The most extreme value of the brightness temperature at each spatial position is shown. These plots capture some of the main trends in the dataset, including the large number of masers in the 1612\,MHz line (mostly evolved stellar sources), the tendency of the main lines to be seen in absorption, and the 1720\,MHz line to be seen in emission. Velocity channel maps are presented in the online Appendices.}
\label{fig_maxminmaps}
\end{figure*}

The data for the Galactic Centre region (see Figure \ref{coverage_fig}) -- taken with higher attenuator settings to avoid saturation -- is processed through \textsc{gridzilla} separately. The resulting mini-cubes are then substituted into the main datacubes by replacing the data over the relevant area ($l$ = 359.0 to 0.9, $b$ = -0.5 to 0.35), with a Gaussian smoothing function of FWHM 6 arcmin (2 pixels) used to generate a weighting mask at the overlap region. The brightness temperature agreement between the original and replacement data is within a few percent over the majority of the region, with the exception of Sgr~A, where the saturation in the original cubes is severe.

\begin{figure*}
\begin{center}
\includegraphics[scale=0.53]{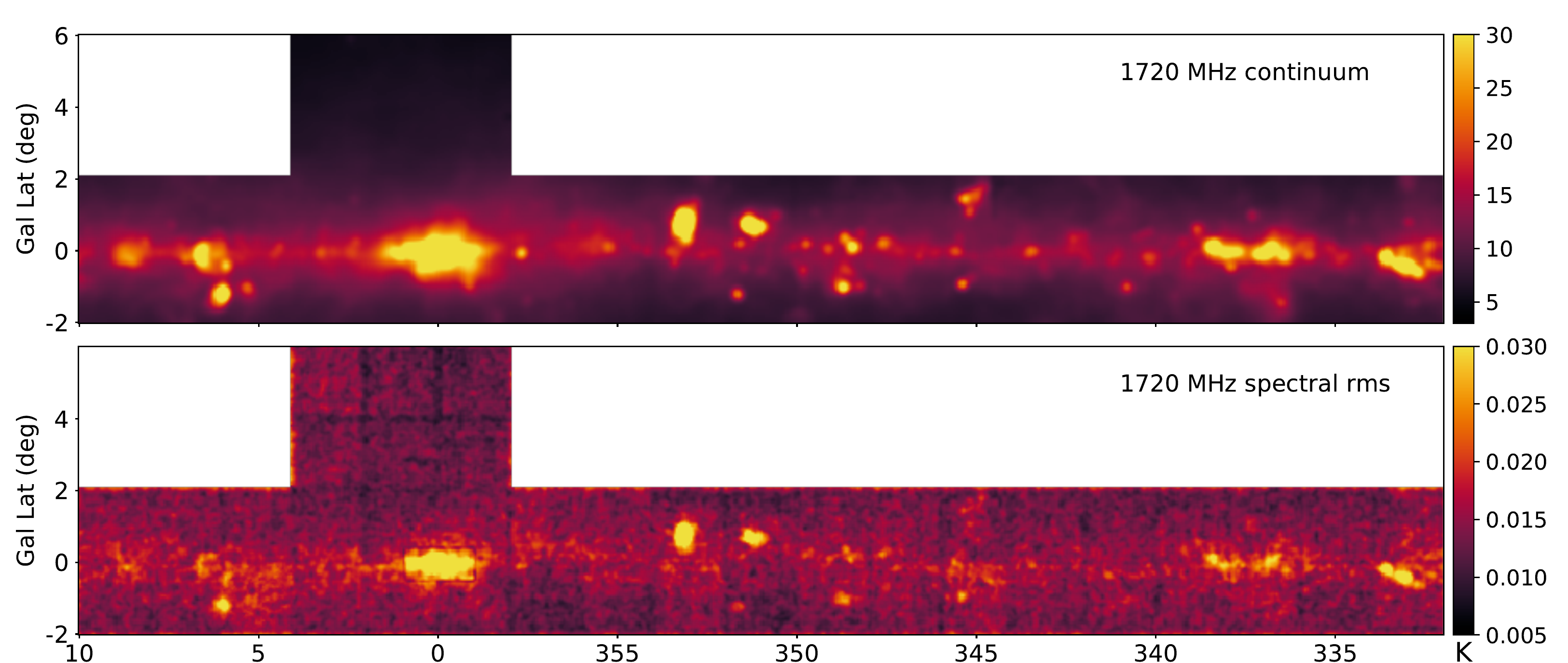}
\end{center}
\caption{Upper panel: Final 1720\,MHz continuum map in main beam brightness temperature units. The data here have been derived from the spectral baselines, averaged over $\sim7$ MHz of bandwidth, and corrected for reference position continuum as described in the text. Bottom panel: spectral rms noise in the 1720\,MHz transition, as derived from line-free portions of the final cube. The effective velocity resolution of this data is $\sim$0.9 \kms (see text). These maps are representative of the data in all four transitions. Note that the dark striping seen at the intersection of tiles indicates reduced noise where the edges of adjacent tiles overlap.}
\label{fig_cont_rms_maps}
\end{figure*}

The final cubes are binned up by a factor of two in the velocity axis and then 3-point hanning smoothed, reducing the noise by a factor or two. The resulting data is Nyquist sampled with 0.35 \kms channels, with an effective velocity resolution of $\sim0.9$ \kms (slightly degraded from the ideal 0.70 \kms due to the factors discussed above). Figure \ref{fig_cont_rms_maps} shows an example map of the per-channel rms noise for the 1720\,MHz datacube. The $1\sigma$ spectral rms noise is $\lesssim16$\,mK for $\sim75$ per cent of positions and $\lesssim20$\,mK for $\sim95$ per cent of positions, and only exceeds 30\,mK towards the 1 per cent of pixels with the strongest continuum emission. Note that since the rms is measured in signal-free portions of the spectral cube where the baseline fits are generally good, it cannot appropriately capture uncertainties due to baseline quality in velocity channels with emission or absorption. This is discussed separately in Section \ref{spec_uncertainty}.

Measurements of the effective resolution of the gridded data are made by fitting 2D Gaussian profiles to bright unresolved masers. Since outlier filtering artificially narrows the point-spread-function of bright compact sources, these measurements are performed on a cube without outlier filtering applied. The effective HPBWs are found to be $15.5\pm{0.2}$, $15.4\pm{0.3}$ and $15.3\pm{0.2}$ arcmin, for the 1612, 1665/1667 and 1720\,MHz lines, respectively (quoting means and standard deviations). This is very close to the theoretical resolution of 16.0, 15.8 and 15.6 arcmin, obtained by convolving ideal Gaussian beams with the truncated Gaussian kernel. 

The final spectral line datacubes are shown in Figure \ref{fig_maxminmaps}, which collapses the data along the velocity dimension by plotting the most extreme value of the brightness temperature at each spatial position. Individual channel maps are provided in the online Appendices. 

\subsection{The Continuum Images}
\label{cont_images}

Continuum images at 1612, 1666 and 1720\,MHz are formed from the baseline fits subtracted from the spectral data in section \ref{baselining_2} (excluding the manual corrections described in \ref{manual_correction}). The baseline fit has an average level that is the difference between the brightness temperatures of the on-source position and the off-source reference position. Fluctuations across the spectral bandpass, even when ``severe'' from the point of view of spectral baseline structure, are a small fraction of the absolute continuum level: $<1$ per cent for 96 per cent of positions (see Figure \ref{fig_auto_baseline_scatter}). The off-source reference positions are never more than 4 degrees offset in elevation from the on-source positions, and are observed within a few minutes in time, so the difference in spill-over and atmospheric contributions to the system temperature is small. However, since the reference positions are close to the Galactic Plane they all have significant (and differing) continuum brightness temperatures. The raw on-source continuum levels extracted from the baseline solutions are therefore too low by an amount $T_\mathrm{C,OFF}^*$, which cannot be recovered directly from the SPLASH data. Note that here, and throughout the rest of the paper, we use the asterisk to denote explicitly a measured, beam-averaged quantity. 

We estimate $T_\mathrm{C,OFF}^*$ at the three observed frequencies using the same method as for the SPLASH pilot region \citep[described in][]{dawson14}. The method uses CHIPASS data at 1395 MHz \citep{calabretta14} and S-PASS data at 2300 MHz \citep{carretti19}, and interpolates between them to obtain the reference position continuum level at 1612, 1666 and 1720\,MHz. The CHIPASS data are on a full beam temperature scale, are absolutely calibrated, and include the 2.7\,K cosmic microwave background (CMB). 
S-PASS is also on a full-beam temperature scale  
and is absolutely calibrated for Galactic emission only, not including the CMB \citep{carretti19}. We therefore subtract 2.73\,K from CHIPASS and rescale to a main-beam temperature scale using a main beam efficiency of $\eta_{\mathrm{mb}}=0.60\pm0.08$. As in \citet{dawson14}, this value and its uncertainties were estimated from the scaling applied in the CHIPASS survey paper (which implies $\eta_{\mathrm{mb}}=0.52$) and the $\eta_{\mathrm{mb}}=0.69$ measured for the central beam of the Parkes multibeam receiver at 1.4\,GHz \citep{staveley96}. The S-PASS data are smoothed to match the lower resolution of the CHIPASS data, and a spectral index map produced from the two, noting that Galactic continuum emission includes both synchrotron ($\alpha \sim -0.7$ where $S_{\nu} \propto \nu^{\alpha}$, so $\beta \sim -2.7$ where  $T_{\rm C} \propto \nu^{\beta}$) and thermal free-free components ($\alpha \sim -0.1$, $\beta \sim -2.1$). We find that measured spectral indices at the off-Plane reference positions range from $\beta=-2.48$ to $-$2.75, with a mean and standard deviation of $-$2.61 and 0.08, respectively.

The Galactic emission at the reference positions is then computed from this image and the original CHIPASS data, and the 2.73\,K CMB added back in to provide absolute continuum levels. These values of $T_\mathrm{C,OFF}^*$ are then added to the retained baseline fits via a lookup table for the appropriate reference position used for each $2\times2$ degree tile. $T_\mathrm{C,OFF}^*$ ranges between 5.88--9.62, 5.63--9.05 and 5.40--8.53\,K for the 1612, 1666 and 1720\,MHz bands, respectively. 
The corrected data is gridded as described above, but without outlier filtering, and then collapsed in the frequency domain with edge channels excluded to obtain final images for the three frequency bands. An example final continuum map is shown (for 1720\,MHz) in Figure \ref{fig_cont_rms_maps}.

\subsection{Comments on Brightness Temperature Accuracy}

The basic relation by which the spectral and continuum data are obtained can be written as
\begin{equation}
S_\mathrm{sp}^*(\nu) + (S_\mathrm{sys,ON}-S_\mathrm{sys,OFF})=\frac{P_{\mathrm{ON}}(\nu)-P_{\mathrm{OFF}}(\nu)}{P_{\mathrm{OFF}}(\nu)}~S_{\mathrm{sys,OFF}},
\label{eq:bandpass2}
\end{equation}
where $S_\mathrm{sp}^*(\nu)$ is the measured spectral line flux density, $P_{\mathrm{ON}}(\nu)$ and $P_{\mathrm{OFF}}(\nu)$ are the on-source and off-source power, $S_{\mathrm{sys,OFF}}$ and $S_{\mathrm{sys,ON}}$ are the on- and off-source system temperatures (in Jy). $S_{\mathrm{sys}}$ in this formulation explicitly includes the contribution from the continuum, $S_\mathrm{C}^*$ such that
\begin{equation}
S_\mathrm{sys,ON}-S_\mathrm{sys,OFF}=S_\mathrm{C,ON}^*-S_\mathrm{C,OFF}^* + \Delta S_{\mathrm{sys}},
\end{equation}
\noindent where $\Delta S_{\mathrm{sys}}$ is the non-astrophysical difference in the system temperature at the on- and off-source positions (for example due to changes in atmospheric contributions or spillover).

As described in Section \ref{intensity_cal}, the flux density scale is tied to the standard calibrator 1934-638 \citep[specifically the flux density model presented in][]{reynolds94}. While the day-to-day system-stability is excellent, the absolute flux density calibration of the 1934$-$638 model is likely to be accurate to within $\sim5\%$ \citep[e.g.][]{ott94}, leading to a systematic uncertainty on our brightness temperature scale. 
The conversion to main-beam brightness temperature is a performed for each frequency band from the measured beam FWHM under the assumption of a Gaussian beam profile: 
\begin{equation}
T_\mathrm{b}^*(\nu) = G_\mathrm{mb}^{-1} ~S^*(\nu) = \frac{\lambda^2}{2k \Omega} S^*(\nu),
\label{eq:snutotb}
\end{equation}
\noindent where $G_\mathrm{mb}$ is the ideal main-beam gain computed in Section \ref{intensity_cal} and $\Omega$ is the beam solid angle. While the measured beam profiles are very well fit by Gaussian functions (as seen in Figure \ref{beamfits}), small deviations could propagate through to the beam solid angle, as could uncertainties (at the few percent level) in the FWHMs themselves. 

Stray radiation (signal entering via the sidelobes) introduces some additional uncertainty in the measured main beam brightness temperatures. We mention it here for completeness, but do not attempt a correction. The effect is most problematic for H{\sc i} surveys, where emission fills the whole sky, and weak off-Plane signal can be contaminated by stray emission from the bright Galactic Plane \citep[e.g.][]{kalberla10}. In contrast, the OH signal is well-confined to the Plane, and we lack both the coverage and the signal-to-noise to detect weak off-Plane emission, should it be present. 

\begin{figure}
\begin{center}
\includegraphics[scale=0.42]{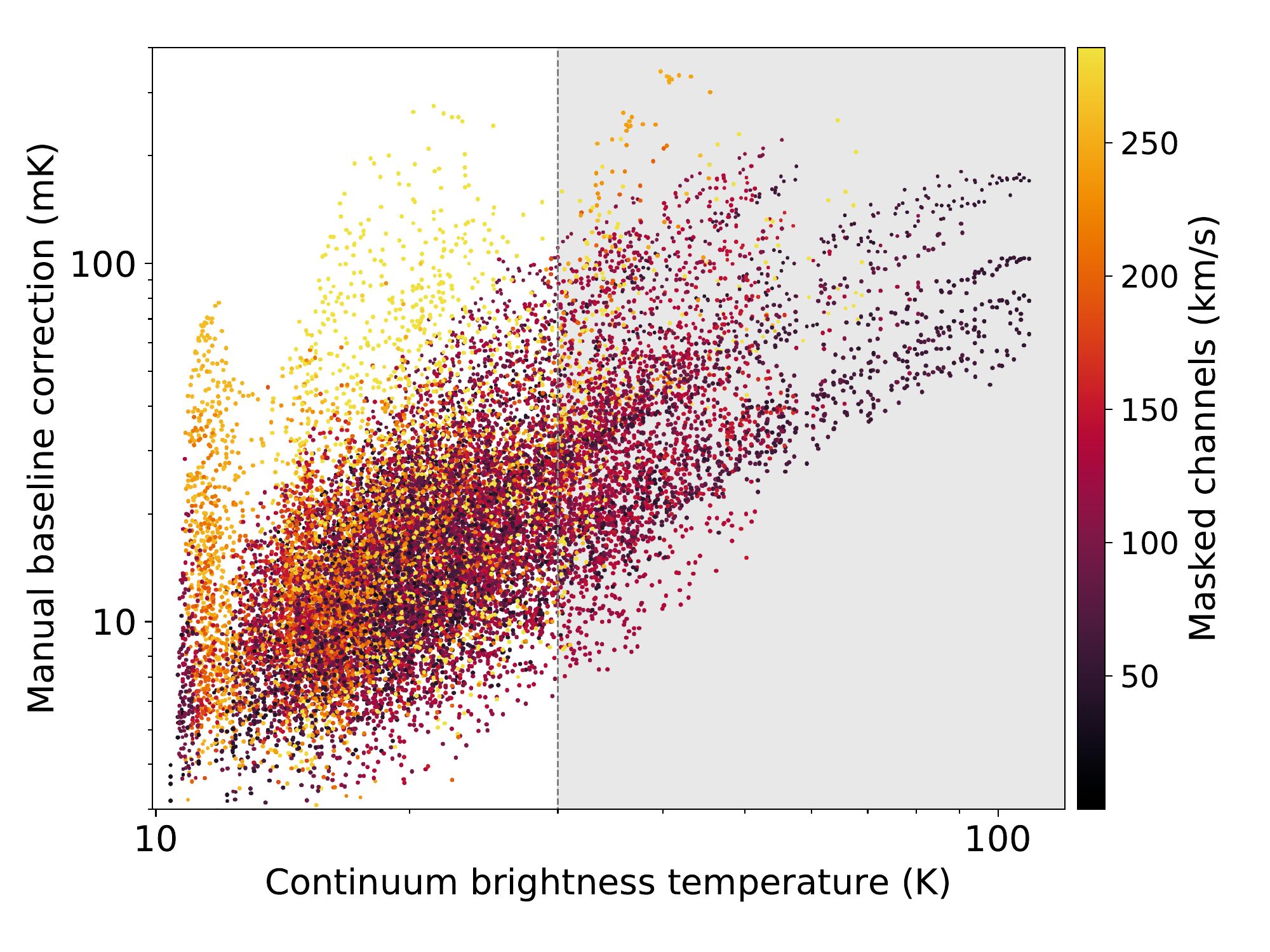}
\vspace{-0.3cm}
\end{center}
\caption{Amplitude of the additional baseline models subtracted during manual baselining (see Section \ref{manual_correction}), plotted as a function of 1666\,MHz continuum brightness temperature. The y-axis shows the maximum absolute value of baseline fit for each line at each gridded pixel, and all transitions are plotted together. The colour scale indicates the total channels in the \textit{Duchamp} mask for each position, in units of \kms. The grey shaded region delimited by the grey dashed line shows the continuum brightness temperature (30K) above which \textit{all} positions in the dataset (outside of the CMZ and GC) had a manual correction performed.}
\label{fig_manual_baseline_scatter}
\end{figure}

\subsubsection{$T_\mathrm{b}$ Uncertainty in the Spectral Line Datacubes}
\label{spec_uncertainty}

For spectral line data, the $(S_\mathrm{sys,ON}-S_\mathrm{sys,OFF})$ term in equation \ref{eq:bandpass2} is subtracted during baseline correction, meaning that the spectral line cubes are unaffected by differences in the on- and off-source system temperatures. Comparisons of individual passes of the same 2-by-2 degree map tiles demonstrate an excellent agreement between data taken at different times, elevations and epochs. 

The main source of uncertainty in the spectral line cubes is the quality of the baseline models, particularly where the signal is broad. This is difficult to quantify in a robust way (the real residual is by its nature unrecoverable), but some insight can be gained from the amplitude of the manual baseline corrections performed in Section \ref{manual_correction}. 

Figure \ref{fig_manual_baseline_scatter} shows the maximum absolute value of the additional baseline corrections as a function of the 1666\,MHz continuum brightness temperature, coloured according to the total number of masked channels. The median values are 16, 20, 12, and 21\,mK for the 1612, 1665, 1667 and 1720\,MHz lines respectively. Outliers of up to $\sim100$\,mK occur for bright continuum and/or spectra where the total mask was $\gtrsim200$\,\kms. Considering only ``good'' positions where $T_\mathrm{b,C}^*(1666) < 20$\,K and $v_\mathrm{msk}<200$\,\kms\ (representative of 90 per cent of the survey data outside of the CMZ and GC), brings these median values down to 10, 12, 8, and 13\,mK, respectively. We therefore consider $\lesssim20$\,mK a reasonable estimate of the baseline fidelity towards such positions in the survey as a whole. Of the remaining 10 per cent of ``bad'' positions with broad lines or bright continuum, half have been corrected in the manual correction step (including every position with $T_\mathrm{b,C}^*(1666)>30$\,K), and we include these in the $\lesssim20$\,mK uncertainty category. The remaining 5 per cent of uncorrected broad-signal positions are generally within $\pm$5 degrees of the Galactic centre, and were excluded because they were difficult to reliably correct. Cubes of the manual corrections are made available in this data release.

\subsubsection{$T_\mathrm{b}$ Uncertainty in the Continuum Images}

The continuum images are generated from the subtracted baseline solutions for the spectral data -- equal to the $(S_\mathrm{sys,ON}-S_\mathrm{sys,OFF})$ term in equation \ref{eq:bandpass2}. The expression for the final continuum brightness temperature, $T_\mathrm{C}^*$, is 
\begin{equation}
T_\mathrm{C}^*=G_\mathrm{mb}^{-1}\times[S_\mathrm{C,ON}^*-S_\mathrm{C,OFF}^* + \Delta S_{\mathrm{sys}}]+T_\mathrm{C,OFF}^*,
\end{equation}
\noindent where $T_\mathrm{C,OFF}^*$ is the main-beam continuum brightness temperature at the reference position (not directly measurable), derived as detailed in Section \ref{cont_images}. The continuum data is thus sensitive to $\Delta S_\mathrm{sys}$, the non-astrophysical difference in system temperature between the on- and off-source positions, as well as any errors in the $T_\mathrm{C,OFF}^*$ estimates. 

Comparing individual passes of the same 2-by-2 degree map prior to the addition of $T_\mathrm{C,OFF}^*$, we find the per-pixel standard deviation for a given position is typically $\sim0.05$--0.6 K, which is within $10$ per cent of the mean brightness temperature for 99 per cent of pixels. 
The maximum possible elevation difference between on-source and off-source positions is $\sim4^{\circ}$. Skydip observations made during the observing period imply that the maximum continuum error introduced as a result of this difference is $\sim0.5$ K, for low elevations. However, we find that different instances of the same region observed at the same elevation can have quite different measured offsets from the mean, and we also find no systematic correlations between the measured values in individual maps and their LST, UTC, Julian date or the azimuth coordinate at the time of mapping. This suggests an effectively random element whose impact is mitigated in the final mean maps. 
\begin{table*}
\centering
\caption{Summary of key survey parameters. Where multiple values are quoted on a single line, they are in order of increasing frequency.} 
\label{table_obs}
\begin{tabular}{ll} 
\hline
\hline
Area Coverage & 332$^{\circ}$ $< l <$ 10$^{\circ}$, $|b|<$ 2$^{\circ}$ \\
& 358$^{\circ}$ $< l <$ 4$^{\circ}$, 2$^{\circ}$ $< b <$ 6$^{\circ}$ \\
Observed frequency bands & 1612, 1666, 1720 MHz \\
Raw beam FWHM & 12.6, 12.4, 12.2 arcmin \\
Main beam gains & 1.22, 1.25, 1.29  Jy/K \\ 
Raw bandwidth per frequency band & 8 MHz \\
Raw frequency channel width & 0.977 kHz \\
Brightness temperature scale & Main beam \\
Absolute calibration uncertainty & $\sim5\%$ (assumed uncertainty on 1934$-$638 flux density model).\\
\hline
Gridding kernel Gaussian FWHM & 10 arcmin \\
Gridding kernel cutoff diameter & 20 arcmin \\ 
Pixel size & 3 arcmin \\
Effective beam FWHM & $15.5\pm{0.2}$, $15.4\pm{0.3}$, $15.3\pm{0.2}$ arcmin \\
\hline
Spectral line rest frequencies & 1612.231, 1665.402, 1667.359, 1720.530 MHz\\
Raw velocity channel width & 0.18, 0.18, 0.18, 0.17 \kms \\
Gridded velocity channel width & 0.36, 0.35, 0.35, 0.34 \kms \\
Gridded velocity resolution & $\sim0.9$ \kms \\
LSRK velocity coverage & $\pm300$ \kms \\
Maximum velocity error at band edges & $\mp0.14$ \kms \\
Mean spectral rms noise & 15 mK (12--20 mK over 95\% of the survey area)\\
Spectral baseline fidelity (excluding CMZ \& GC) & $\lesssim20$\,mK over 95\% of positions\\
\hline
Continuum frequency bands & 1612, 1666, 1720 MHz \\
Estimated uncertainty (see text) & $\lesssim$ 2\% \\
CMB and diffuse background included? & Yes \\
\hline
\hline
\end{tabular}
\label{table:data_pars}
\end{table*}

Once $T_\mathrm{C,OFF}^*$ is added in and the data gridded, 
we perform a further check by comparing our maps to the 1612, 1666 and 1720\,MHz models derived by interpolating between CHIPASS and S-PASS data in Section \ref{cont_images} (here referred to as model A). Whereas previously these model maps were used only to estimate $T_\mathrm{C,OFF}^*$ at each reference position, we may also directly compare the model Galactic Plane continuum emission to that measured by SPLASH. We also perform a check on model A 
by generating a second continuum model (B) using the free-free component, assumed $\beta = -2.1$, from \citet[][also derived from CHIPASS]{alves+2015}, and derived residual synchrotron with $\beta = -2.7$, to extrapolate from CHIPASS to the SPLASH frequencies. 
The mean pixel-by-pixel offset between models A and B is 0.1 K with a standard deviation of $\sim 0.1$~K (or a ratio mean and standard deviation of 2 and 3 percent respectively). 


Comparing the SPLASH continuum images to model A 
we find that the SPLASH images have mean positive offsets of 0.05--0.17 K, (0.4--1.4 per cent) 
with a standard deviation of $\sim0.2$\,K. Much of this scatter arises from slight jumps across tile boundaries, providing us with a rough estimate of the error on $T_\mathrm{C,OFF}^*$ of around $\sim0.2$\,K. 
Overall, given that some of the errors will be in the model, not in the SPLASH data, we interpret this to suggest that the SPLASH data is accurate (with respect to the CHIPASS scale) at the $\sim 0.1$~K and $\sim 1$ percent level. 
We note that the uncertainty in the CHIPASS absolute intensity scale itself is $\sim30$\,mK \citep{calabretta14}, and in S-PASS is around $\sim70$\,mK. 
We adopt a final conservative uncertainty estimate for the SPLASH continuum data of around 2 per cent.




\subsection{Data Summary and Availability}
Table \ref{table:data_pars} summarises the key parameters of the SPLASH survey data products. FITS format spectral line datacubes for each of the four OH lines (1612.231, 1665.402, 1667.359 and 1720.530\,MHz), and continuum images for for the 1612, 1666 and 1720\,MHz bands, are publicly available via AAO Data Central\footnote{https://docs.datacentral.org.au/splash/}, 
Also included are auxilliary data products including spectral cubes of manual baseline corrections. Full-velocity-resolution spectral line cubes are available upon request. 

Note that while the CMZ and Galactic Centre are included in this data release, the spectral baseline fidelity in these regions is subject to much greater uncertainty than the $\lesssim20$\,mK appropriate to the rest of the survey. The continuum brightness temperature, on the other hand, is reliable, with the possible exception of the central pixels of Sgr A$^*$. 

\begin{figure*}
\begin{center}
\includegraphics[scale=0.52]{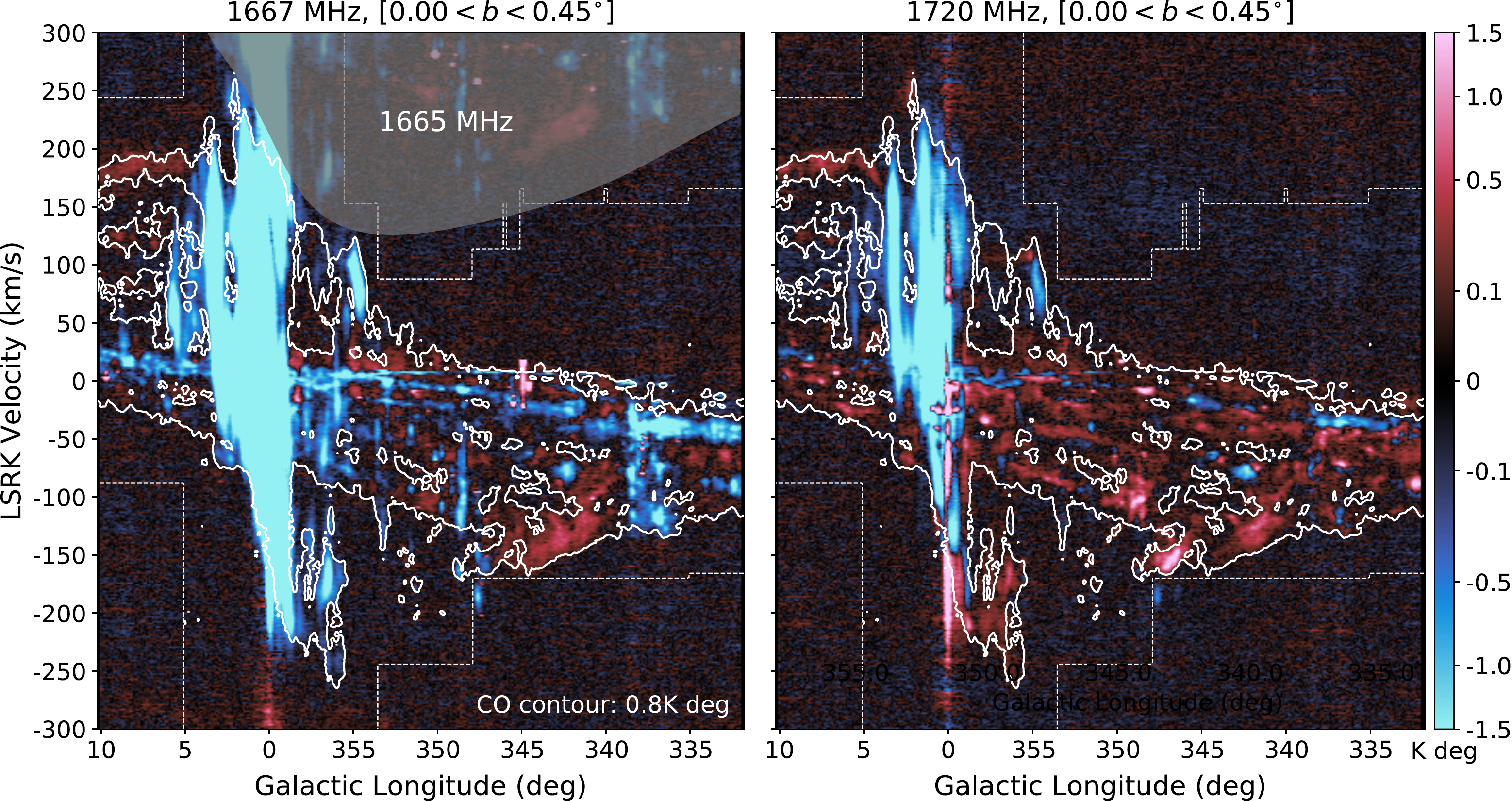}
\end{center}
\caption{Example longitude-velocity maps of the 1667\,MHz and 1720\,MHz lines, integrated over a latitude range of 0.5 degrees, overlaid with a single $^{12}$CO(J=1--0) contour from \citet{dame01}. The contour is drawn at the $4\sigma$ level of the noisiest portion of the CO image (noting that noise levels vary in different parts of the composite survey). The dashed white line indicates the edges of the CO data. The zone where signal from the 1665\,MHz line runs into the 1667\,MHz velocity range is shown in grey. Note that OH emission and absorption extending beyond the CO contour around $l=0^{\circ}$ is mostly residual baseline.}
\label{fig_lv_main}
\end{figure*}

\section{Discussion}
\label{discussion}

\subsection{Preliminary comparison with CO}

A major reason for observing OH is that it can trace diffuse molecular gas in which CO is not detected. This has been demonstrated for local clouds \citep[e.g.][]{barriault10, allen15, xu16}, off-Plane sightlines \citep{li18}, and in the Outer Galaxy \citep[e.g.][]{wannier93,engelke20,busch21}. While deeper CO observations -- unsurprisingly -- tend to decrease the measured `CO-dark' gas fraction \citep{li18,donate19a,donate19b}, there is still a diffuse, low-$A_V$ regime in which CO is not an optimal tracer \citep{wolfire10}. It was therefore initially surprising that \citet{dawson14} found no evidence of OH envelopes extending beyond the CO-bright regions of molecular cloud complexes in the SPLASH pilot region ($334 < l < 344^{\circ}$). While our sensitivity ($\sim15$\,mK) 
is not as high as some smaller, deeper surveys, 
a naive comparison with the brightness temperatures of past detections ($\lesssim50$\,mK) implied that at least some CO-dark gas should be detected in SPLASH. 
In \citet{dawson14} we suggested that the similarity of the OH excitation temperature to the continuum background brightness temperature in the inner Galaxy was at least partly responsible for this lack of diffuse gas detection, since the low contrast renders the lines more difficult to detect. 
However, the expectation is also that there is simply a lower fraction of diffuse, CO-poor gas in the inner Galaxy, where the metallicity, ambient pressure and mean gas density are all higher \citep{pineda13,langer14a}. This may imply thinner cloud envelopes that are more difficult to resolve spatially, particularly at the low resolution of SPLASH.  

Future work will carry out a detailed analysis of SPLASH in the context of CO-dark gas. However, it is worth making a simple preliminary comparison here. Comparing with the $^{12}$CO(J=1--0) Galactic Plane survey of \citet{dame01}, we find that OH rarely extends outside of CO-bright regions in our dataset. 
While some voxels do show weak OH where CO is not significantly detected, the majority of these either have marginal CO emission, or represent broad line wings where the CO signal appears to have dropped into the noise before the OH. Figure \ref{fig_lv_main} shows an example longitude-velocity plot of the 1667 and 1720\,MHz lines 
overlaid with a single $\sim 4\sigma$ $^{12}$CO(J=1--0) contour, illustrating the generally good agreement between the extent of the two tracers. The remaining $l$-$v$ plots are shown in the online Appendices. (See the $0.50^{\circ} < b < 0.95^{\circ}$ plot for the most convincing example of CO-dark OH at these sensitivities, in the main line emission feature around $l\approx345^{\circ}$, $v\approx-150$\,\kms.) 

What may be important is that the brightness temperature ratio of the two species varies considerably, and includes velocity components that are weak in CO but strong in OH. While such differences are partially driven by excitation and radiative transfer effects (as we will discuss more in Section \ref{noh_is_hard}), it is reasonable to suppose that differences in column density also contribute. 
In principle, careful analysis 
can recover some of these differences, allowing us to tease out the OH-bright molecular gas fraction not represented by CO. 
This requires a good model of the distribution of OH and continuum-emitting components in 3D space, as we will now discuss. 

\subsection{Column densities from the OH lines: Cautions and Considerations}
\label{noh_is_hard}



The column density of OH molecules, $N_\mathrm{OH}$, is a key physical measurement, and is particularly important if OH is to be used to estimate CO-dark molecular gas masses along sightlines where both are detected.  
In principle it is relatively straightforward to estimate $N_\mathrm{OH}$ from observations of the 18-cm line brightness temperatures and continuum background levels. In practice, there are complicating factors that can cause errors at the order-of-magnitude level or greater. In this section we will recap the basic physical relations underpinning column density estimation, summarise some common approaches, and consider the appropriate treatment of the SPLASH data. 

\subsubsection{Basic relations and common simplifications}

The total column density of OH is related to the excitation temperature ($T_\mathrm{ex}$) and optical depth ($\tau_v$) of the 18-cm lines by: 
\begin{equation}
N_\mathrm{OH}=\frac{8\pi\nu_0^2}{A_{ul}c^3}\frac{\Sigma g_i}{g_u}\frac{k T_\mathrm{ex}}{h}\int_{-\infty}^{\infty}{\tau_v}~dv.
\label{eq:noh_general}
\end{equation}

\noindent Here $\nu_0$ is the rest frequency of the line, 
$A_{ul}$ is the Einstein A coefficient for the line, $g_u$ is the degeneracy of the upper level and $g_i$ are the degeneracies of the remaining ground-state sub levels. We have made two important assumptions: (1) there is no significant population in the first rotationally excited state (at $E/k=121$\,K), (2) $|T_\mathrm{ex}| \gg h\nu_0/k \approx 0.08$. These assumptions are generally good for quasi-thermal OH in molecular clouds, although the latter may be weakly violated for the satellite lines in cases of extremely anomalous excitation. Evaluating the constants (see \citealt{destombes1977} for the Einstein coefficients), we may simplify to:
\begin{equation}
 \frac{N_{\mathrm{OH}}}{[\mathrm{cm}^{-2}]}=C \cdot \frac{T_\mathrm{ex}}{[K]} \cdot \int_{-\infty}^{\infty}{\tau_v}~\frac{dv}{[\mathrm{km~s}^{-1}]},
 \label{eq:noh_consts}
\end{equation}
where the constant $C$ is equal to $2.2\times10^{14}$ and $4.0\times10^{14}$ for the 1667 and 1665\,MHz lines, respectively. (Note that the satellite lines are generally not used in column density estimation since their excitation temperatures can vary wildly.)

We now wish to measure $T_\mathrm{ex}$ and $\tau_v$. For an isothermal, homogeneous cloud of OH molecules located in front of a radio continuum source, these are related to the observed line brightness temperature by:
\begin{equation}
   T_\mathrm{b,line}^*(v)=(T_\mathrm{ex}-T_\mathrm{C}^*)(1-e^{-\tau_v}),
\label{eq:basic_rt}
\end{equation}
\noindent where $T_\mathrm{b,line}^*(v)$ is the continuum-subtracted line brightness temperature, and $T_\mathrm{C}^*$ is the continuum background brightness temperature, including the CMB. The asterisks on these terms indicate beam-averages. This expression assumes that the OH gas fills the beam, but we return to the question of sub-beam structure in Section \ref{noh_hard_version}. 

\textit{The small optical depth case:} 
If $\tau_v \ll 1$ across the whole line, $(1-e^{-\tau_v})\approx\tau_v$ in Equation \ref{eq:basic_rt}, and Equation \ref{eq:noh_consts} becomes
\begin{equation}
 N_{\mathrm{OH}}=C~ \frac{T_\mathrm{ex}}{T_\mathrm{ex}-T_\mathrm{C}^*}~ \int_{-\infty}^{\infty}T_\mathrm{b,line}^*(v)~dv.
 \label{eq:noh_small_tau}
\end{equation}
\noindent For quasi-thermal OH, the assumption of a small optical depth is likely to be reasonable. \citet{dawson14} demonstrate that $\tau_v$ is always small in the SPLASH pilot region, albeit averaged over the Parkes beam. Measurements of main line optical depths via absorption against bright, compact continuum sources find typical values of $\tau\sim0.01$ in diffuse, off-Plane gas, and $\sim0.1$--$0.2$ for denser and/or in-Plane signtlines \citep[e.g.][]{dickey81,liszt96,li18,rugel18}. The highest measurement in the literature is (to our knowledge) $\tau\sim0.5$, in the inner Galactic Plane \citep{rugel18}. Even in this case, the error introduced by assuming $\tau \ll 1$ is only $\sim20$ per cent. 


\textit{The bright continuum case:} 
Another common simplification is to assume $T_\mathrm{C}^* \gg |T_\mathrm{ex}|$, 
allowing $\tau_v$ to be measured directly from the line and continuum brightness temperatures as $\tau_v={\rm ln}\left[T_\mathrm{C}^*/(T_\mathrm{b,line}^*(v)+T_\mathrm{C}^*) \right]$. 
This approach is appropriate for absorption measurements against very bright background sources, particularly in interferometric observations \citep{rugel18}. It is generally \textit{not} appropriate for the SPLASH dataset, where $T_\mathrm{C}^*\sim10$--20\,K over much of the survey field. 
We note that unsuitable use of this expression by \citet{engelke19} appears to be largely responsible for the factor of 10--100 underestimates they find for column density as measured from OH absorption lines. 


\textit{The `on-off' method:} Observing the same OH cloud against different continuum background levels yields two simultaneous instances of Equation \ref{eq:basic_rt}, that can be solved directly for both $T_\mathrm{ex}$ and $\tau_v$. Ideally, distinctly different continuum levels must be measured along close-by sightlines for which the OH properties vary minimally. A small beam and a bright, compact continuum source in an unconfused region of the sky are therefore preferred. Most literature measurements of OH excitation temperature have been obtained from this approach \citep[see e.g.][]{rieu76,dickey81,liszt96,li18}. In the SPLASH dataset the prospects for the `on-off' method are limited, but may be possible towards some select sources. 

\begin{figure}
\begin{center}
\includegraphics[scale=0.55]{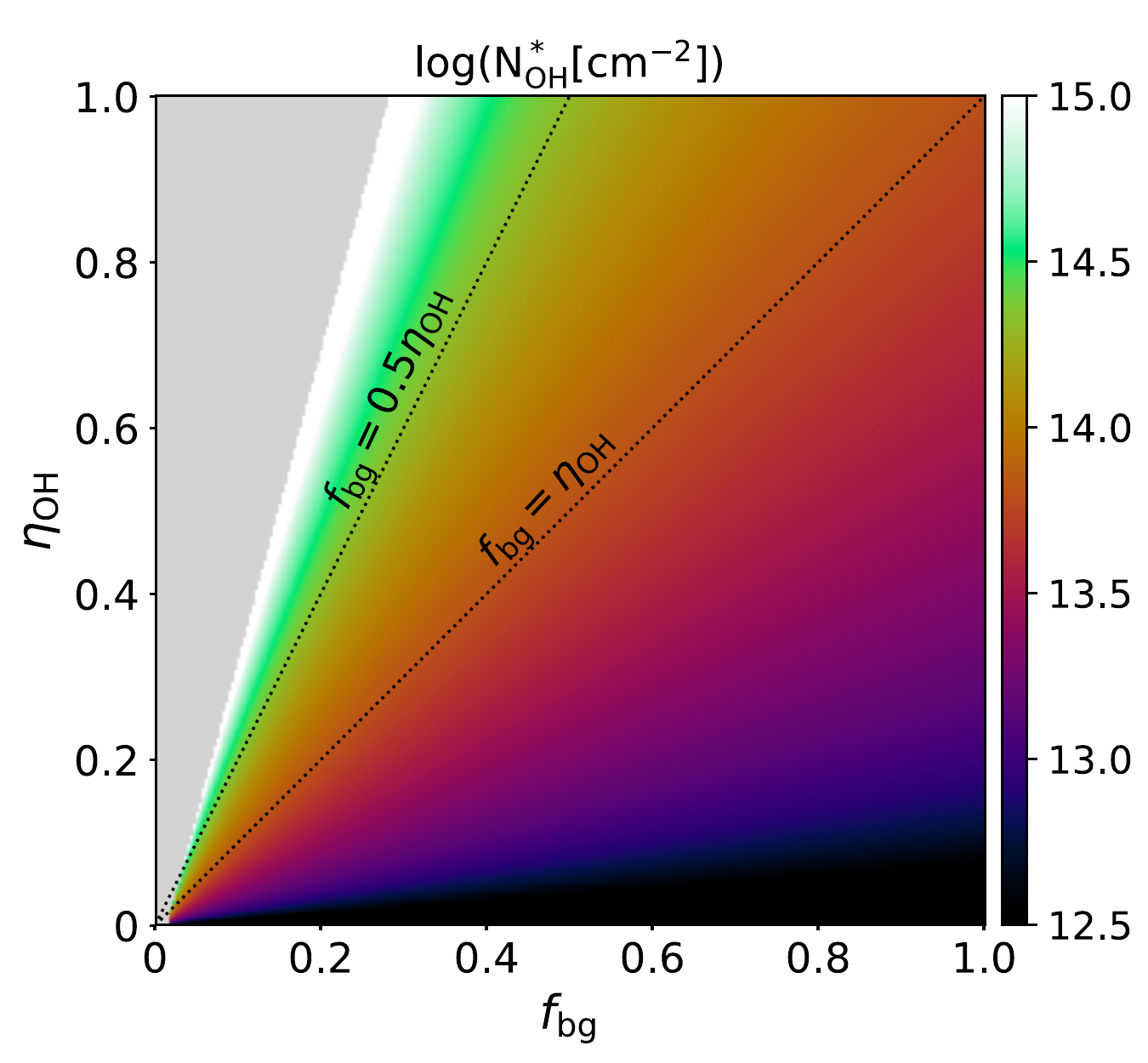}
\vspace{-0.6cm}
\end{center}
\caption{Behavior of the beam averaged column density, $N_{\mathrm{OH}}^*$, as a function of $\eta_\mathrm{OH}$ 
and $f_\mathrm{bg}$. 
This model takes typical measured values of $T_\mathrm{C}^*=15$\,K, peak $T_\mathrm{b,line}^*(v)=-0.2$\,K, a linewidth of $2.0$\,\kms\ and assumes $T_\mathrm{ex,1667}=6.0$\,K, and $T_\mathrm{CMB}=2.7$\,K. The colour scale shows log($N_{\mathrm{OH}}^*$), and the grey area shows the zone where there are no physical solutions. The dotted lines illustrate two canonical cases where the continuum distribution is completely smooth and (a) all continuum is located behind the OH cloud ($f_\mathrm{bg}=\eta_\mathrm{OH}$), and (b) half of the continuum is located behind the OH cloud ($f_\mathrm{bg}=0.5\eta_\mathrm{OH}$). 
}
\label{fig_noh}
\end{figure}

\subsubsection{Estimating excitation temperature}
\label{tex}

This brings us to an important point: in the absence of direct measurements, we have no choice but to estimate the excitation temperature. In general, the OH ground state lines are anomalously excited, even outside of strongly masing regions \citep[see][]{crutcher79,dawson14}. The main line excitation temperatures are comparatively well-behaved, but cannot be assumed to be equal, though the difference is relatively small, around 1--2\,K \citep[e.g.][]{crutcher79,li18,engelke18}. Typically, $T_\mathrm{ex,mains}<15$\,K, with measurements suggesting a distribution peak at $\sim5$\,K, and a tendency for $T_\mathrm{ex,1665}$ to be the higher of the two \citep{crutcher79,engelke18,li18}. 

The majority of literature measurements have been made towards off-Plane sightlines and the Outer Galaxy, with a dearth of observations towards the regions mapped by SPLASH. It is not clear the degree to which we might expect the excitation conditions to vary in the environment of the inner Galaxy. There are some prospects for estimating $T_\mathrm{ex}$ directly in the SPLASH dataset, by noting the continuum background level at which lines switch from emission to absorption (i.e. $T_\mathrm{ex}=T_\mathrm{C}^*$). However, the distance and structure of the continuum-emitting gas is a complicating factor here, as we will discuss. In any case, column densities for SPLASH will be sensitive to the exact choice of $T_\mathrm{ex}$, since $T_\mathrm{ex}\sim T_\mathrm{C}^*$ throughout much of the survey volume (c.f. Equation \ref{eq:noh_small_tau}).



\subsubsection{Sub-beam structure and continuum source placement}
\label{noh_hard_version}

The primary complication in SPLASH is that we are observing in the confused and continuum-bright inner Galaxy. Multiple OH components exist along the line of sight, often overlapping in velocity space, and with degenerate kinematic distances within the solar circle. Continuum emission is everywhere, from the relatively smooth Galactic synchrotron through to discrete and highly-structured H{\sc ii} regions and SNR, which are distributed at various (and a-priori unknown) distances along a sightline. Unlike the simple case of off-Plane measurements against the CMB, the continuum cannot be assumed to be either smooth within the beam or to lie behind the OH gas. Clearly this presents some challenges. To obtain good column density estimates across the whole cube requires coupled modelling of both the molecular and continuum-emitting gas in 3D space, and some sensible estimates of sub-beam structure in both -- a significant undertaking far beyond the scope of this paper.

\begin{figure*}
\begin{center}
\includegraphics[scale=0.55]{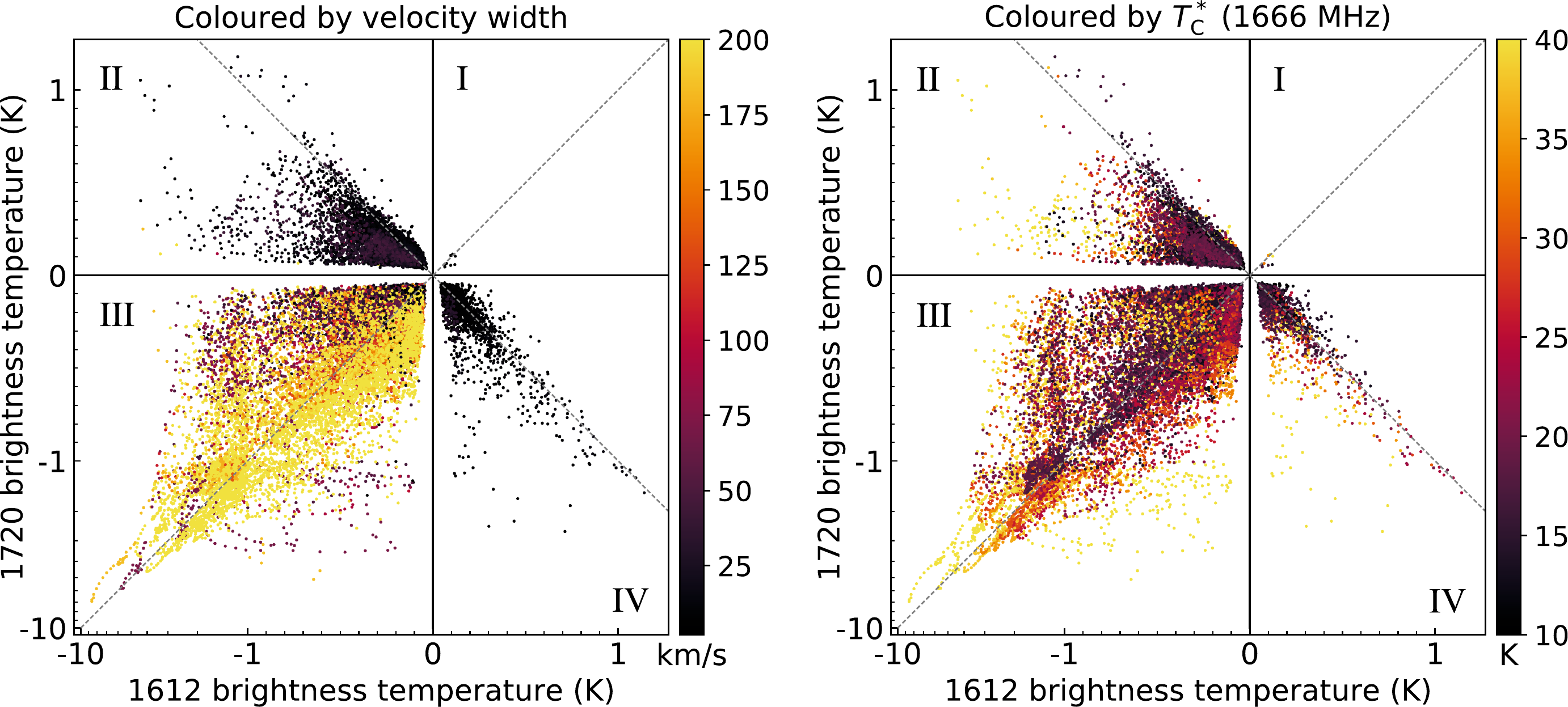}
\end{center}
\caption{Behavior of the satellite lines across the full survey region (excluding $-0.5<l<+0.5^{\circ}$ and $-0.35<b<+0.35^{\circ}$), with compact maser sources excluded. Each point represents a pair of brightness temperatures, where both lines are detected above the $5\sigma$ level in a 1.45\kms\ channel, sampled at 3 pixel (9 arcminute) intervals across the cubes. The colour scale on the left shows the number of contiguous channels across which the sense of the line ratio remains consistent (a rough analogue for the velocity width of the feature); on the right it shows the continuum brightness temperature at the position of the detection. The grey dashed lines show $T_\mathrm{b,1612}^*=T_\mathrm{b,1720}^*$ and $T_\mathrm{b,1612}^*=-T_\mathrm{b,1720}^*$. 
}
\label{fig_sat_scatter}
\end{figure*}

However, we may explore the sensitivity of $N_\mathrm{OH}$ to assumptions about source structure and line-of-sight ordering. 
Considering a single OH velocity component, and assuming no variation of $T_\mathrm{ex}$ and $\tau_v$ within the OH cloud, we can obtain a useful simplification in terms of two parameters, the OH beam filling factor, $\eta_\mathrm{OH}$, and the fraction of the measured (non-CMB) continuum flux participating in OH radiative transfer, $f_\mathrm{bg}$. Equation \ref{eq:basic_rt} then becomes:
\begin{equation}
    T_\mathrm{b,line}^*(v)=\left[\eta_\mathrm{OH}\left(T_\mathrm{ex}-T_\mathrm{CMB}\right)-f_\mathrm{bg}\left(T_\mathrm{C}^*-T_\mathrm{CMB}\right)\right](1-e^{-\tau_v})
    \label{eq:nasty_rt}
\end{equation}
Here, as in the SPLASH continuum data, $T_\mathrm{C}^*$ explicitly includes the CMB. However, we may separate the CMB conceptually, since it is always smooth and distant, meaning that the fraction of CMB photons passing through the OH cloud is described exactly by $\eta_\mathrm{OH}$. The $f_\mathrm{bg}$ parameter is then the fraction of \textit{non-CMB} photons passing through OH gas, and combines two concepts: the fraction of continuum emission physically more distant than the OH cloud, and a covering fraction describing how much of this physically distant continuum actually overlaps with the OH gas along the sightline. It can be flexibly varied to model many combinations of these two variables. 

For the small optical depth case the \textit{beam averaged} column density is then: 
\begin{equation}
 N_{\mathrm{OH}}^*= \frac{\eta_\mathrm{OH}CT_\mathrm{ex}}{\eta_\mathrm{OH}\left(T_\mathrm{ex}-T_\mathrm{CMB}\right)-f_\mathrm{bg}\left(T_\mathrm{C}^*-T_\mathrm{CMB}\right)} \int_{-\infty}^{\infty}T_\mathrm{b,line}^*(v)~dv.
 \label{eq:nasty_noh_small_tau}
\end{equation}
\noindent Figure \ref{fig_noh} shows the sensitivity of the derived beam-averaged column density to different assumptions of $f_\mathrm{bg}$ and $\eta_\mathrm{OH}$, for illustrative values of $T_\mathrm{b,line}^*(v)$, $T_\mathrm{C}^*$ and $T_\mathrm{ex,1667}$. We see that the same measurements can return beam-averaged column densities that vary by up to two orders of magnitude depending on the assumed filling factors and continuum placement. This basic conclusion remains unchanged for a wide range of measured $T_\mathrm{b,line}^*(v)$ and $T_\mathrm{C}^*$, and realistic excitation temperatures. 

Fortunately not all regions of parameter space are equally plausible. For example, the combination of a small OH filling factor and high $f_\mathrm{bg}$ corresponds to a compact continuum source located precisely behind a compact OH cloud, and is in general improbable. Some properties of the data itself also inform the most likely configurations, even in the absence of additional information -- e.g. a very strong absorption line likely indicates a high continuum background, and hence high $f_\mathrm{bg}$. Good models will incorporate this information. Nevertheless, care is warranted when deriving column densities and gas masses from the SPLASH dataset. 



\subsection{Excitation Patterns in the Satellite Lines}
\label{satellites}

The OH satellite lines at 1612 and 1720\,MHz are often highly anomalously excited, even in typical molecular cloud conditions (i.e. outside of bright, high-gain masers). When the upper level population of one line is enhanced, corresponding to a high or negative $T_\mathrm{ex}$, the upper level in the other is underpopulated, corresponding to a low, sub-thermal $T_\mathrm{ex}$ \citep{elitzur76, guibert78}. This results in characteristic conjugate profiles, where one line is seen in emission and the other in absorption. 
The parameter space that gives rise to each sense of this anomalous excitation 
is complex, with dependencies on density, column density, velocity width, and the gas and dust temperatures. However, enhanced 1612\,MHz emission in normal molecular clouds is often (though not exclusively) associated with high dust temperatures, while enhanced emission in the 1720\,MHz transition is more widespread, and is readily produced with lower gas temperatures or column densities \citep{elitzur76, guibert78, vanlangevelde95, turner82}. Measurements of the brightness temperature alone offer limited ability to discriminate between a high positive excitation temperature and weak masing (negative $T_\mathrm{ex}$), but the sense of the emission/absorption still has high diagnostic power, and can provide good constraints on the physical state of the ISM \citep[see e.g.][]{ebisawa19,ebisawa20,petzler20}. 

Figure \ref{fig_sat_scatter} plots the brightness temperatures of the satellite lines on a voxel-by-voxel basis, excluding high-gain masers. 
Known masers were removed using the catalogues of \citet{qiao16,qiao18,qiao20} and \citet{ogbodo20}, and the unpublished catalogue of Ogbodo et al. (in prep)\footnote{This catalogue presents all OH maser sources from the MAGMO survey \citep[see][]{green11}, which were not observed independently in the SPLASH maser followup observations.}.
The datapoints are colour coded by the 1666\,MHz continuum brightness temperature at the location of the voxel, and by the `velocity extent' of its parent spectral feature, defined as the number of contiguous channels over which the sense of the emission/absorption remains unchanged. 
This metric is not a direct analogue to a fitted linewidth: it treats blended features as one, will break very wide lines up if a different sense of the inversion is imposed upon them. It also tends to run much wider than a FWHM. However, it is a useful proxy in the absence of formal fitting. We will now briefly examine some of the trends shown in these plots.

\subsubsection{Disabling of anomalous excitation by line overlap in the CMZ}

On a voxel-by-voxel basis, by far the most common profile pattern is both the 1612 and 1720\,MHz lines in absorption. This is initially a surprising result, and all the more so because it shows no correlation with the level of continuum emission at the location of the spectra. There is, however, a striking correlation with velocity width: the mean velocity extent of double-absorption features is 47\,\kms, compared to only 5.6\,\kms and 3.7\,\kms\ in the 1720\,MHz-emission and 1612\,MHz-emission cases. It quickly becomes clear that double-absorption profiles are located 
almost exclusively in extremely velocity-broadened gas located within 6 degrees in longitude of the Galactic Centre. This comprises the CMZ ($-2.0\lesssim l\lesssim 1.5^{\circ}$) as well as broad velocity features 
thought to arise from dynamical interactions of gas within the Galactic Bar potential \citep[see e.g.][]{Liszt08,sormani19}. The large number of channels across these wide spectral features is responsible for the apparent dominance of double-absorption in the figure.

The fact that both lines are seen in absorption, and that this is not correlated with abnormally high background continuum, suggests that anomalous excitation is disabled (or at least weakened) in these regions. The most likely explanation is line overlap. Population inversions in the ground state OH hyperfine levels are driven by asymmetries in the excitation and de-excitation pathways in and out of the excited rotational states. When the Doppler broadening of the IR lines connecting these levels is greater than the hyperfine separation between rotational state sub-levels, photons emitted in one transition can be absorbed in another, effectively shuffling the level populations. The degree of line overlap can be critical; for example, modest line widths ($\sim1$\kms) are implicated in producing the ubiquitous main line anomalies discussed in Section \ref{tex} \citep{bujarrabal80}. 
However the asymmetries are washed out for linewidths exceeding $\sim40$\,\kms\, causing the lines to thermalise again \citep{lockett08}. 
This mechanism provides a convincing explanation for the lack of anomalous excitation in the extremely velocity broadened gas of the CMZ.

\subsubsection{Conjugate emission/absorption in the inner Galaxy}

When extremely velocity broadened components are excluded, conjugate emission/absorption in the satellite lines is the norm. 
Counting by spectral feature as opposed to by voxel, 1720\,MHz emission paired with 1612\,MHz absorption accounts for 71 per cent of detections in the SPLASH survey region, with the reverse pattern accounting for only 19 per cent. The remaining 10 per cent of features are the double-absorption spectra just discussed, and we see no examples where both lines are seen in emission. 

\begin{figure}
\begin{center}
\begin{subfigure}{0.4\textwidth}
\includegraphics[scale=0.65]{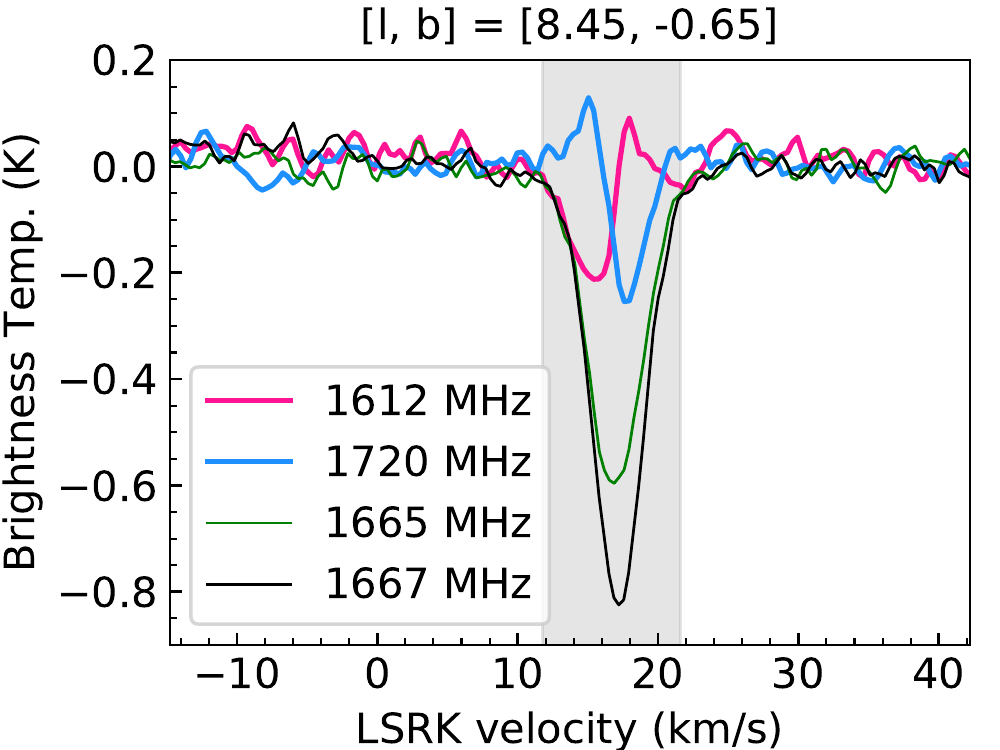}
\end{subfigure}\vspace*{0.2cm}
\end{center}

\begin{center}
\begin{subfigure}{0.4\textwidth}
\includegraphics[scale=0.65]{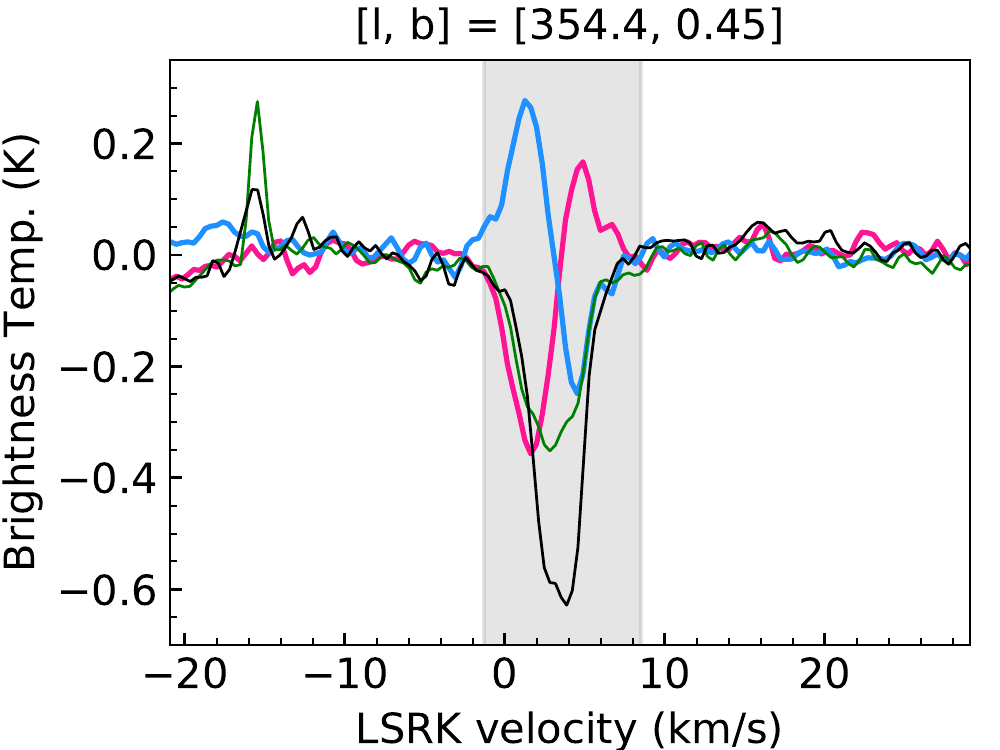}
\end{subfigure}
\end{center}
\caption{Two illustrative examples of satellite line `flips' discussed in the text, defined as profiles in which the sense of the conjugate 1612 and 1720 MHz lines reverses within a closely-blended main line feature. The remaining profiles are presented in the online Appendices.}
\label{fig_flips}
\end{figure}

In both senses of the conjugate profiles, the absorption is generally stronger than the emission, and the satellite lines are sometimes seen in the absence of the main lines (despite being intrinsically weaker), indicating significant departures from the main line $T_\mathrm{ex}$, and providing an alternative means of mapping otherwise undetectable gas. 
All these findings confirm the results of \citet{turner82} in the same region, but at an order of magnitude improvement in sensitivity. 
It is unclear whether the anomalous emission is dominated by weakly masing gas or elevated excitation temperatures. Some components appear to brighten with stronger $T_\mathrm{C}^*$, suggesting maser amplification; for some, the opposite is true, suggesting a high positive $T_\mathrm{ex}$. In any case, the low resolution and uncertainties in line-of-sight placement limit our ability to assess this reliably; likely both cases are common. 

Given the propensity of enhanced 1612\,MHz emission to occur in the presence of warm dust, we might hope to observe a correlation with $T_\mathrm{C}^*$, as a proxy for H{\sc ii} regions. Zones of 1612\,MHz emission are indeed more compact on the sky and do show some evidence of by-eye association with discrete continuum sources. This is not evident in Figure \ref{fig_sat_scatter}, however, which shows no preference for any excitation pattern with continuum brightness temperature. This is perhaps not surprising: high $T_\mathrm{C}^*$ alone is a very limited indicator of the H{\sc ii} region population, many of which will not rise significantly above the high background continuum levels at this resolution, and line-of-sight confusion also washes out correlations. 
We will return to this question in future work, making use of H{\sc ii} region catalogues with velocity information \citep[e.g.][]{wenger21} to more reliably associate sources. 

\subsubsection{Satellite line `flips' as a tracer of H{\sc ii} region expansion?}

We finally look briefly at satellite line `flips' in the SPLASH datacube. These are a common profile shape in which the sense of the satellite line conjugate profile reverses -- with one line flipping from emission to absorption, and the other the reverse -- across closely-blended double feature \citep{vanlangevelde95,rugel18}. Illustrative examples are shown in Figure \ref{fig_flips}. This configuration can occur in any place where the excitation conditions change sufficiently in two adjacent velocity components. However, \citet{petzler20} have recently suggested that the majority of flips can be explained by a specific astrophysical scenario: a shock driven into a molecular cloud that is being irradiated by the warm dust of an H{\sc ii} region. In this picture, the shock raises the density in a thin (low column density) layer of accelerated gas, switching off the IR-pumped 1612\,MHz emission 
and inverting the 1720\,MHz line instead. This model rests on two observational findings drawn from literature spectra: (a) flips appear to have a preferred velocity orientation -- the 1720\,MHz emission component is more blueshifted in 90 per cent of cases; (b) the majority of known examples overlap with H{\sc ii} regions on the sky and in velocity space. The velocity orientation can be explained if we preferentially see only the foreground portion of the parent cloud in contrast against the continuum-bright interior of the H{\sc ii} region, meaning that the shocked material always appears blueshifted. 


The SPLASH data shows that satellite line flips are widespread in this part of the Galaxy. While H{\sc ii} region association must wait for more detailed analysis, we may examine their velocity orientation over a larger sample of sightlines than the 30 known from the literature \citep[see references in][]{petzler20}. Restricting ourselves to clean and unambiguous examples (which excludes many complex spectra towards very bright continuum sources), we identify 38 unique structures in $l$-$b$-$v$ space, most of which extend over multiple resolution elements. A single example spectrum from each of the 38 regions is presented in the online Appendices. Of these 32/38 (84 per cent) show the 1720\,MHz emission component blueshifted, consistent with the H{\sc ii} region expansion model. 

For the 6 counter examples, we notice no obvious differences in spatial distribution or characteristic brightness, but do note that they appear to be associated with atypical excitation patterns in the main lines -- either emission in one of the lines, or (in one case) abnormally strong 1665\,MHz absorption. 

\subsection{Brief Comment on the Outer Galaxy Thick Molecular Disk}

\citet{busch21} recently reported the discovery of a very diffuse ($n_\mathrm{H2}\sim5\times10^{-3}$\,cm$^{-3}$), thick, CO-dark molecular disk in the 2nd quadrant of the outer Galaxy, based on ultra-deep observations of the OH main lines with the Green Bank Telescope. The effective integration time of their data was $\sim80$ hours (for an rms sensitivity of $\sim300\mu$K), and the brightness temperature of the emission feature was $\lesssim10$\,mK. The emission was very broad ($\Delta v \sim150$\,\kms) and pervasive, following the H{\sc i} distribution closely across degrees on the sky.

We consider whether it might be possible to recover a similar signal from the SPLASH dataset. With a total on-source time of $\sim6$ hours per square degree, effective integration times of 10s of hours may be achieved by judicious stacking over large areas. However, such weak emission (should it have been present) would not have been identified in the \textsc{Duchamp} masking stage, and its broad velocity width means it would not have survived the automatic baselining of the survey data. We therefore produce non-baselined 1667\,MHz cubes, and experiment with stacking to effective integration times of up to $\sim100$ hours ($\sim16$ square degrees) to seek for evidence of weak emission or absorption at positive velocities in the fourth quadrant. This analysis, while admittedly limited, reveals no evidence of OH signal. However, it is unclear whether we could even expect to achieve the required baseline fidelity with receiver and backend combinations used in SPLASH. Future work plans deep observations with the new Ultra-Wideband Low receiver on Parkes, which has superior noise and bandpass characteristics \citep{hobbs20}. 

\section{Conclusions}
\label{conclusions}

We have presented the first data release for the Southern Parkes Large-Area Survey in Hydroxyl. SPLASH covers 176 square degrees of the inner Galaxy, including the Galactic Centre and CMZ, in all four 18-cm ground-state transitions of OH. It is the largest deep, unbiased survey of OH to-date, achieving a characteristic main beam brightness temperature sensitivity of $\sim15$\,mK for a velocity resolution of $\sim0.9$\kms, and a spatial resolution of $\sim15$ arcmin. While the data presented here are optimised for extended, quasi-thermal OH, the cubes contain numerous maser sources, and SPLASH has already discovered over 400 new OH maser sites, confirmed and localised by interferometric followup with the Australia Telescope Compact Array  \citep{qiao16,qiao18,qiao20}.

The publicly available SPLASH data products include spectral line cubes of the full survey region in the four transitions (at 1612.231, 1665.402, 1667.359 and 1720.530\,MHz), together with matched continuum images in each of the observed bands: 1612, 1666 and 1720\,MHz. The continuum emission is an essential component of the dataset, and is required for many astrophysical interpretations of the spectral line data. We have carefully quantified uncertainties on all of the data products, and these may be found in Table \ref{table:data_pars}, together with the key properties of the cubes and images. While the reliability of the spectral baselines in the Central Molecular Zone and Galactic Centre is lower than the rest of the dataset, the data are well-calibrated and may be used for science applications, with appropriate caution. 


While the main focus of this paper is the data description, we have also presented some preliminary analysis of extended quasi-thermal OH using the full survey cubes. In keeping with the findings of \citet{dawson14} for the SPLASH pilot region, 
we find that OH rarely extends outside CO cloud boundaries in our data. 
However, there are large variations in OH and CO line ratios, at least some of which may arise from differences in the total gas column density traced by each. A SPLASH-based search for CO-dark molecular gas must therefore be made not only on the basis of differences in spatial distribution, but on careful comparisons of the column density along sightlines (and within components) where both are detected. 
However, we demonstrate that in these complex inner Galaxy fields, 
failure to appropriately model the line-of-sight source distribution (or sub-beam structure) of the continuum-emitting and OH-bearing gas can result in errors of up to two orders of magnitude in the beam-averaged column density -- with obvious consequences for any scientific analysis. 
Reliable column density estimation will require coupled modelling of the distribution of continuum and OH in 3D space. 

We have also briefly examined the statistics of the 1612 and 1720\,MHz satellite line emission and absorption, confirming that anomalous excitation is the norm throughout the inner Galaxy, with 1720\,MHz emission (paired with 1612\,MHz absorption) being the dominant pattern. The important exception is the Central Molecular Zone (and other extremely velocity broadened gas in the vicinity of the Galactic Bar), in which line overlap disables the relevant pumping mechanisms, causing the lines to thermalise. We also identify numerous new examples of satellite line `flips' -- a characteristic profile shape in which one line flips from emission to absorption, and the other the reverse -- across closely-blended double feature. We confirm a preferred velocity orientation for these profiles, with the a 1720-emission/1612-absorption component seen at more negative velocities in the majority of cases. 
This is consistent with recent models explaining the flip as a signature of H{\sc ii} region expansion \citep{petzler20}.

Interferometric observations are an important counterpart to single dish surveys such as SPLASH. 
As well as the interferometric followup of maser candidates already carried out for SPLASH \citep{qiao16,qiao18,qiao20}, we will also be looking to 
GASKAP-OH, the OH portion of the Galactic Australia Square Kilometre Array Pathfinder survey \cite[GASKAP,][]{dickey13}, and to 
the Galactic Centre extension to THOR \citep[The H{\sc i}, OH, Recombination line survey of the Milky Way,][]{beuther16, rugel18}. Both of these surveys will make untargeted observations that overlap the SPLASH region,  
detecting both high-gain masers and quasi-thermal OH. 
In the latter case, the target is absorption against the bright and relatively compact continuum structures well imaged by an interferometer. Combined analysis will be important here: SPLASH sees extended diffuse emission, OH absorption against both compact and extended continuum (including the Galactic synchrotron background), and achieves much higher surface brightness sensitivities than an interferometer. Interferometric surveys, on the other hand, can provide direct optical depth measurements, aid in placing components along the line of sight (by their presence or absence in absorption against continuum sources of known distance), can be paired with matching H{\sc i} absorption spectra to study the statistics of the atomic and molecular gas, and can and facilitate detailed high-resolution studies of the gas associated with individual H{\sc ii} regions. Together, these suggest good prospects for a global census of the OH-bearing ISM, including its large-scale relationship to other ISM phases. 

\section{Acknowledgements}
J.R.D. acknowledges the support of an Australian Research Council (ARC) DECRA Fellowship (project number DE170101086). H.I. is supported by JSPS KAKENHI Grant Number JP21H00047. J. F. G. acknowledges support
from the State Agency for Research (AEI/10.13039/501100011033)
of the Spanish MCIU, through grant PID2020-114461GB-I00 and
the “Center of Excellence Severo Ochoa” award for the Instituto de Astrof\'{\i}sica de Andaluc\'{\i}a (SEV-2017-0709). This research made use of the Duchamp source finder, produced at the Australia Telescope National Facility, CSIRO, by M. Whiting. The Parkes radio telescope is part of the Australia Telescope National Facility (grid.421683.a) which is funded by the Australian Government for operation as a National Facility managed by CSIRO. We acknowledge the Wiradjuri people as the traditional owners of the Observatory site.

\section{Data Availability Statement}
The data underlying this article are available from AAO Data Central (https://datacentral.org.au/) via the SPLASH project page at https://docs.datacentral.org.au/splash/.

\footnotesize{
\bibliography{splashbib2}
}

\newpage

\noindent Online only material

\section*{Appendix 1: Velocity Channel Maps, Figure Set 1}
Integrated intensity maps of the full SPLASH survey in all four transitions. Each map is integrated over a velocity range of 5.8 \kms, with central velocities as labelled in the top right-hand corners of the plots. Note that since the main lines are only separated by $\sim350$ \kms, there are places where negative velocities in the 1665\,MHz line run into the positive velocity space of the 1667\,MHz transition and vice-versa.

\section*{Appendix 2: Longitude-Velocity Maps,  Figure Set 2}
Longitude-velocity maps of the 1667\,MHz and 1720\,MHz lines, overlaid with a single $^{12}$CO(J=1--0) contour from Dame, Hartmann \& Thaddeus (2001). The contour is drawn at the $4\sigma$ level of the noisiest portion of each image. The dashed white line indicates the edges of the CO data. Each map is integrated over a latitude range of 0.5 degrees. Note that since the main lines are only separated by $\sim350$ \kms, there are places where negative velocities in the 1665\,MHz line run into the positive velocity space of the 1667\,MHz transition. This zone is marked in grey on one of the panels, for clarity.

\section*{Appendix 3: Satellite line `Flips', Figure Set 3}
Satellite line `flips' in the SPLASH survey region, defined as profiles in which the sense of the conjugate 1612 and 1720\,MHz lines reverses within a closely-blended main line feature. Each panel shows a single example sightline towards a unique instance of the flip in the 3D cubes. Each feature typically extends over multiple beams on the sky.

\begin{figure*}
\begin{subfigure}{0.48\textwidth}
\includegraphics[width=\linewidth]{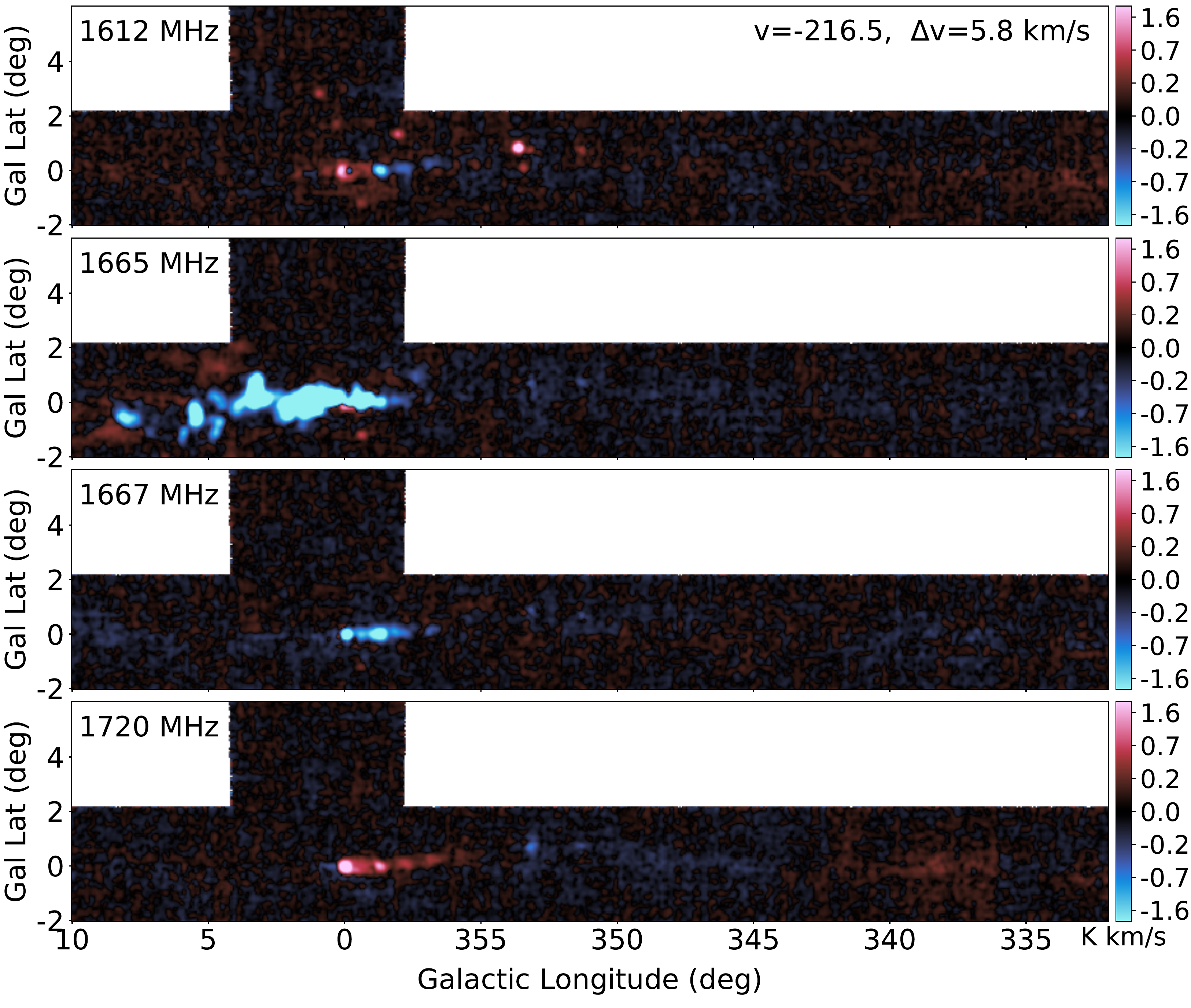}
\end{subfigure}\hspace*{\fill}
\begin{subfigure}{0.48\textwidth}
\includegraphics[width=\linewidth]{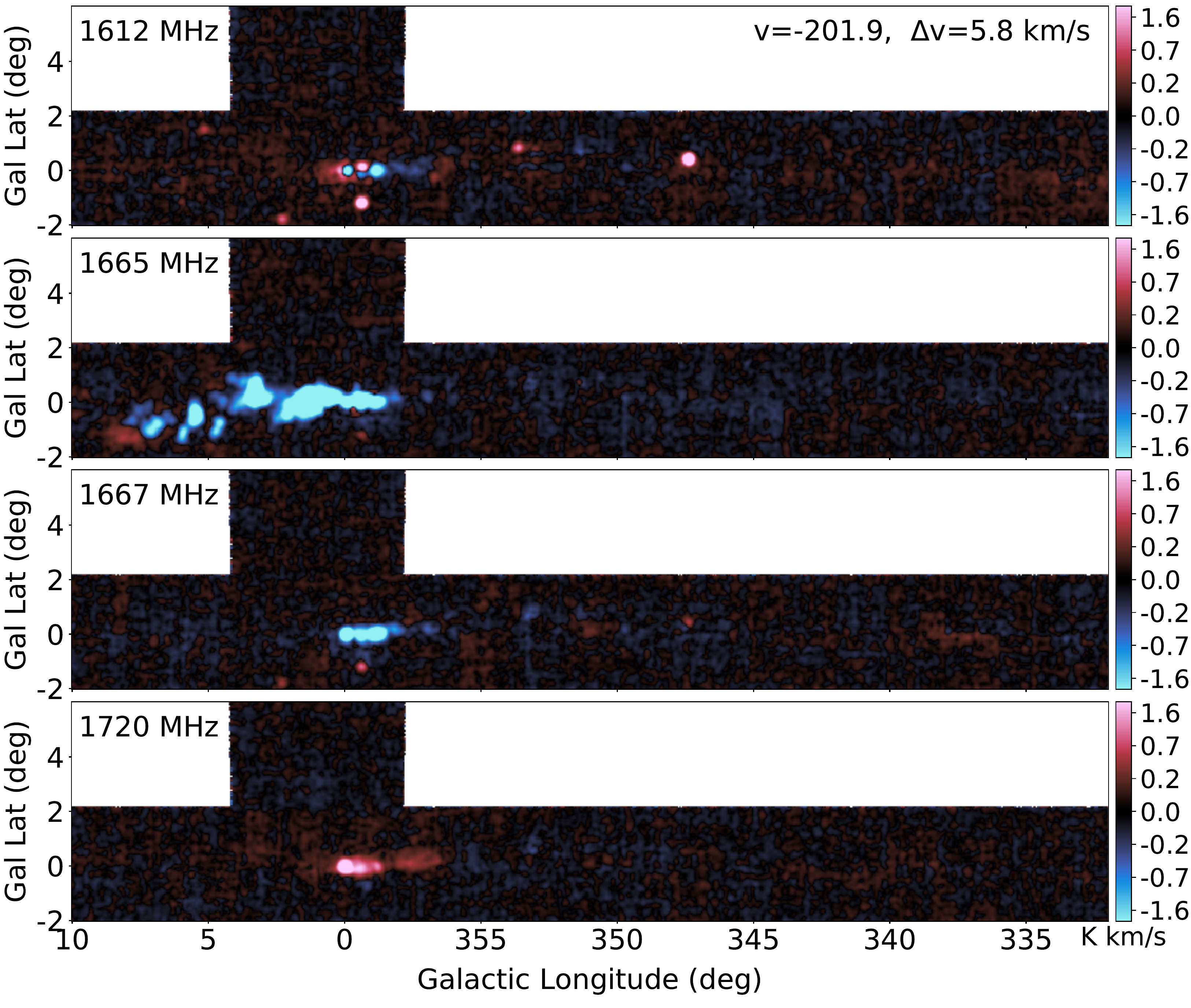}
\end{subfigure}

\medskip
\begin{subfigure}{0.48\textwidth}
\includegraphics[width=\linewidth]{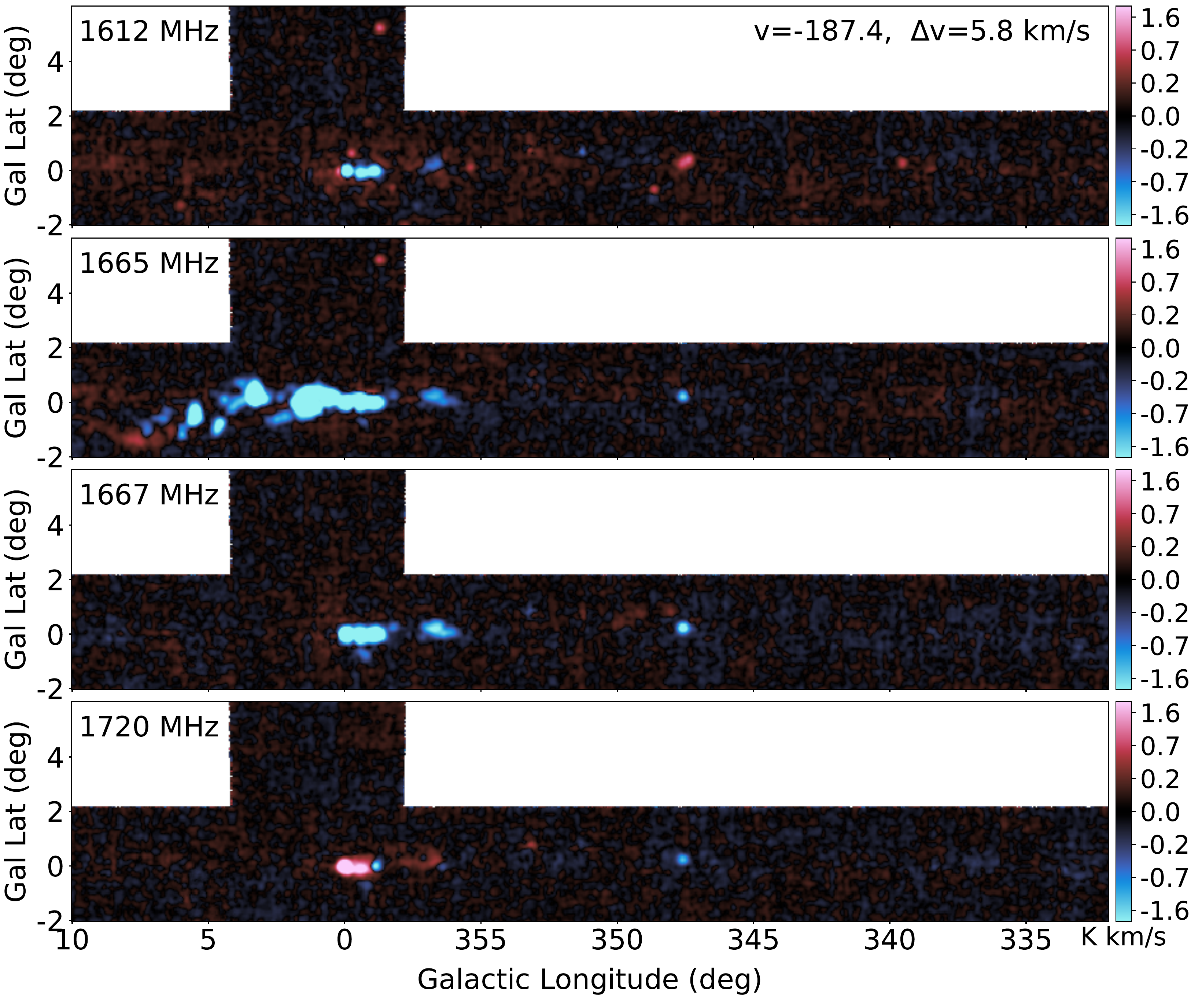}
\end{subfigure}\hspace*{\fill}
\begin{subfigure}{0.48\textwidth}
\includegraphics[width=\linewidth]{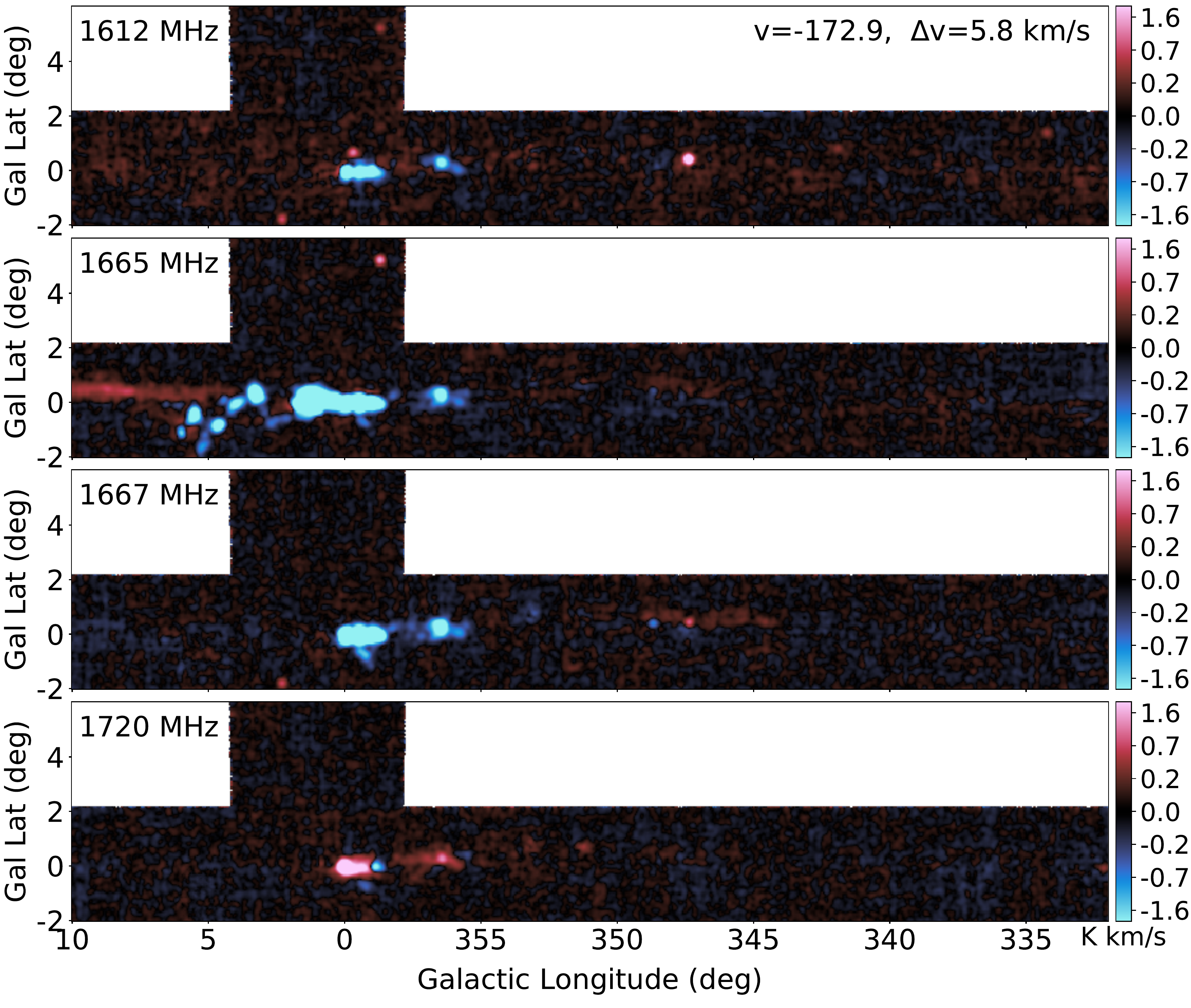}
\end{subfigure}

\medskip
\begin{subfigure}{0.48\textwidth}
\includegraphics[width=\linewidth]{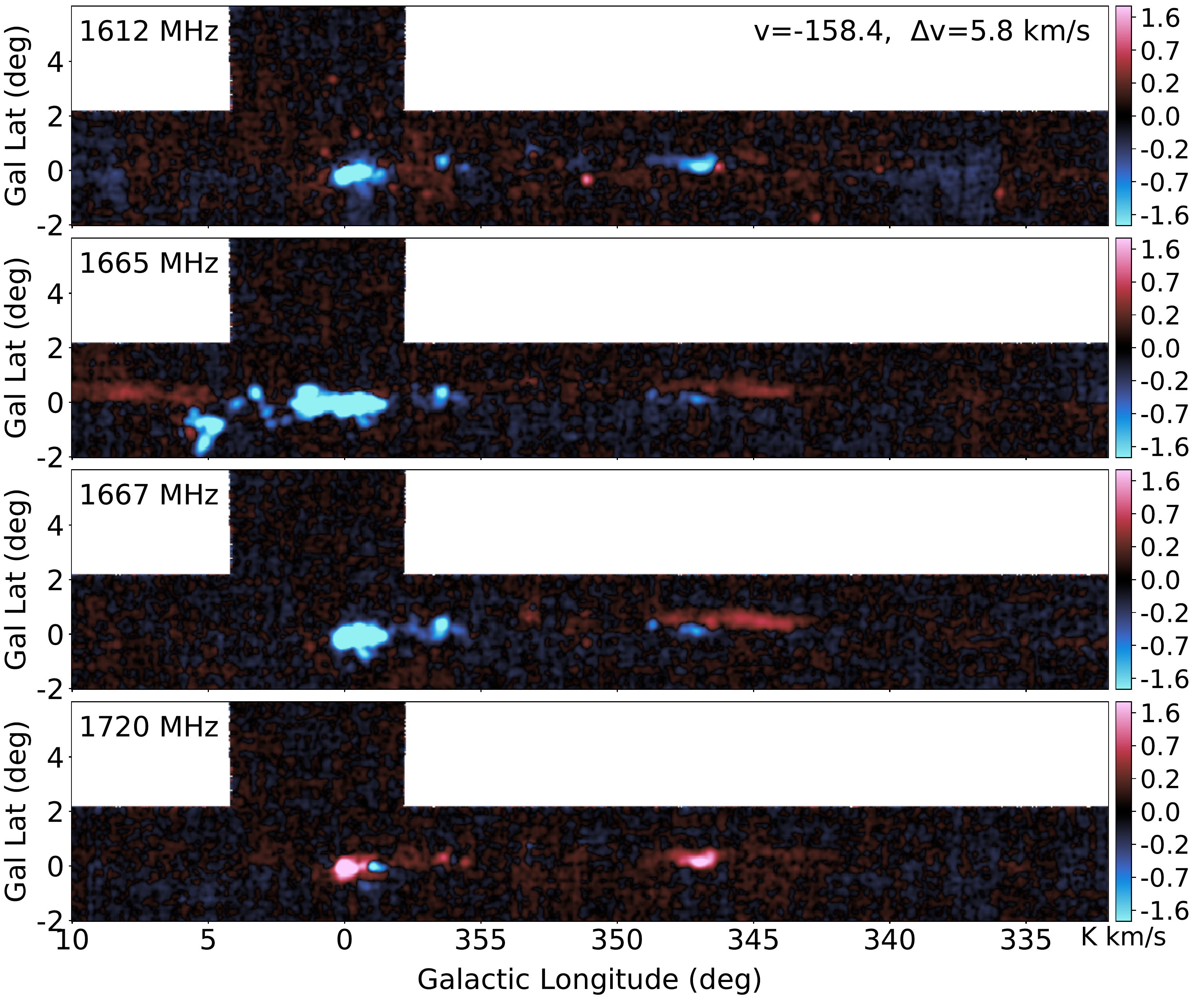}
\end{subfigure}\hspace*{\fill}
\begin{subfigure}{0.48\textwidth}
\includegraphics[width=\linewidth]{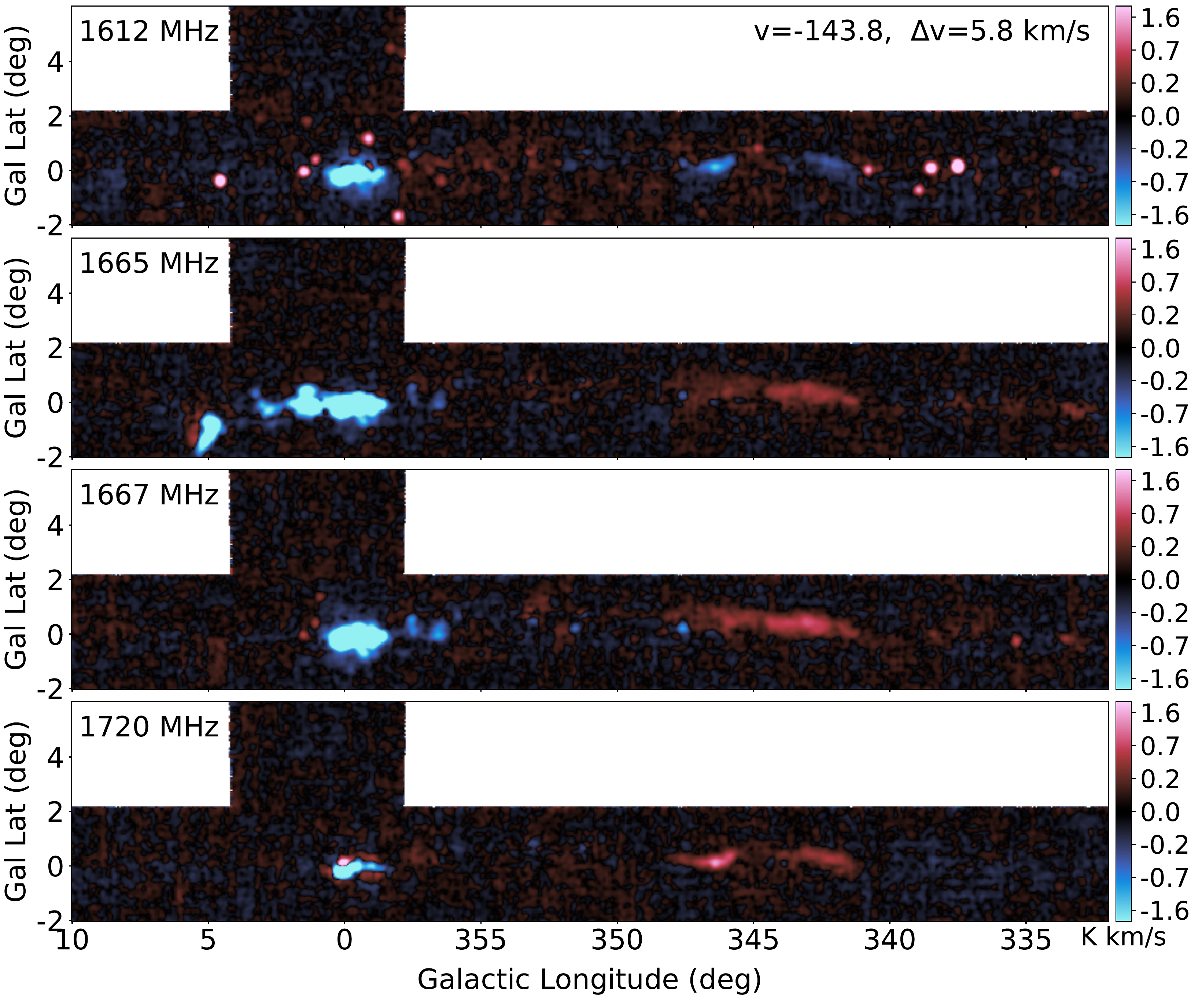}
\end{subfigure}

\caption{Integrated intensity maps of the full SPLASH survey in all four transitions. Each map is integrated over a velocity range of 5.8 \kms, with central velocities as labelled in the top right-hand corners of the plots. Note that since the main lines are only separated by $\sim350$ \kms, there are places where negative velocities in the 1665\,MHz line run into the positive velocity space of the 1667\,MHz transition and vice-versa.}
\end{figure*}

\addtocounter{figure}{-1}
\begin{figure*}
\begin{subfigure}{0.48\textwidth}
\includegraphics[width=\linewidth]{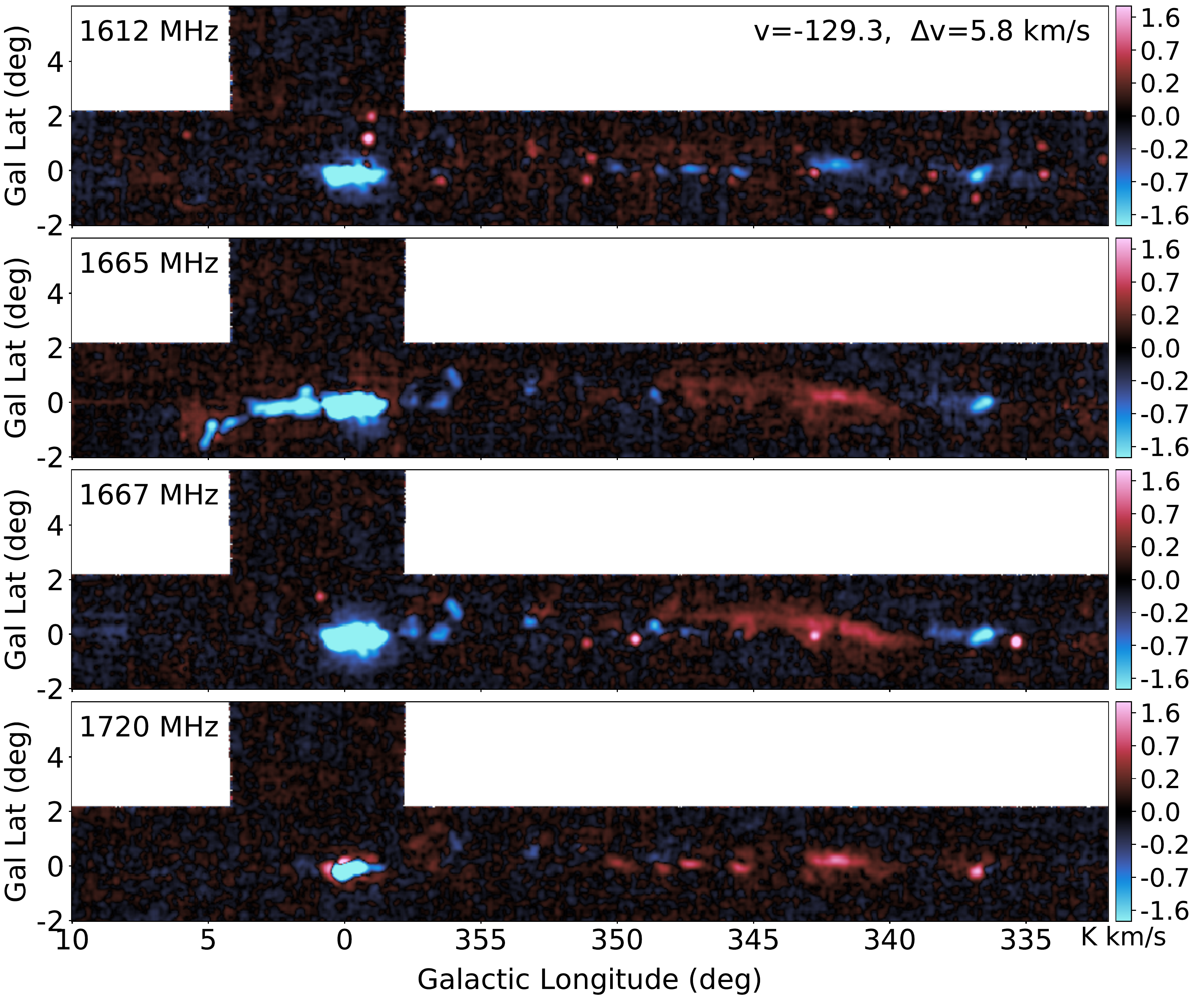}
\end{subfigure}\hspace*{\fill}
\begin{subfigure}{0.48\textwidth}
\includegraphics[width=\linewidth]{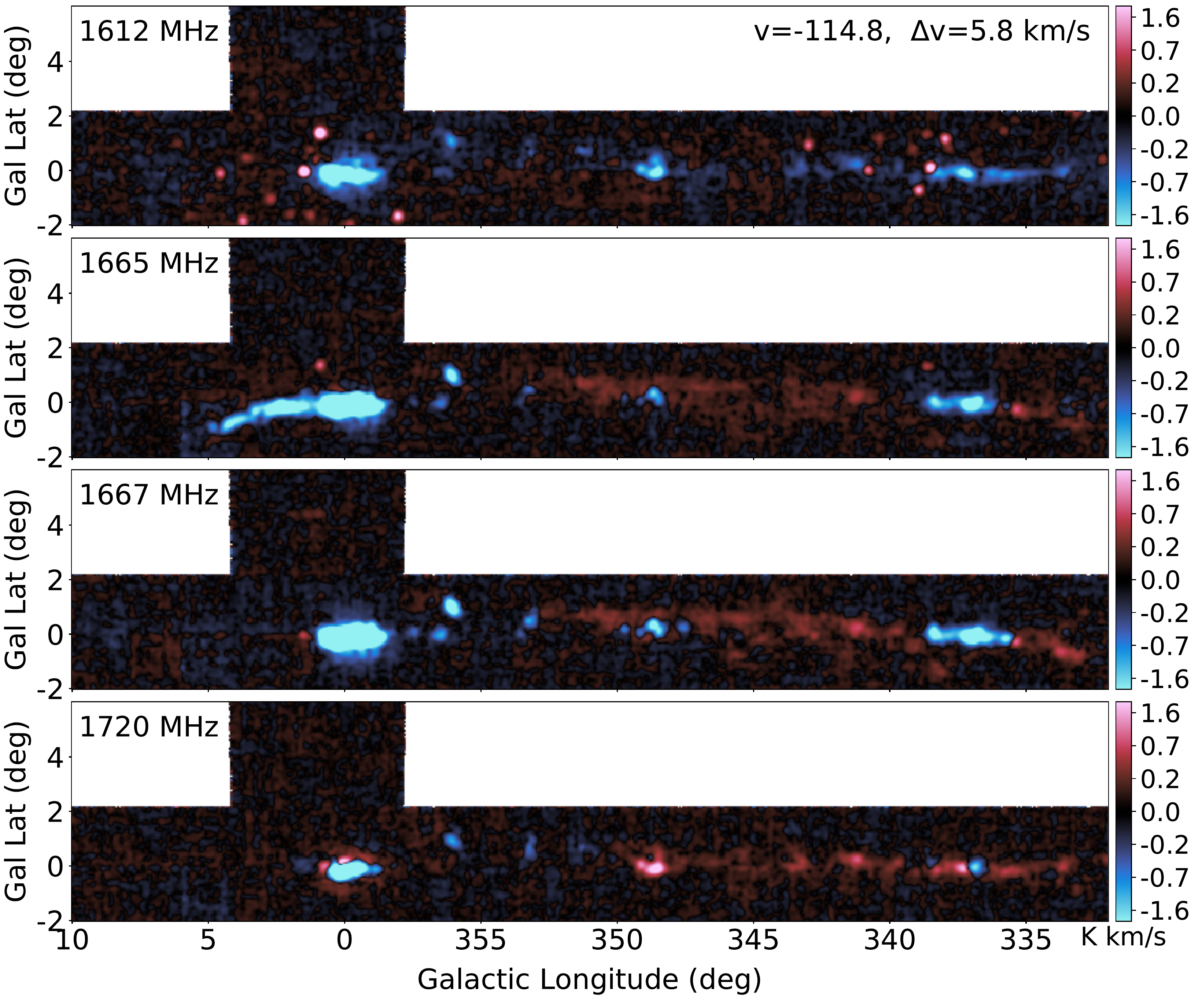}
\end{subfigure}

\medskip
\begin{subfigure}{0.48\textwidth}
\includegraphics[width=\linewidth]{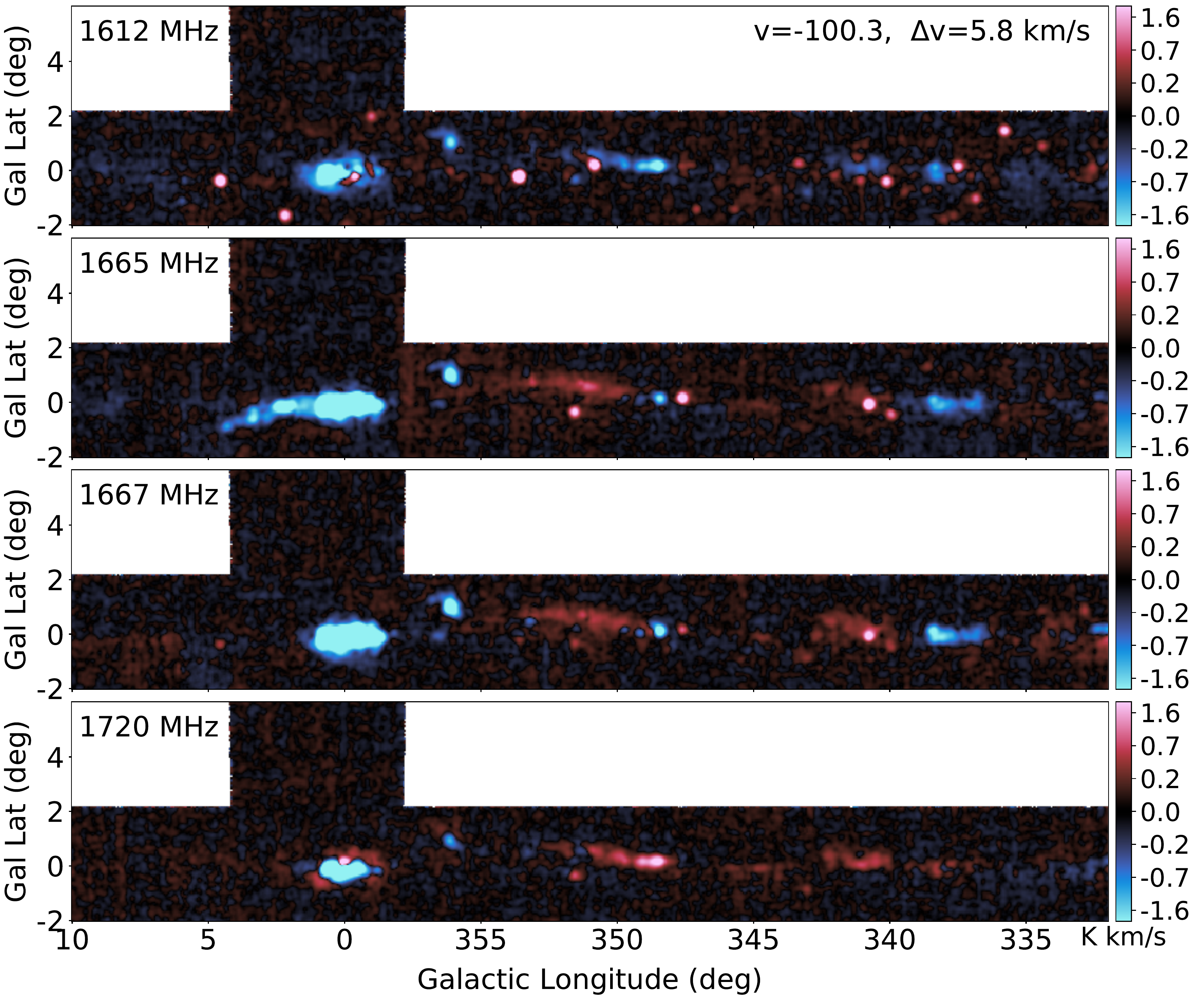}
\end{subfigure}\hspace*{\fill}
\begin{subfigure}{0.48\textwidth}
\includegraphics[width=\linewidth]{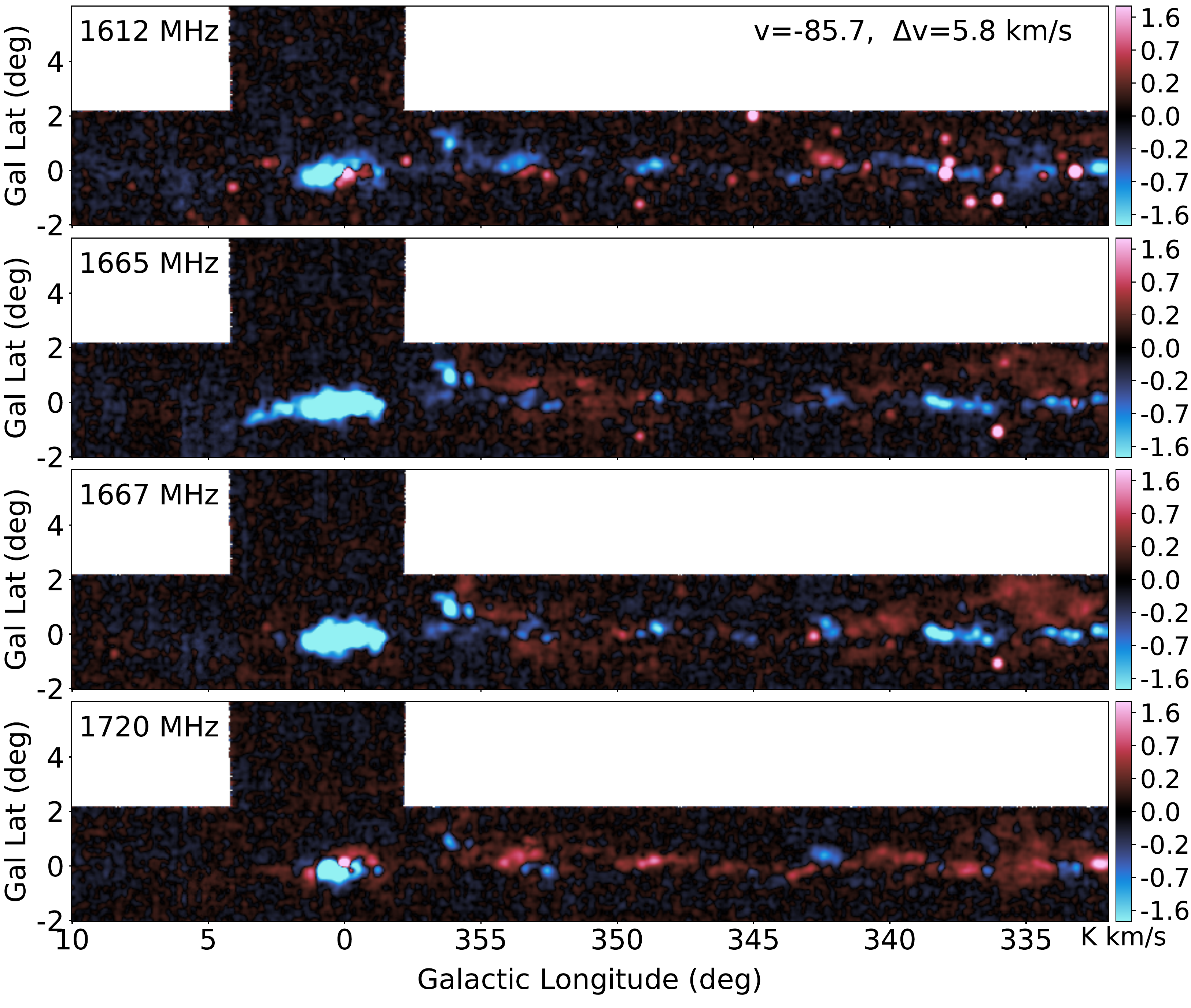}
\end{subfigure}

\medskip
\begin{subfigure}{0.48\textwidth}
\includegraphics[width=\linewidth]{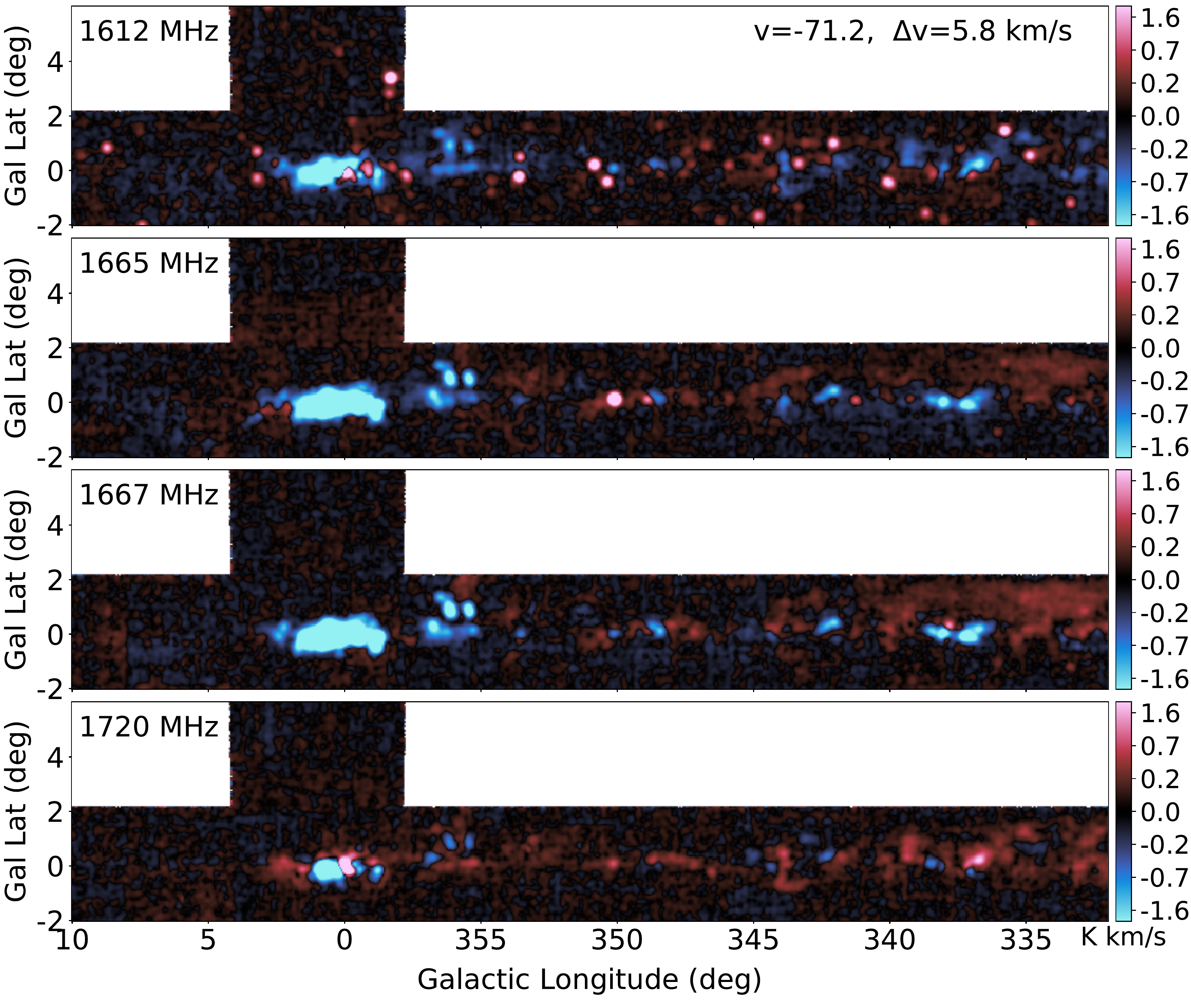}
\end{subfigure}\hspace*{\fill}
\begin{subfigure}{0.48\textwidth}
\includegraphics[width=\linewidth]{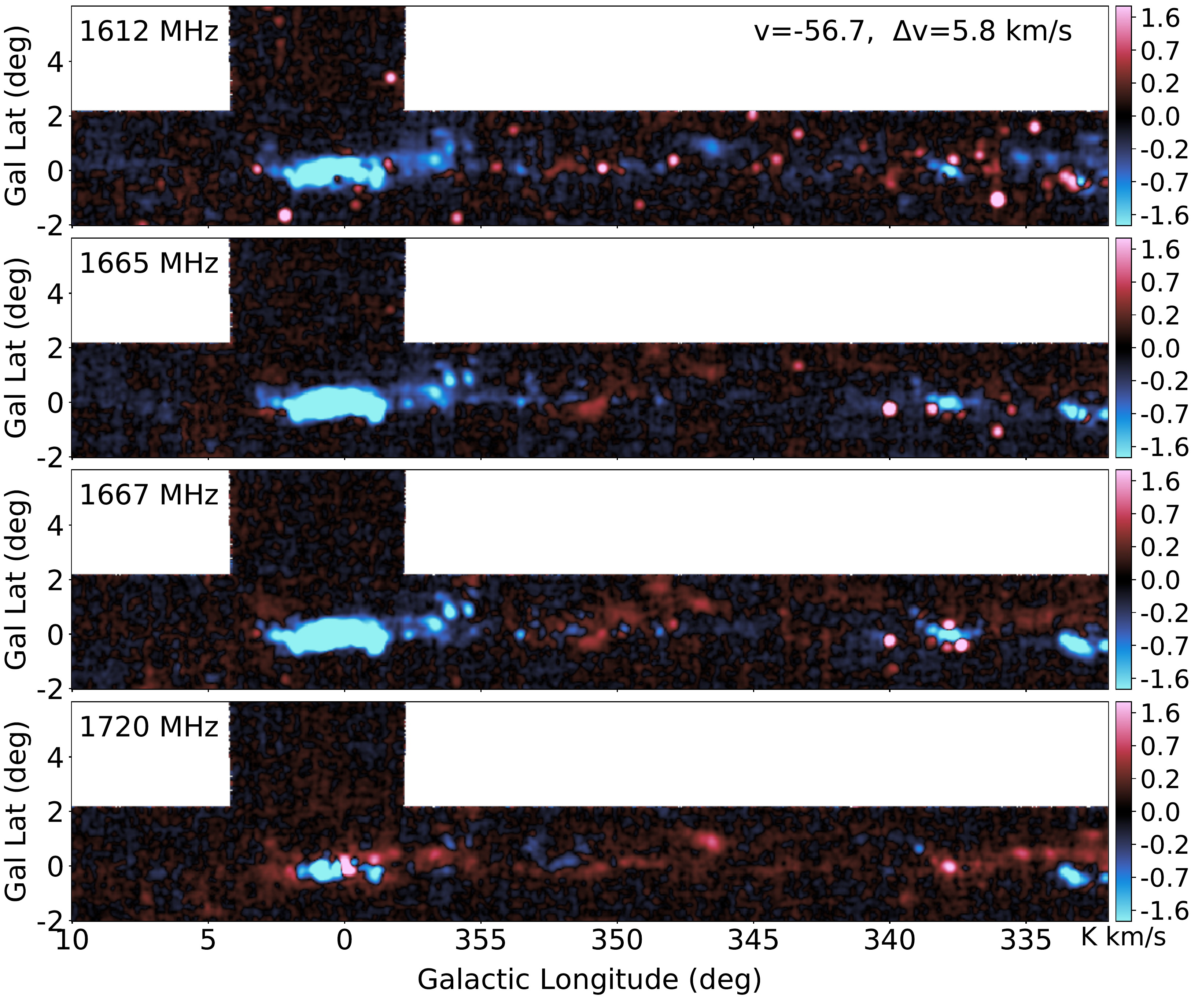}
\end{subfigure}
\caption{(cont.)} 
\end{figure*}

\addtocounter{figure}{-1}
\begin{figure*}
\begin{subfigure}{0.48\textwidth}
\includegraphics[width=\linewidth]{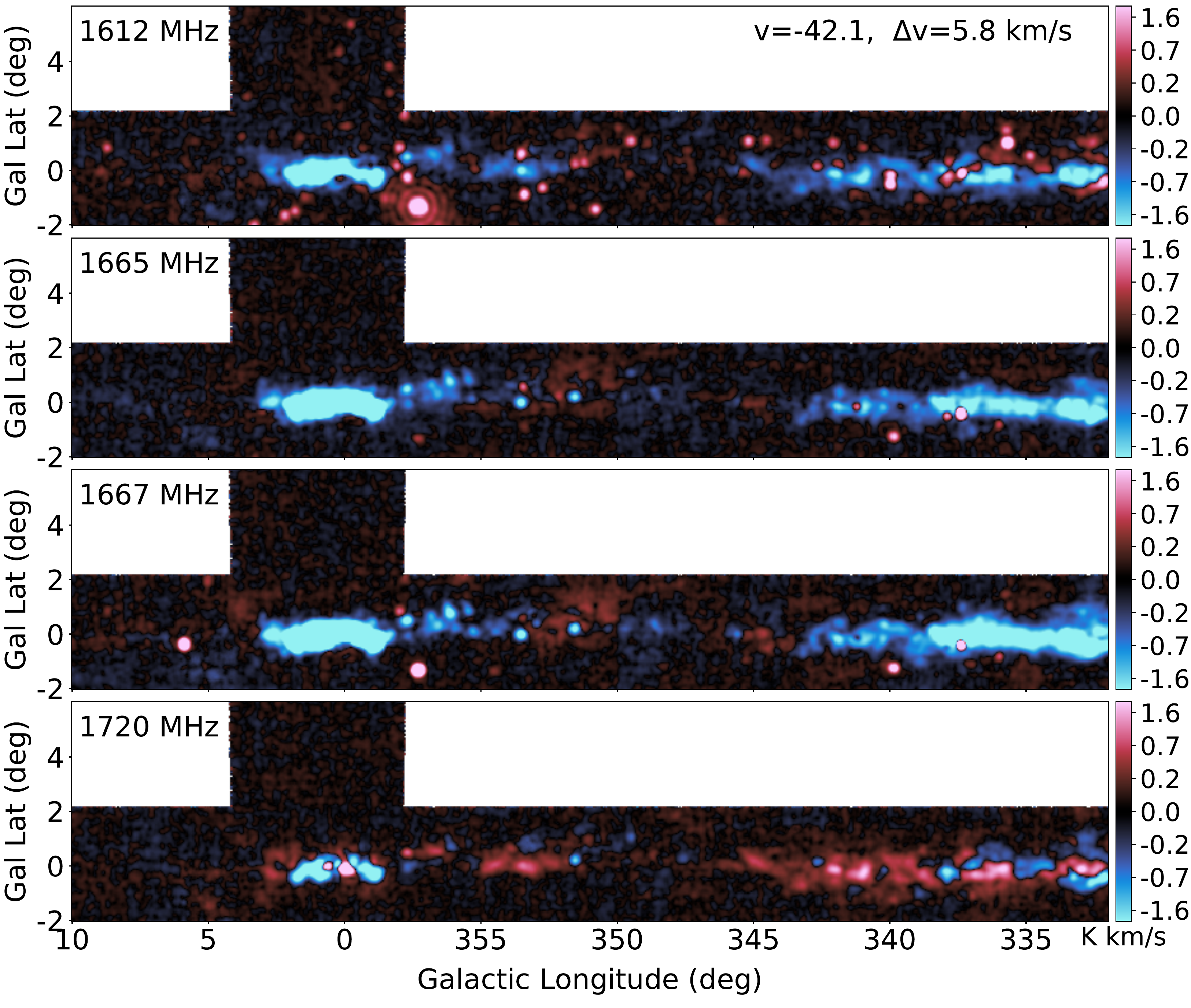}
\end{subfigure}\hspace*{\fill}
\begin{subfigure}{0.48\textwidth}
\includegraphics[width=\linewidth]{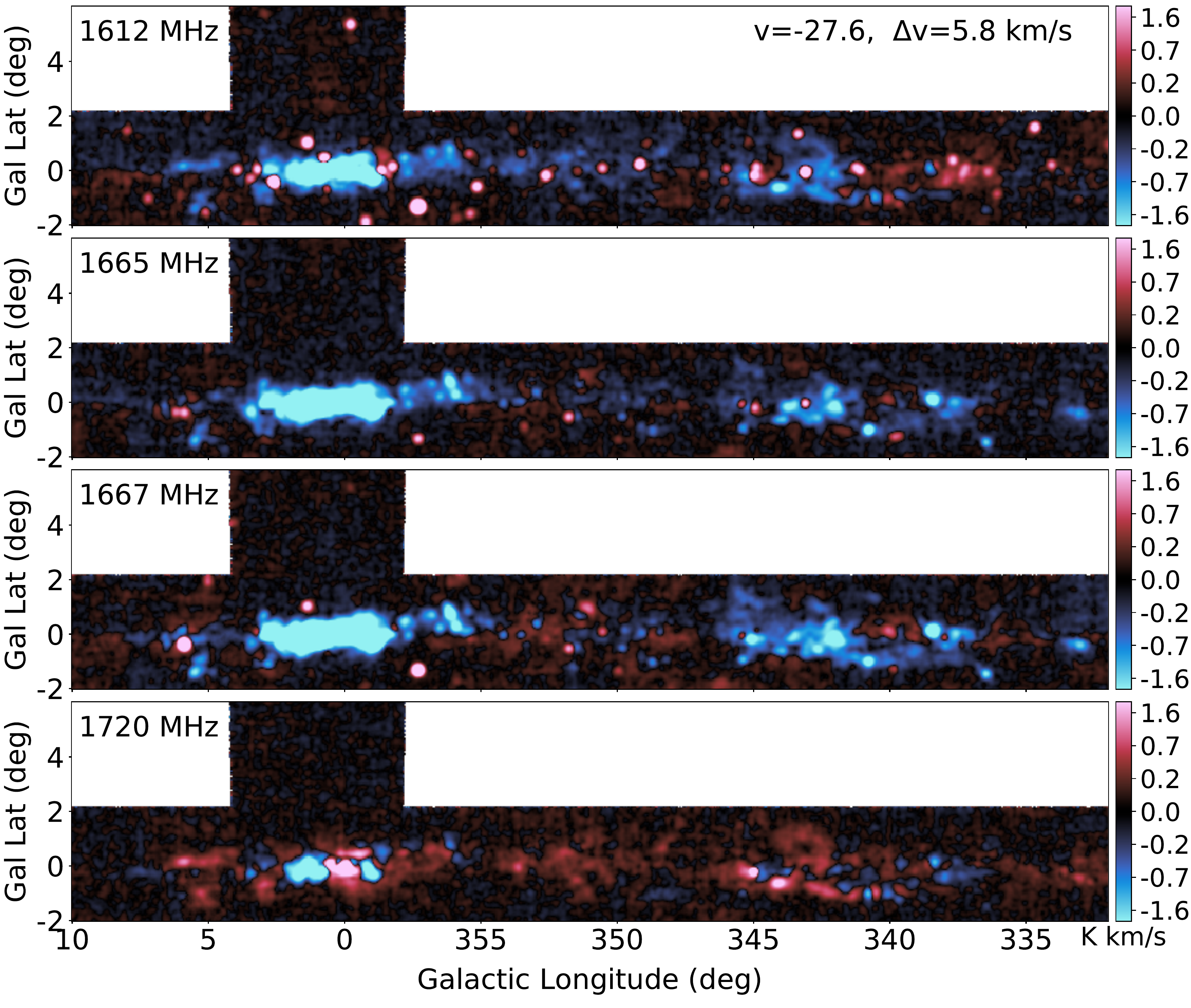}
\end{subfigure}

\medskip
\begin{subfigure}{0.48\textwidth}
\includegraphics[width=\linewidth]{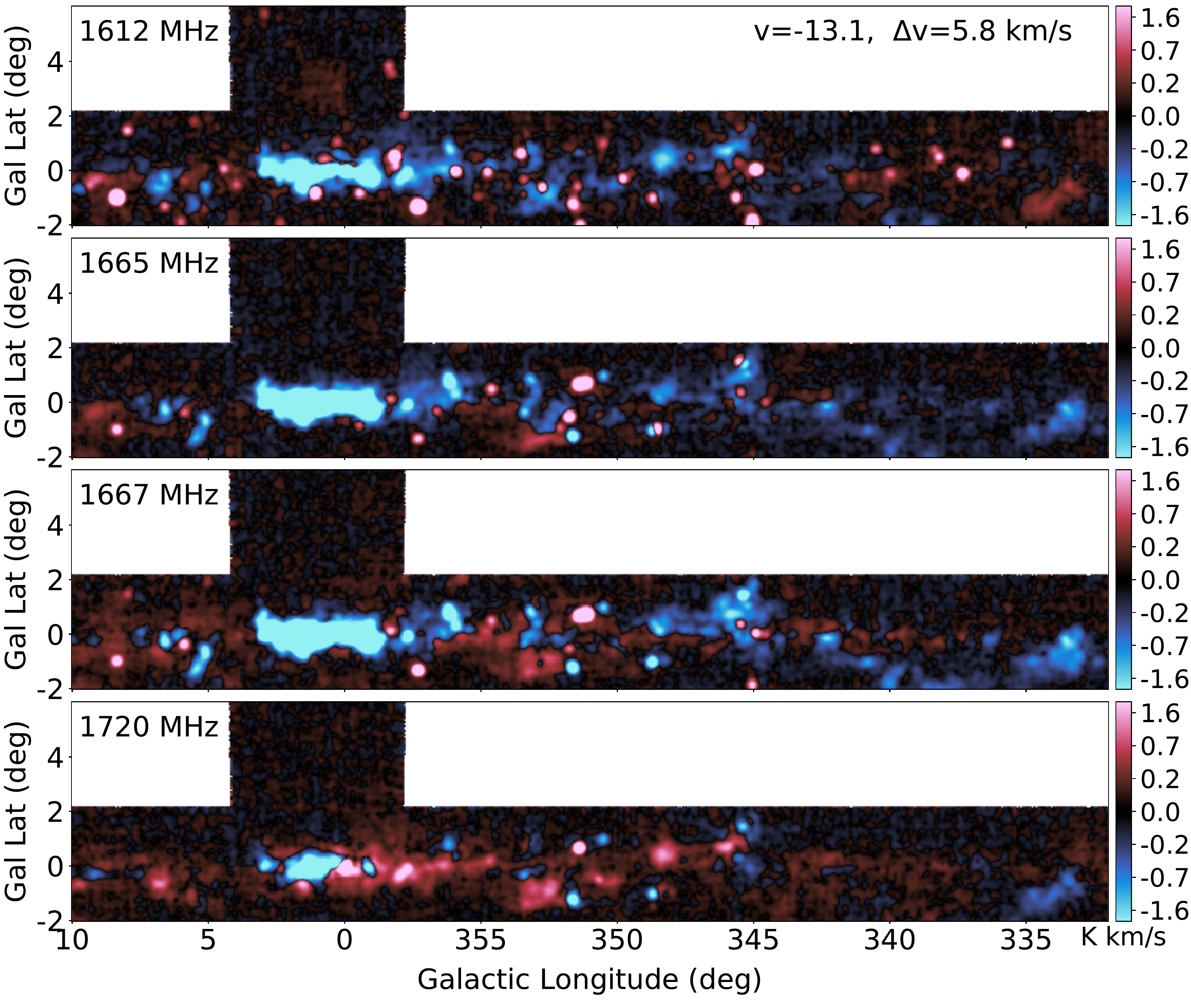}
\end{subfigure}\hspace*{\fill}
\begin{subfigure}{0.48\textwidth}
\includegraphics[width=\linewidth]{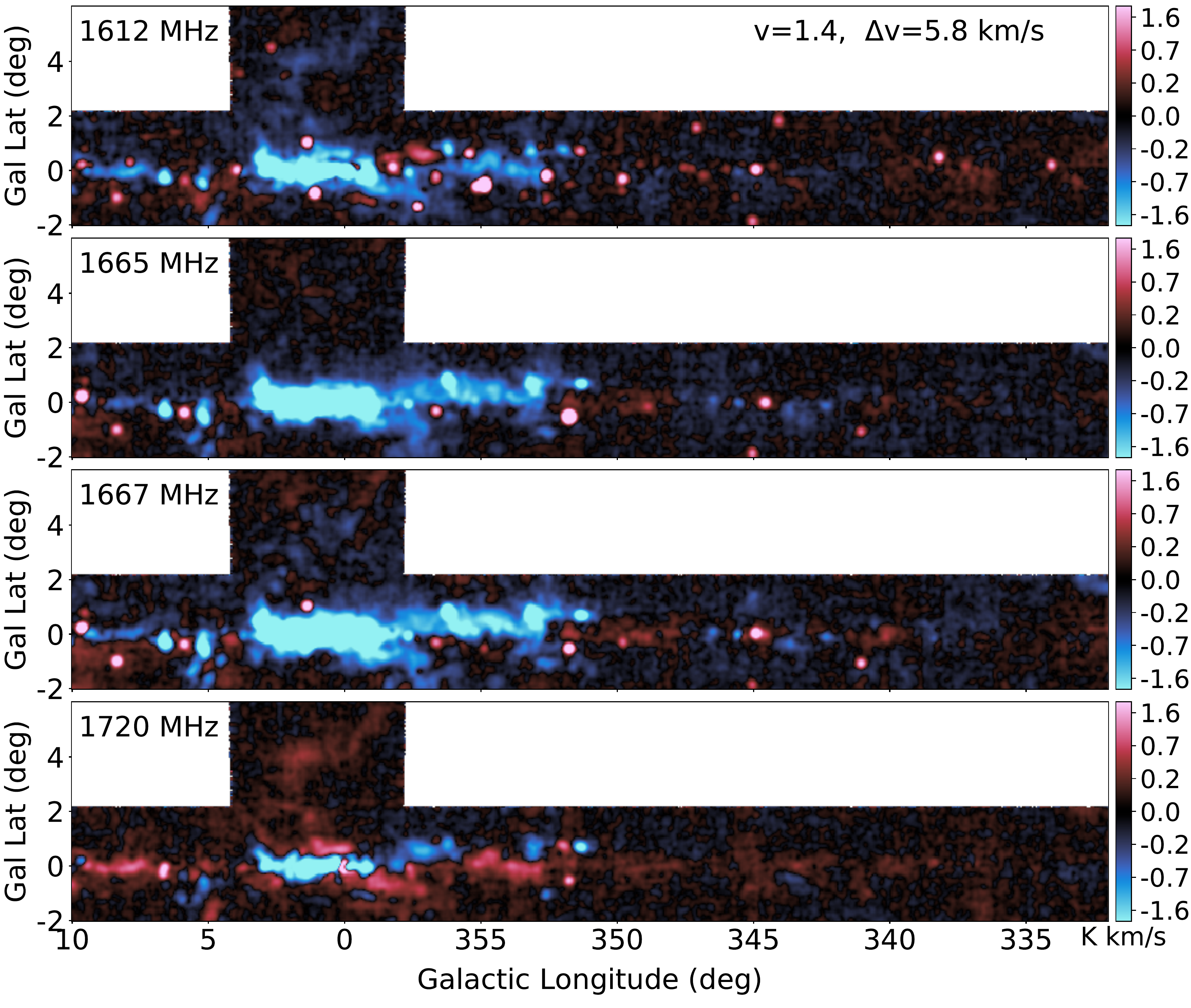}
\end{subfigure}

\medskip
\begin{subfigure}{0.48\textwidth}
\includegraphics[width=\linewidth]{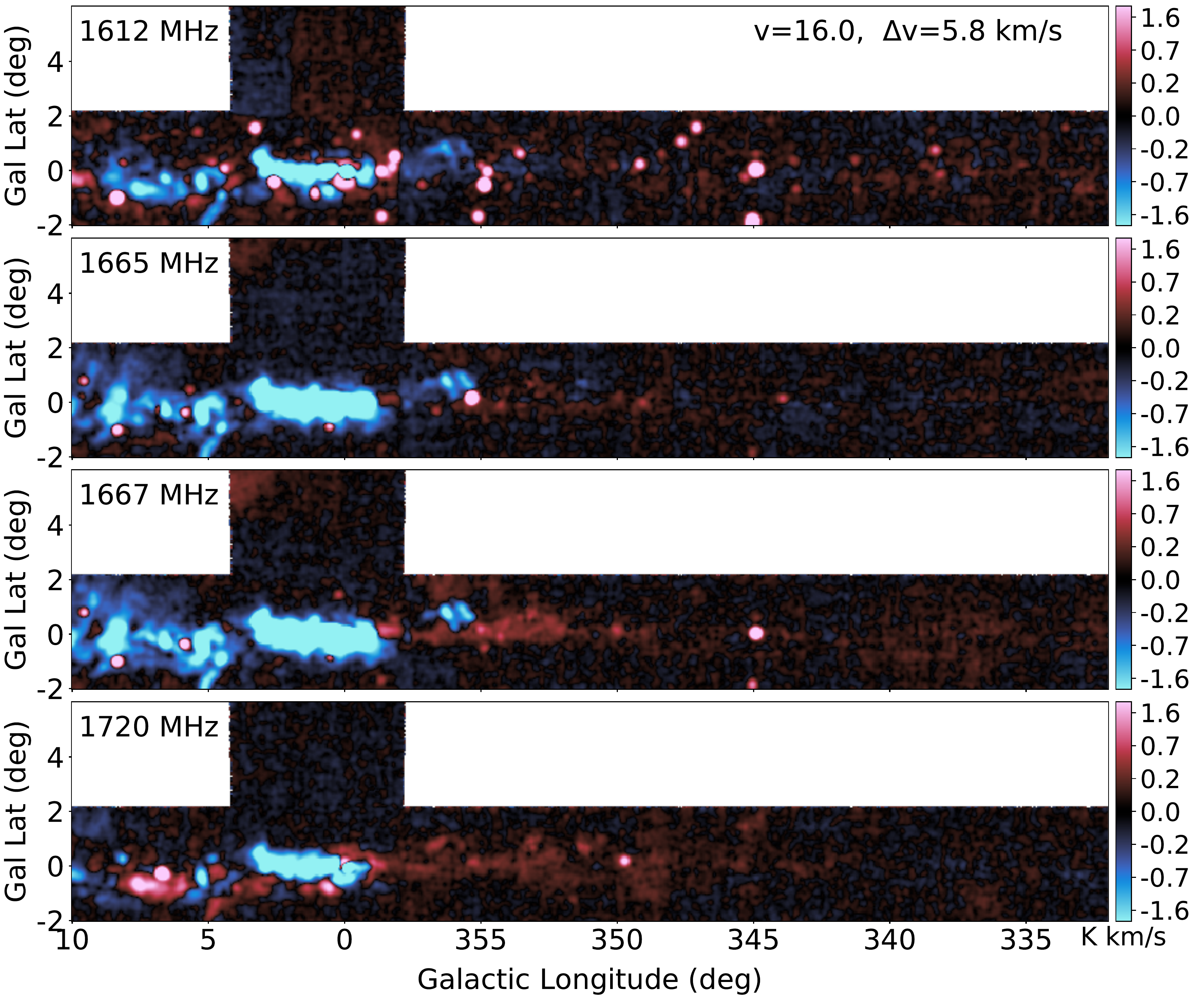}
\end{subfigure}\hspace*{\fill}
\begin{subfigure}{0.48\textwidth}
\includegraphics[width=\linewidth]{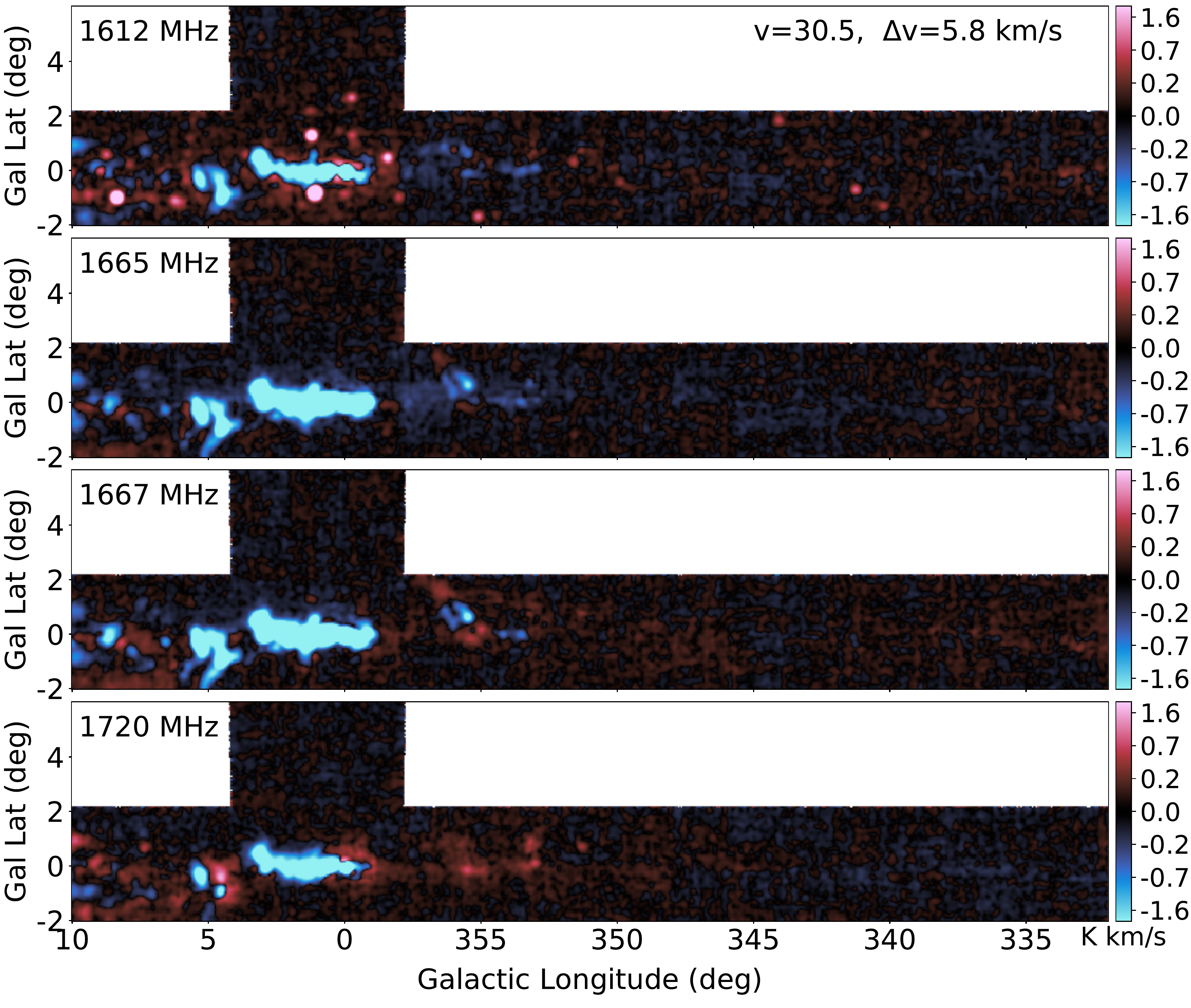}
\end{subfigure}

\caption{(cont.)} 
\end{figure*}

\addtocounter{figure}{-1}
\begin{figure*}
\begin{subfigure}{0.48\textwidth}
\includegraphics[width=\linewidth]{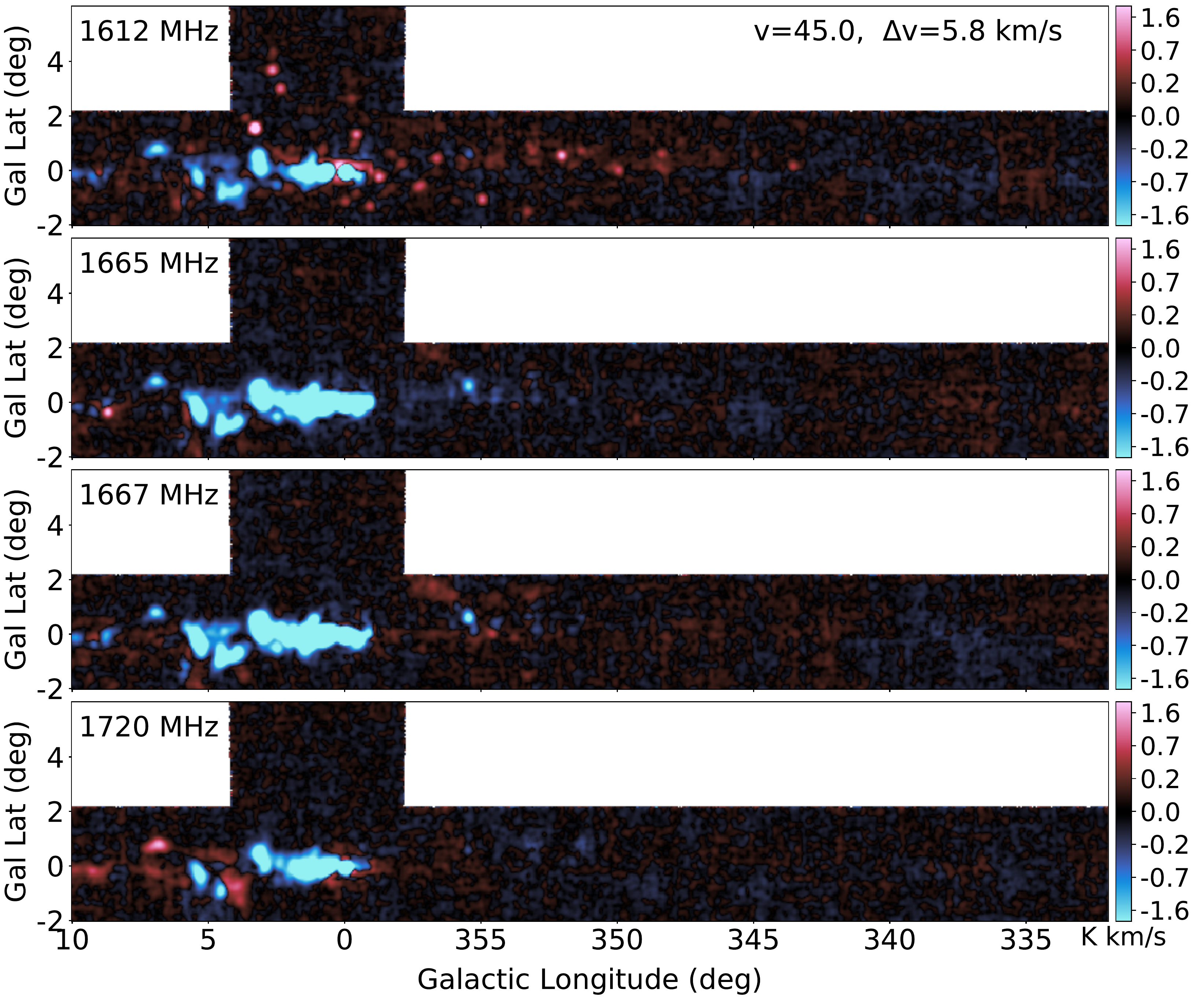}
\end{subfigure}\hspace*{\fill}
\begin{subfigure}{0.48\textwidth}
\includegraphics[width=\linewidth]{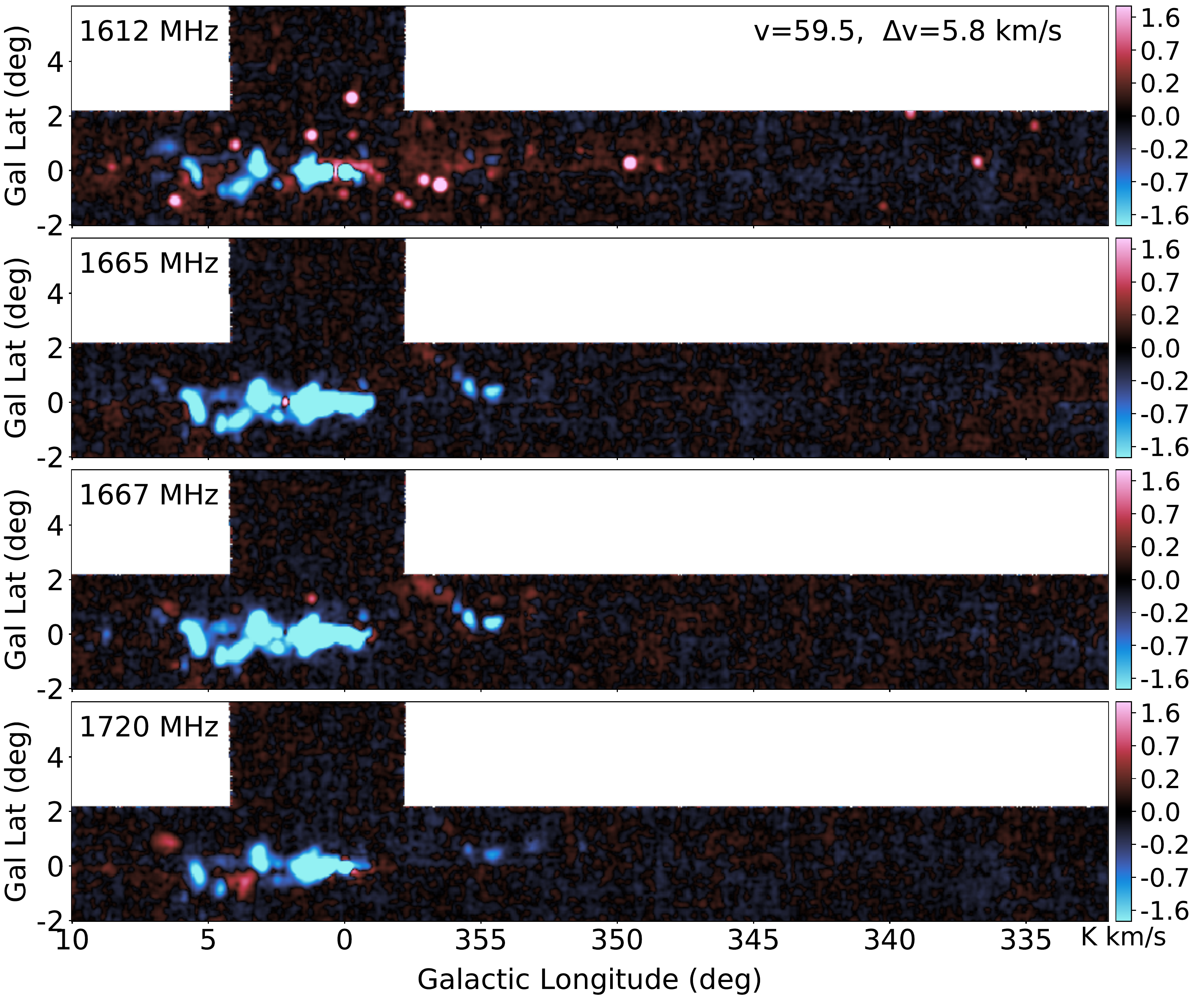}
\end{subfigure}

\medskip
\begin{subfigure}{0.48\textwidth}
\includegraphics[width=\linewidth]{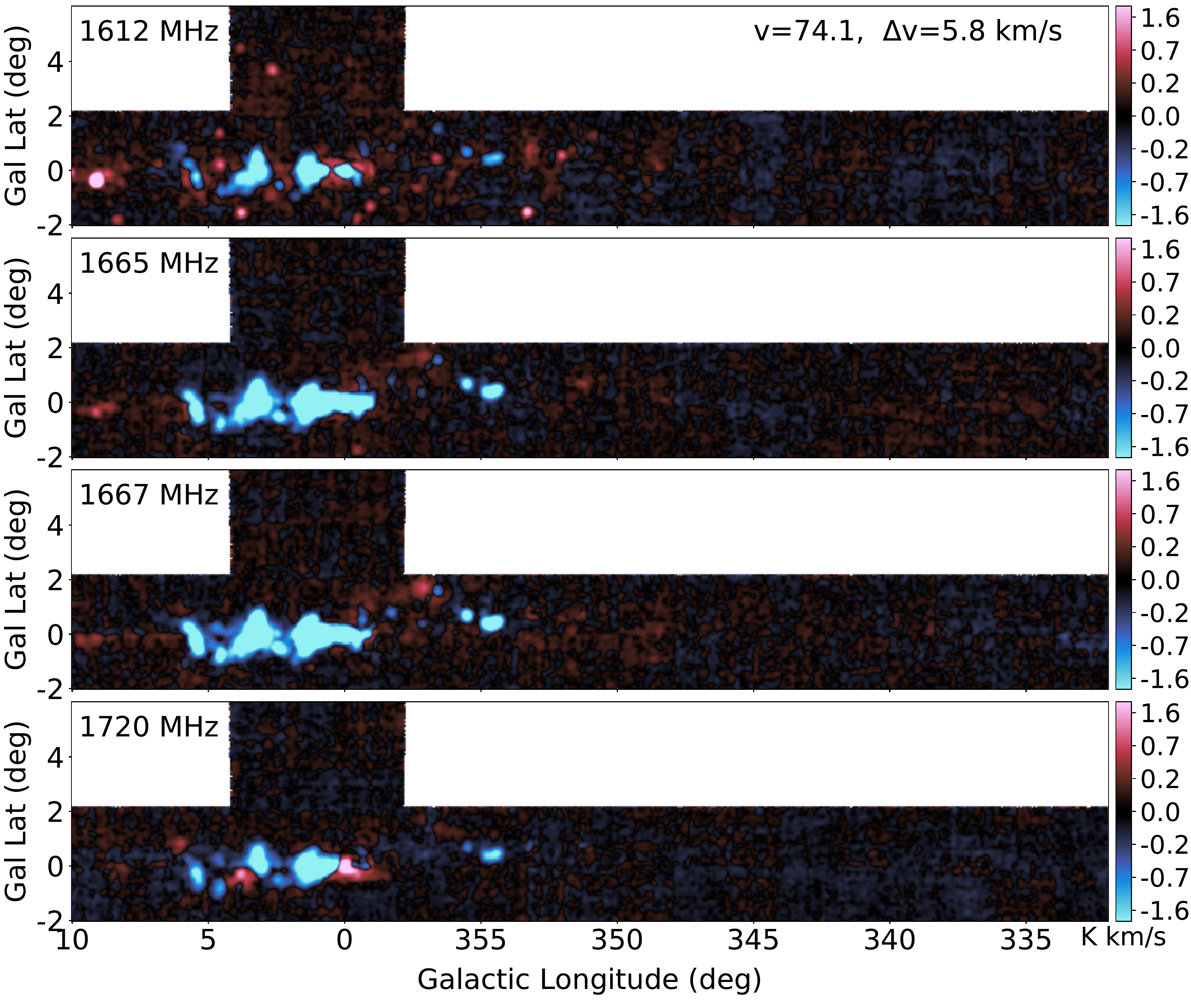}
\end{subfigure}\hspace*{\fill}
\begin{subfigure}{0.48\textwidth}
\includegraphics[width=\linewidth]{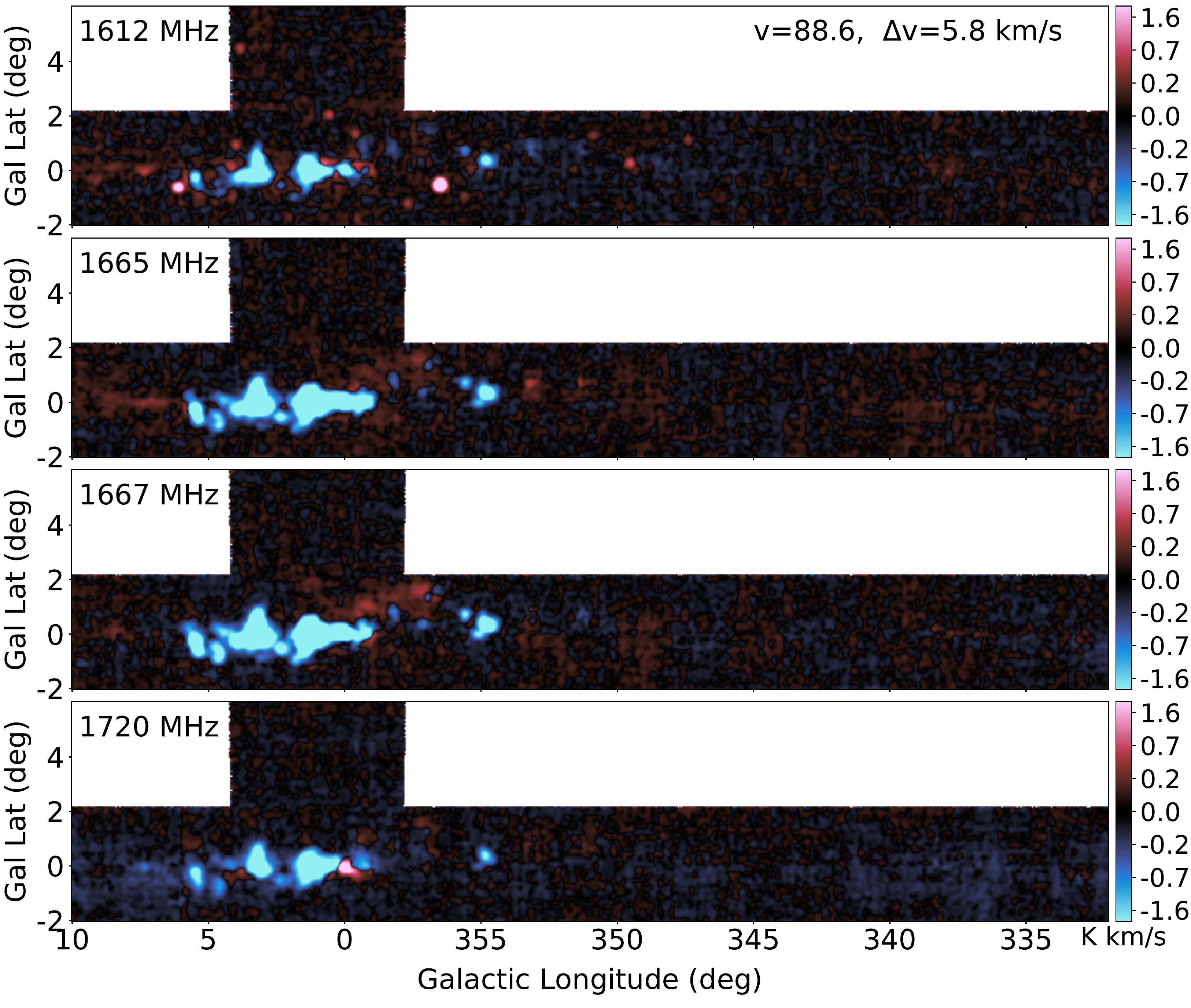}
\end{subfigure}

\medskip
\begin{subfigure}{0.48\textwidth}
\includegraphics[width=\linewidth]{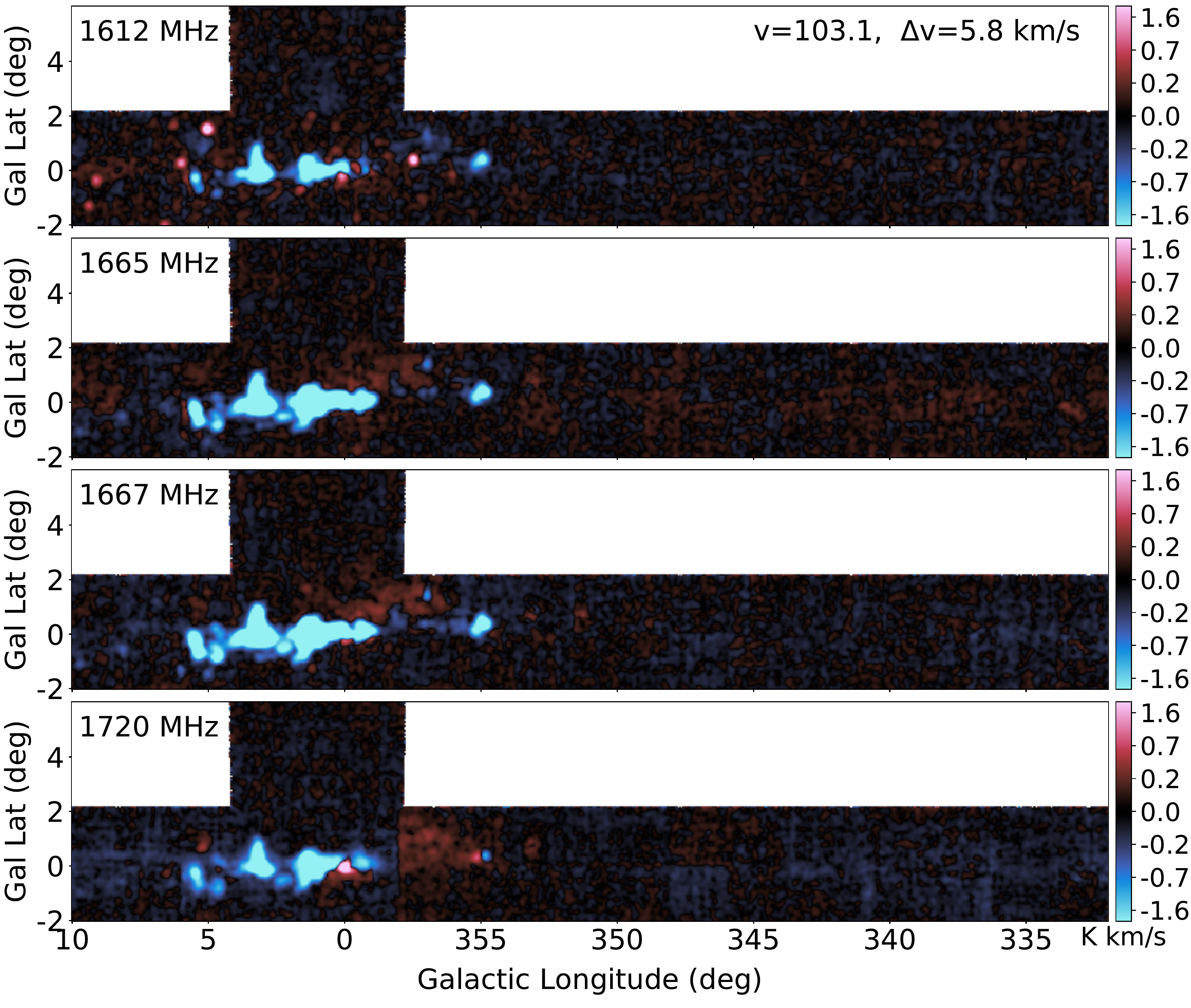}
\end{subfigure}\hspace*{\fill}
\begin{subfigure}{0.48\textwidth}
\includegraphics[width=\linewidth]{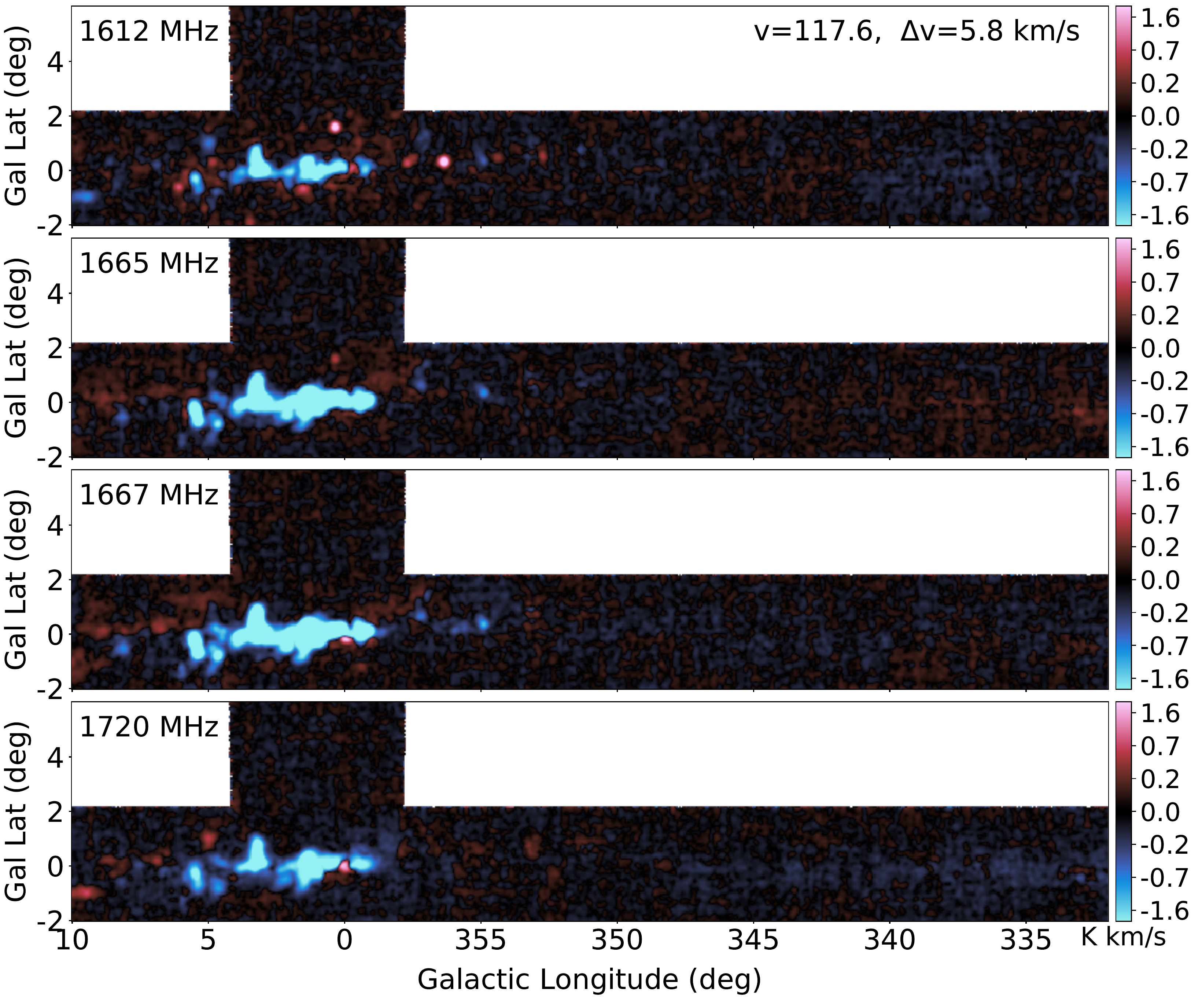}
\end{subfigure}
\caption{(cont.)} 
\end{figure*}

\addtocounter{figure}{-1}
\begin{figure*}
\begin{subfigure}{0.48\textwidth}
\includegraphics[width=\linewidth]{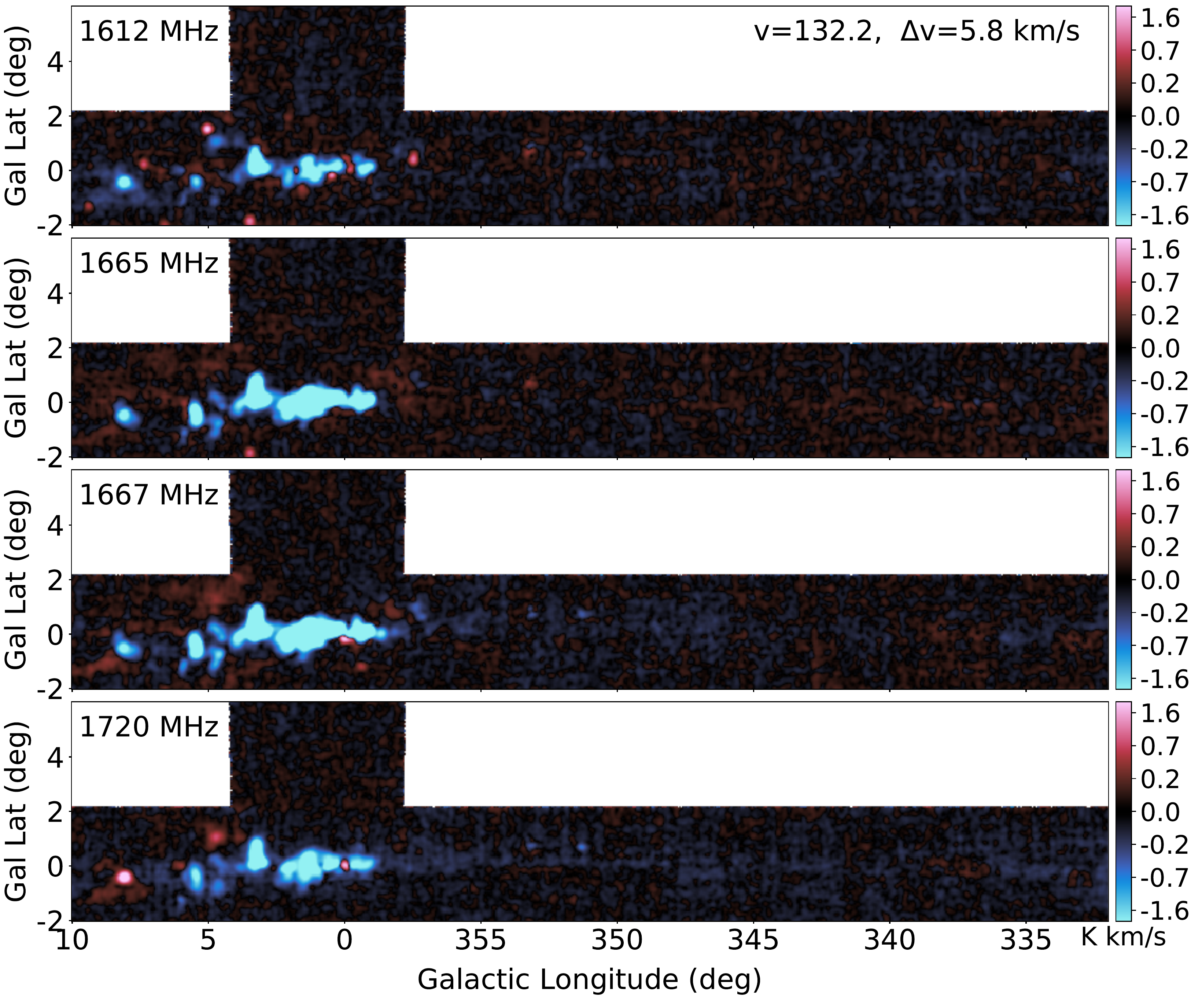}
\end{subfigure}\hspace*{\fill}
\begin{subfigure}{0.48\textwidth}
\includegraphics[width=\linewidth]{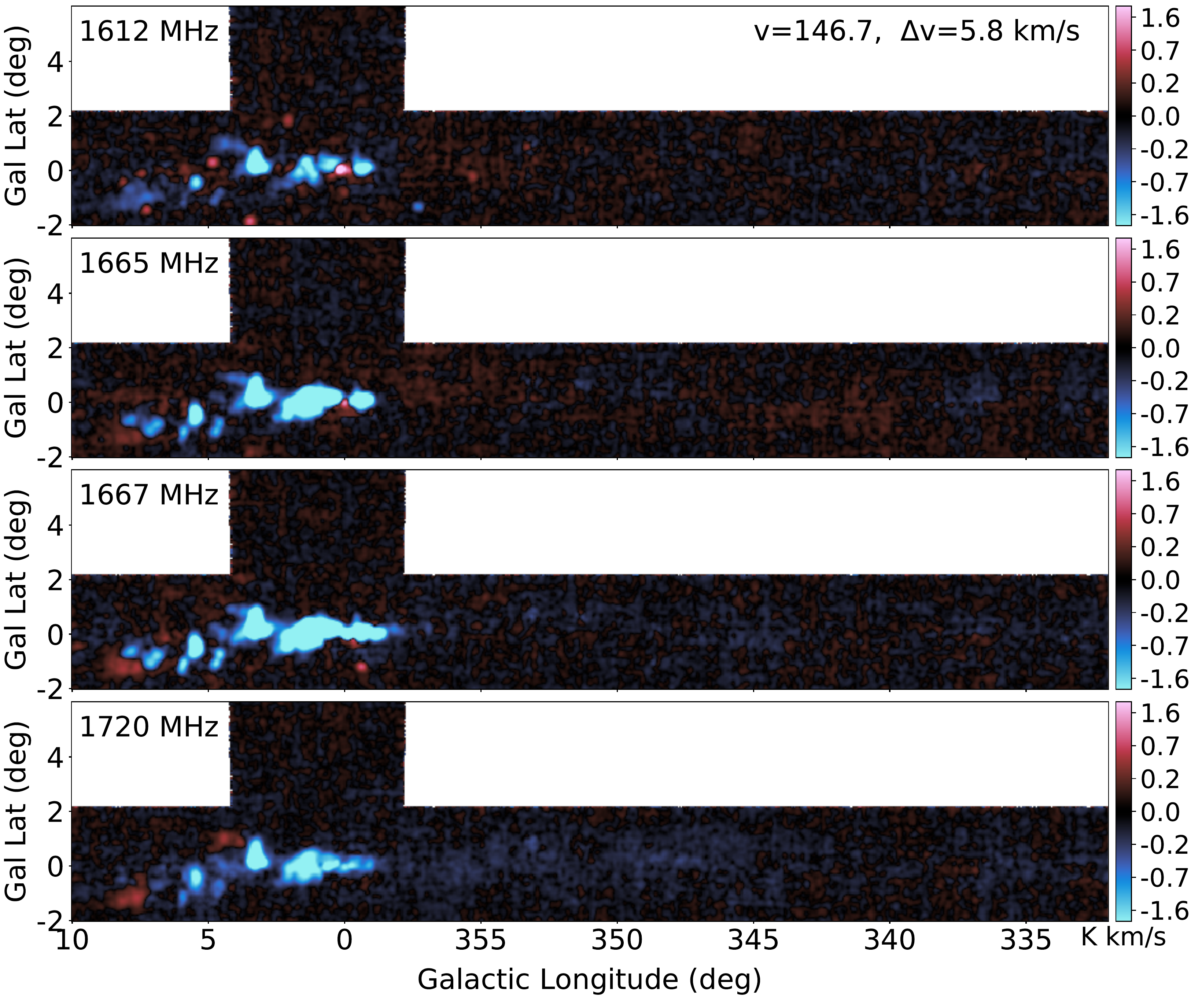}
\end{subfigure}

\medskip
\begin{subfigure}{0.48\textwidth}
\includegraphics[width=\linewidth]{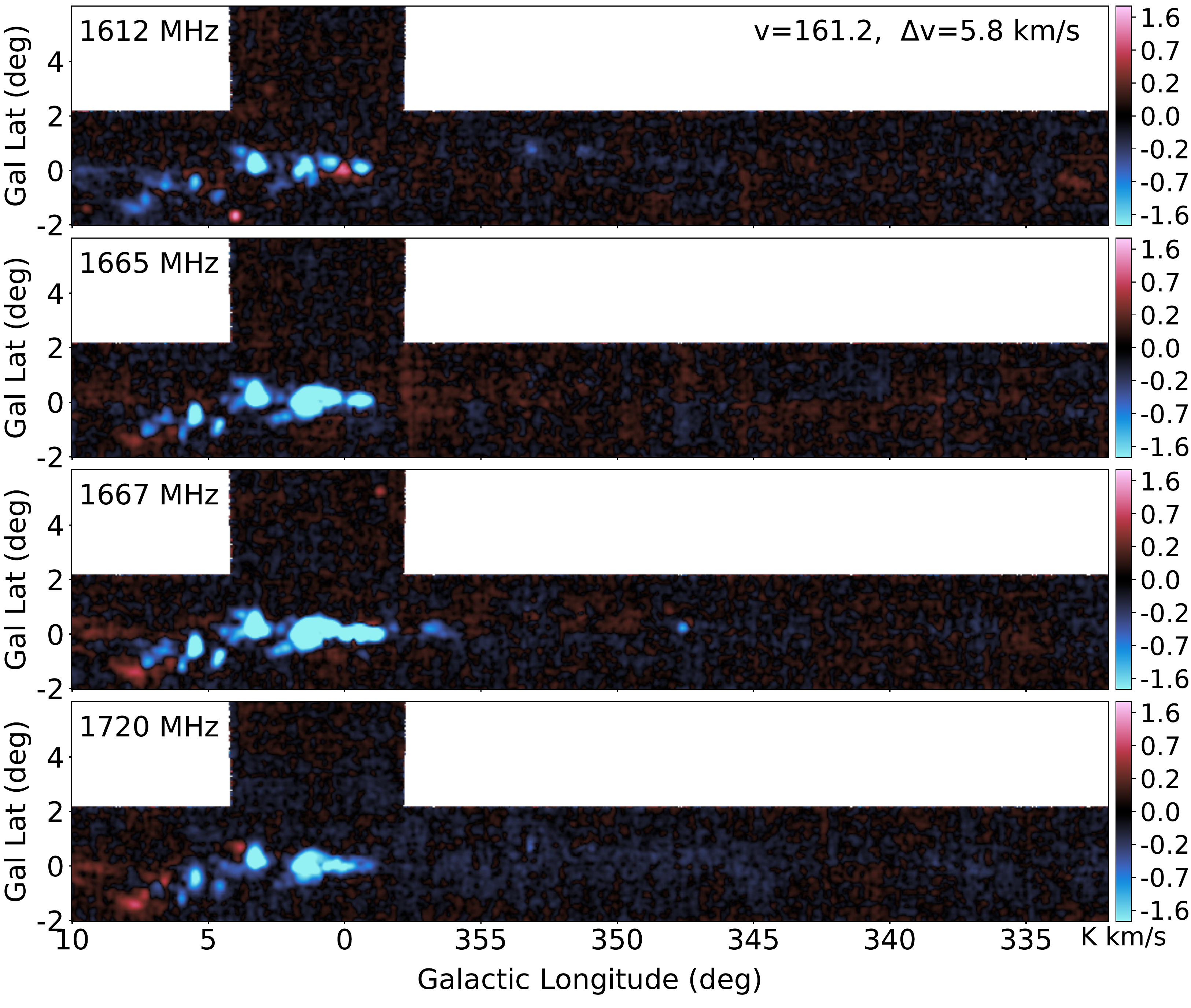}
\end{subfigure}\hspace*{\fill}
\begin{subfigure}{0.48\textwidth}
\includegraphics[width=\linewidth]{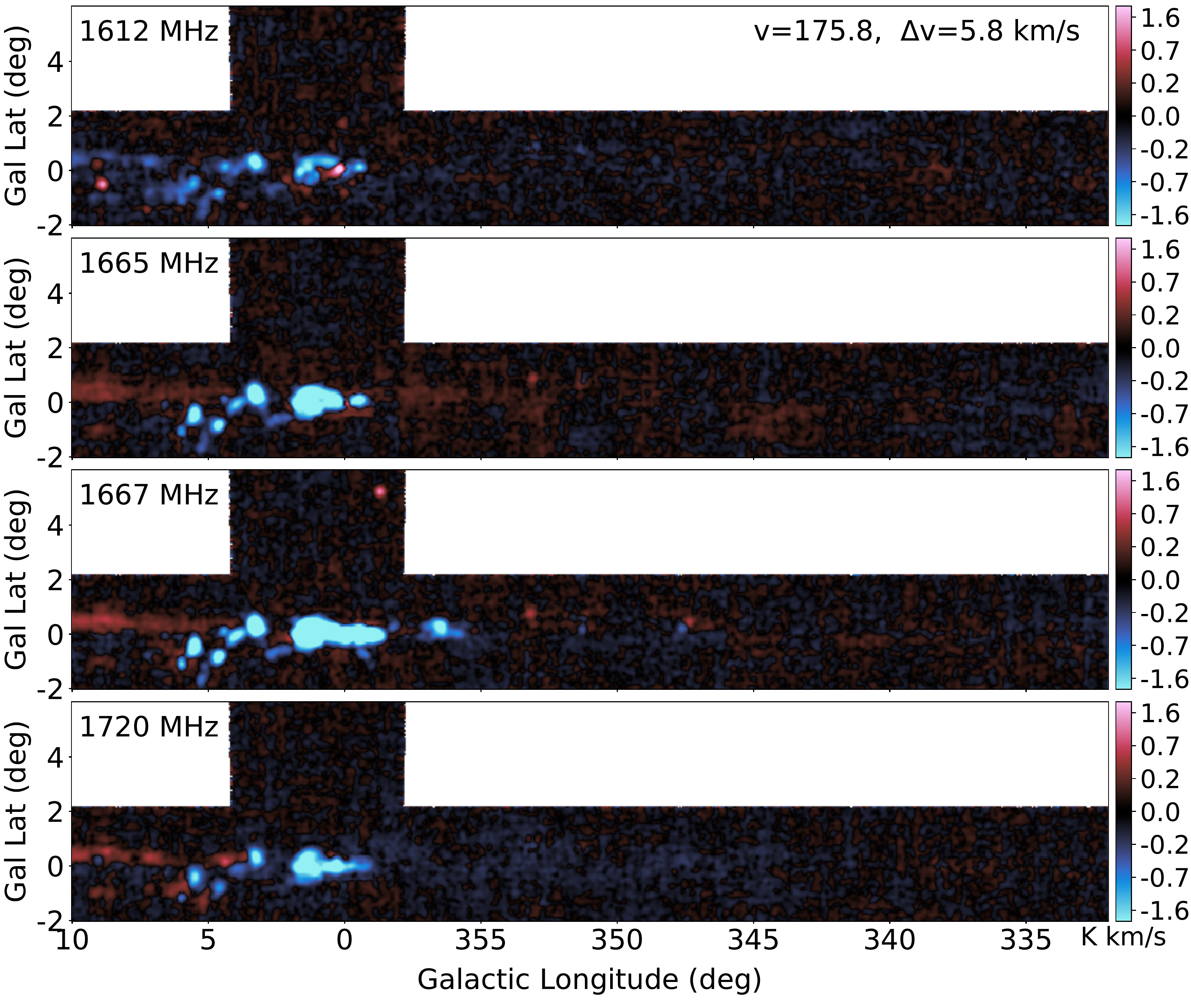}
\end{subfigure}

\medskip
\begin{subfigure}{0.48\textwidth}
\includegraphics[width=\linewidth]{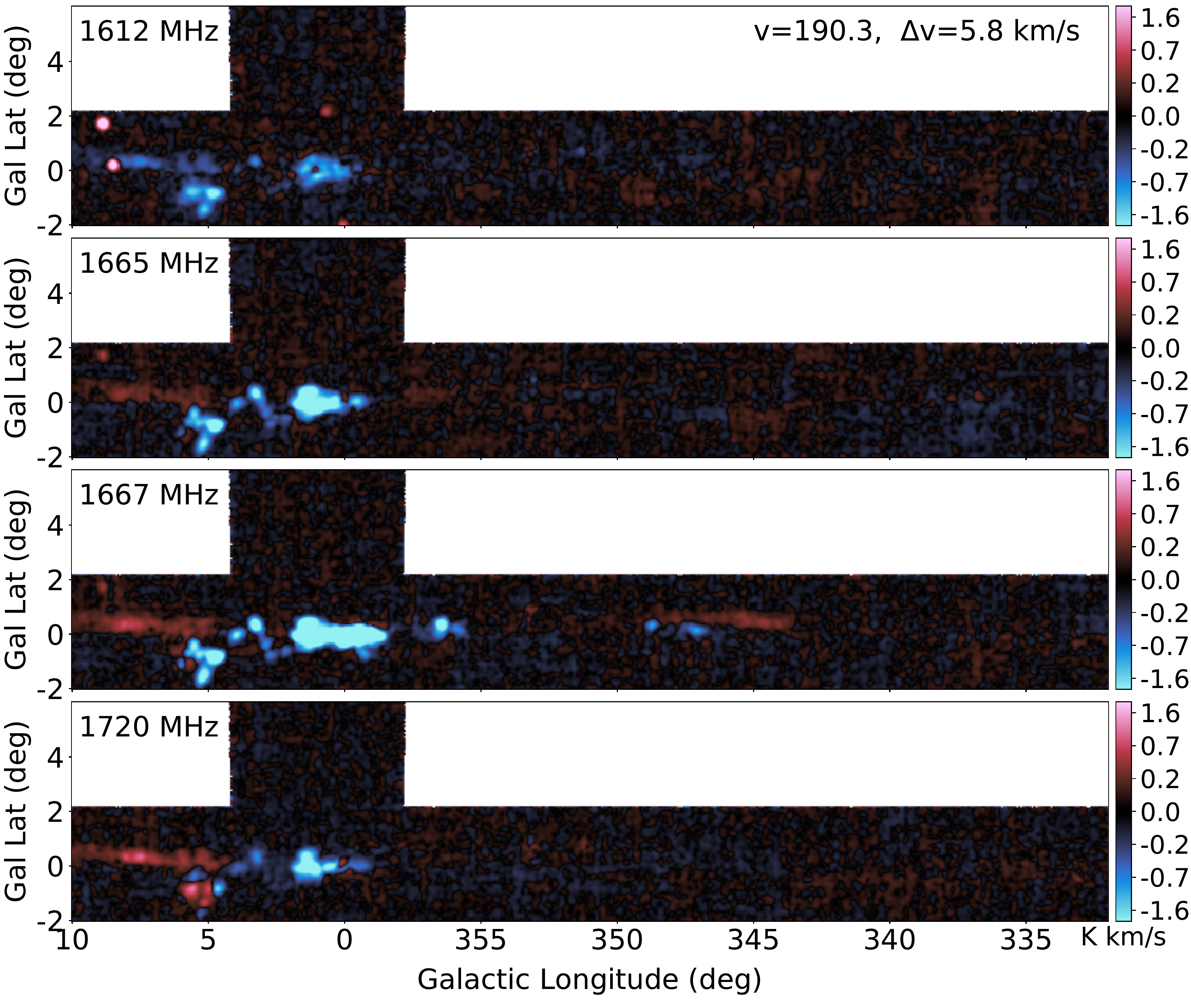}
\end{subfigure}\hspace*{\fill}
\begin{subfigure}{0.48\textwidth}
\includegraphics[width=\linewidth]{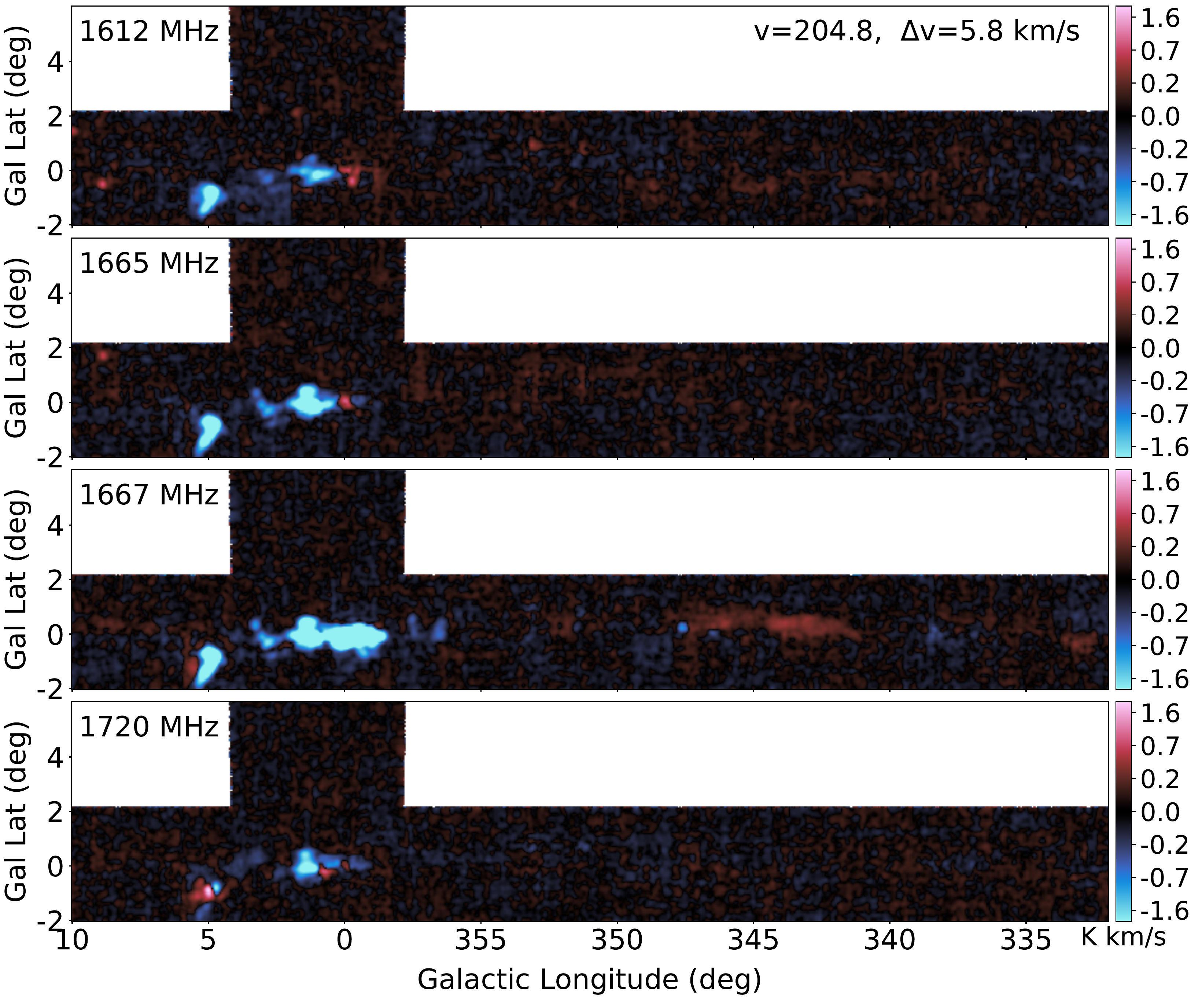}
\end{subfigure}
\caption{(cont.)} 
\end{figure*}


\newpage

\begin{figure*}

\begin{subfigure}{1.0\textwidth}
\centering
\includegraphics[scale=0.43]{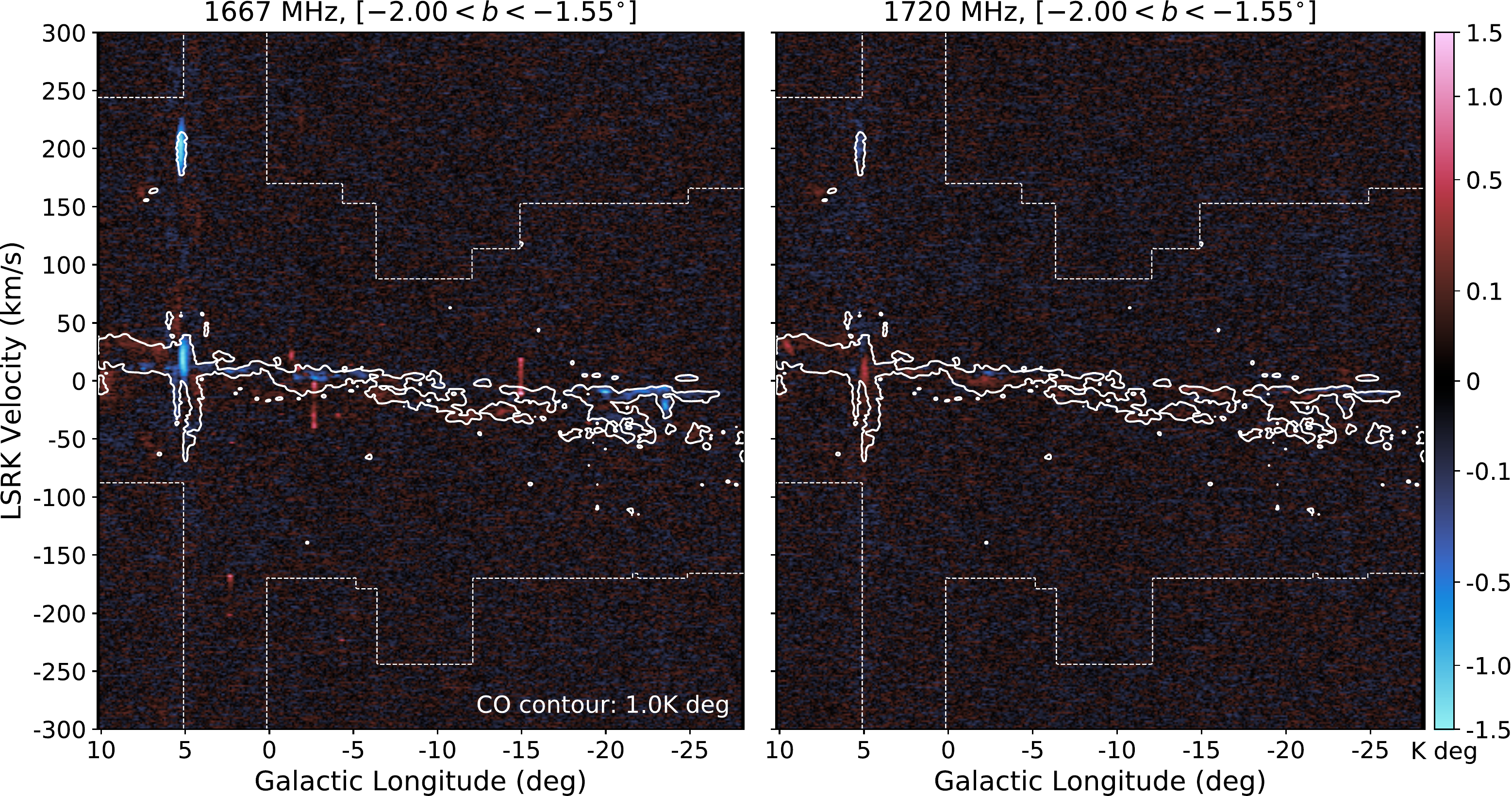}
\end{subfigure}\vspace*{0.25cm}

\begin{subfigure}{1.0\textwidth}
\centering
\includegraphics[scale=0.43]{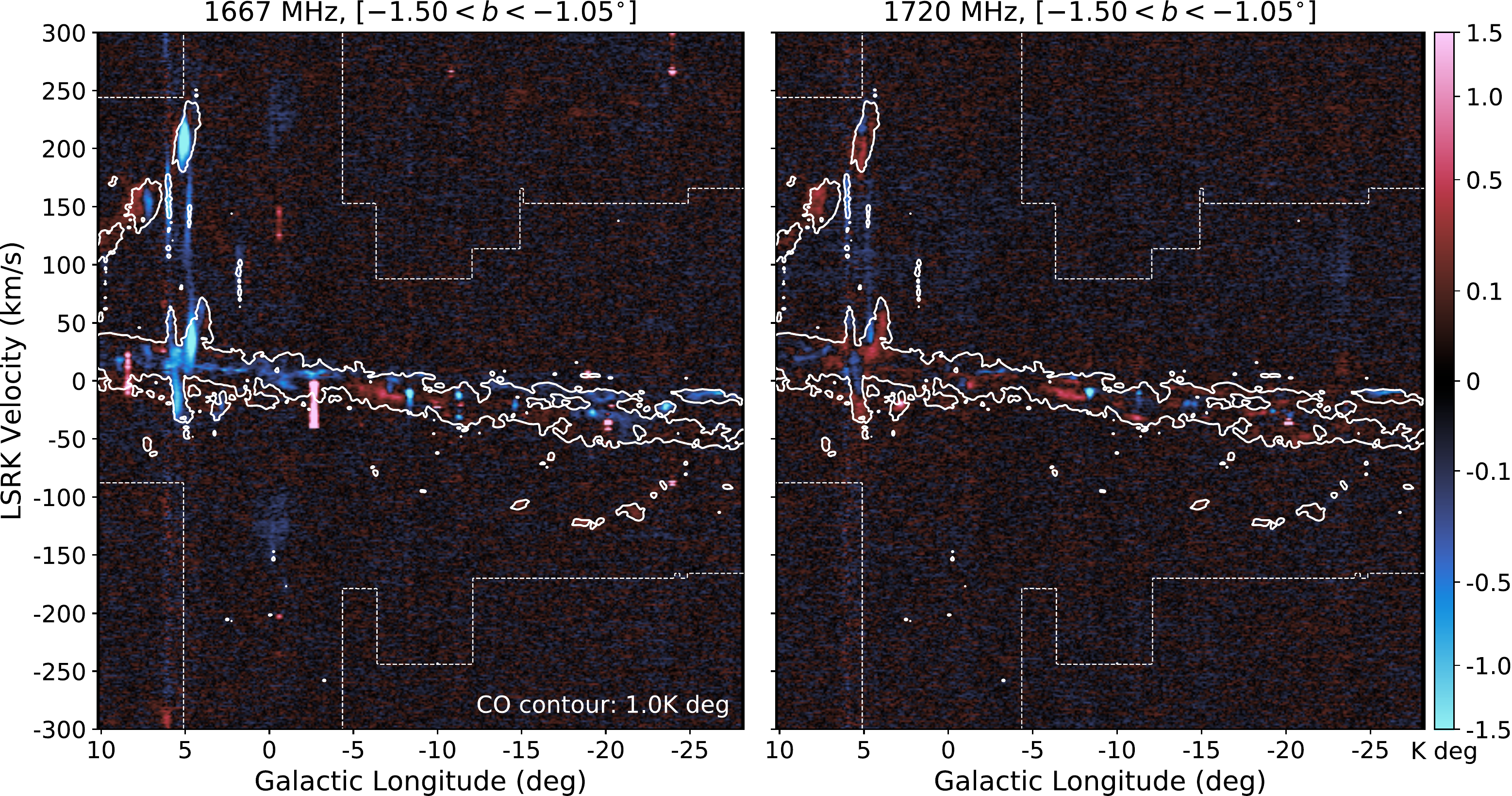}
\end{subfigure}\vspace*{0.25cm}

\begin{subfigure}{1.0\textwidth}
\centering
\includegraphics[scale=0.43]{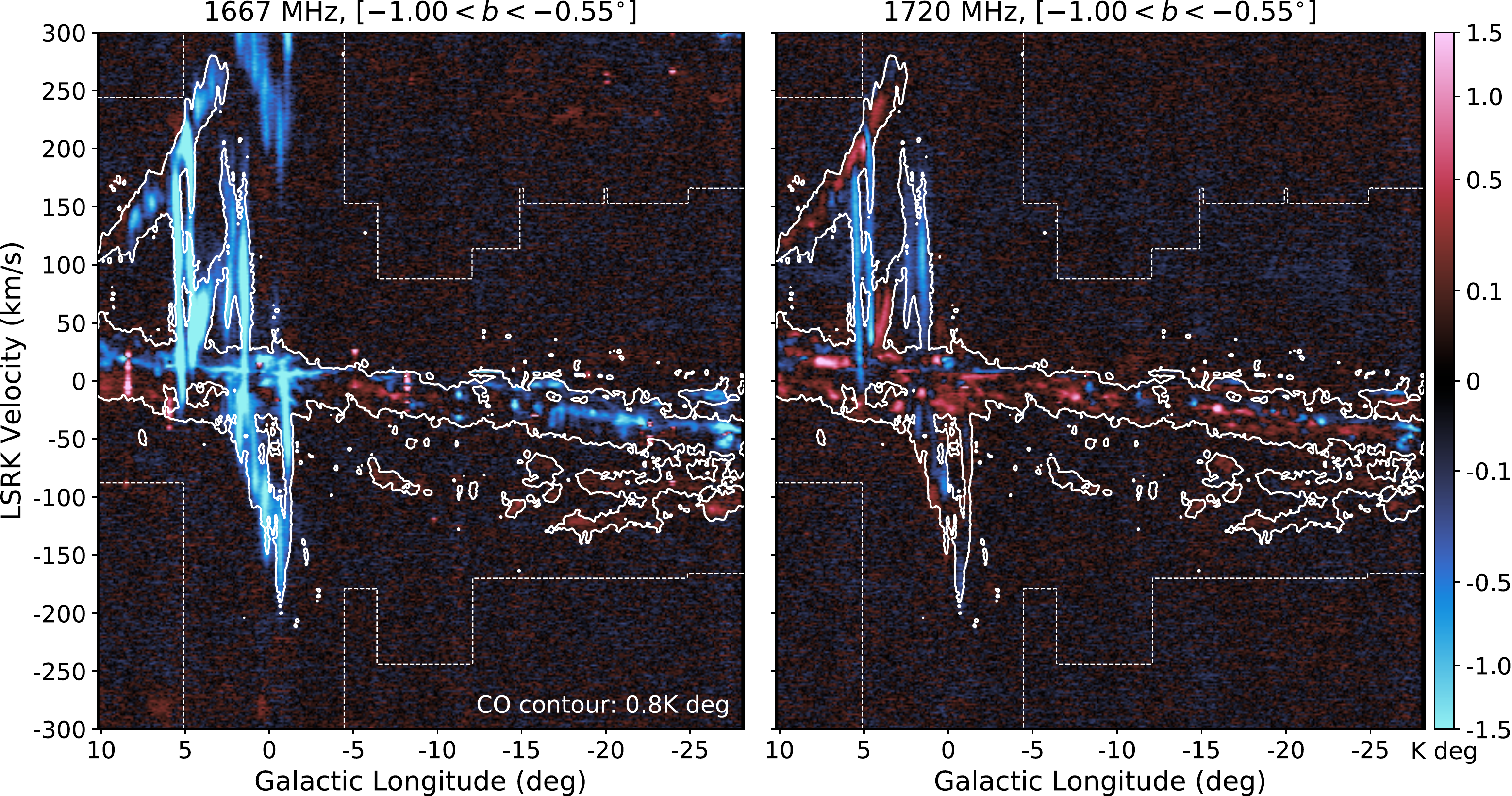}
\end{subfigure}
\caption{Longitude-velocity maps of the 1667\,MHz and 1720\,MHz lines, overlaid with a single $^{12}$CO(J=1--0) contour from Dame, Hartmann \& Thaddeus (2001). The contour is drawn at the $4\sigma$ level of the noisiest portion of each image. The dashed white line indicates the edges of the CO data. Each map is integrated over a latitude range of 0.5 degrees. Note that since the main lines are only separated by $\sim350$ \kms, there are places where negative velocities in the 1665\,MHz line run into the positive velocity space of the 1667\,MHz transition. This zone is marked in grey on one of the panels, for clarity.}
\end{figure*}

\newpage
\addtocounter{figure}{-1}
\begin{figure*}

\begin{subfigure}{1.0\textwidth}
\centering
\includegraphics[scale=0.43]{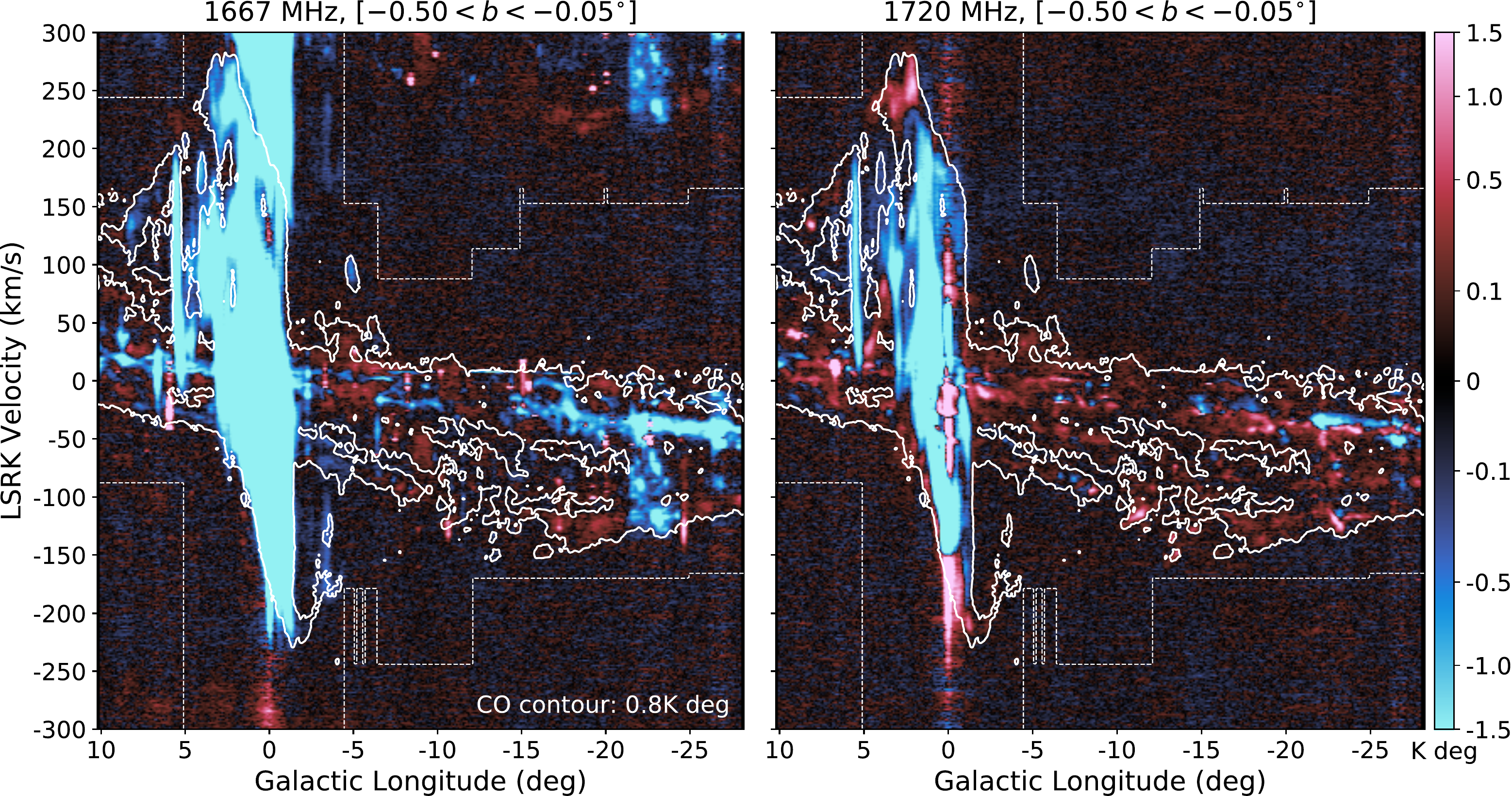}
\end{subfigure}\vspace*{0.25cm}

\begin{subfigure}{1.0\textwidth}
\centering
\includegraphics[scale=0.43]{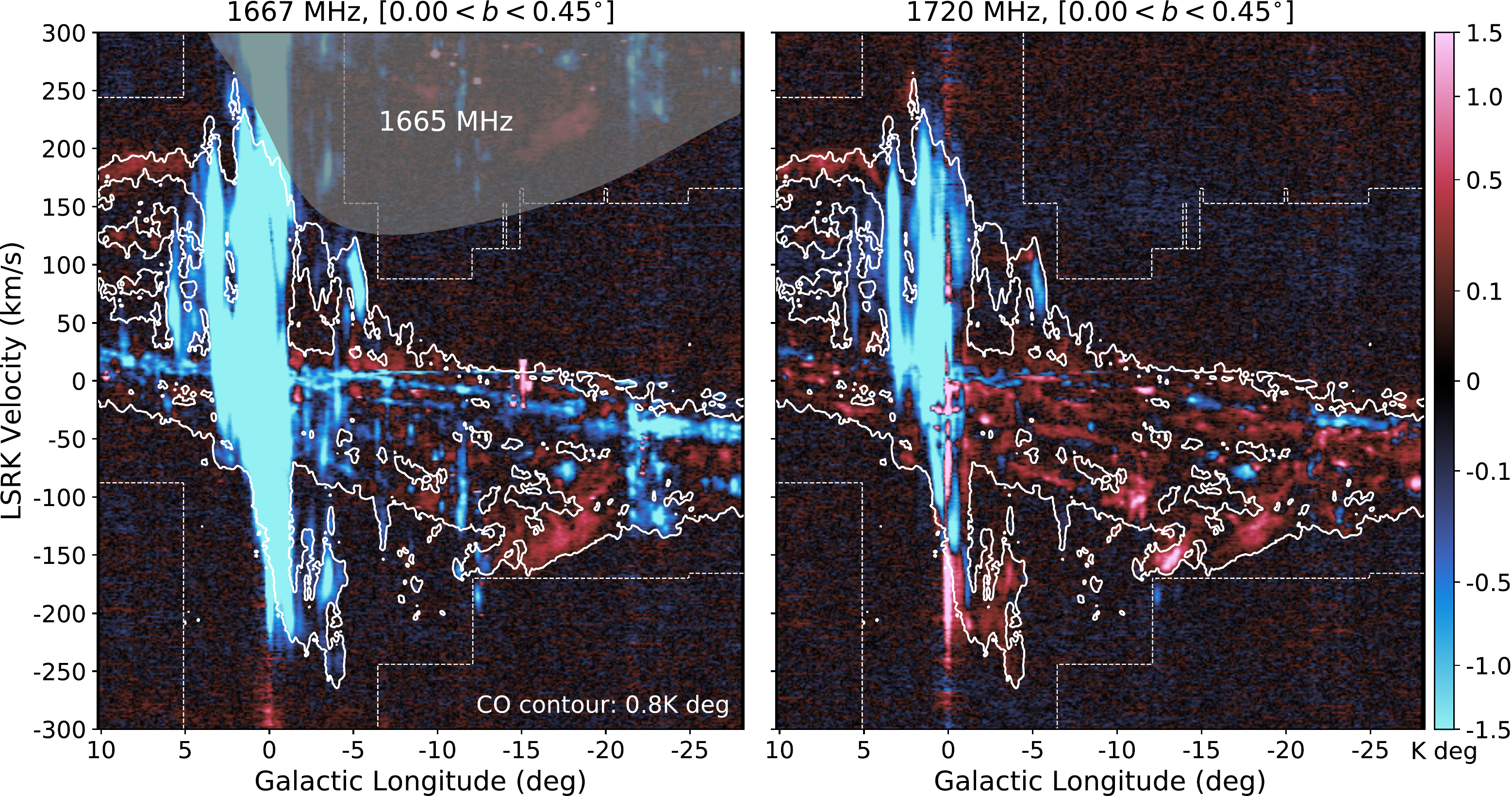}
\end{subfigure}\vspace*{0.25cm}

\begin{subfigure}{1.0\textwidth}
\centering
\includegraphics[scale=0.43]{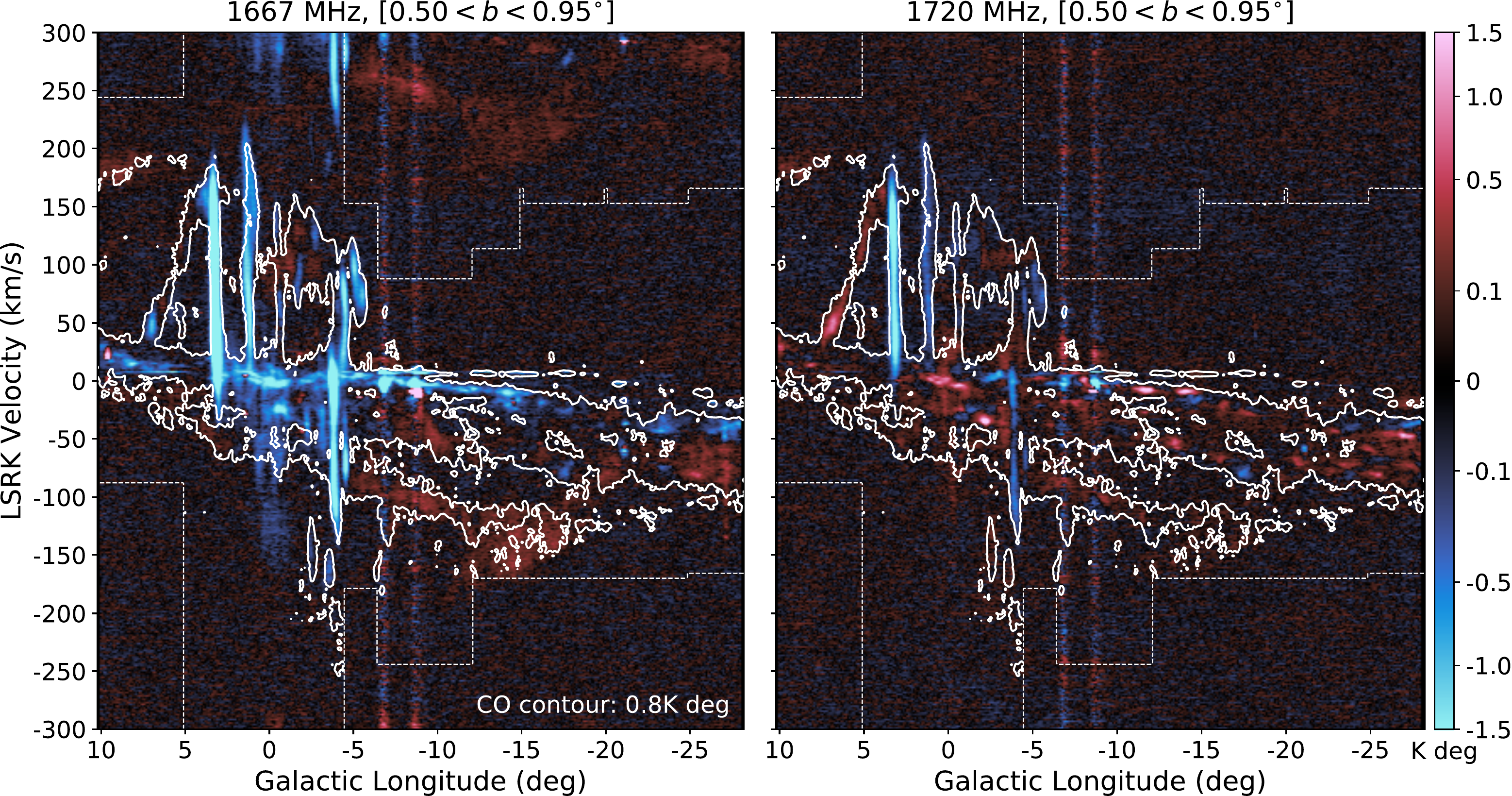}
\end{subfigure}
\caption{(cont.)}
\end{figure*}

\newpage
\addtocounter{figure}{-1}
\begin{figure*}

\begin{subfigure}{1.0\textwidth}
\centering
\includegraphics[scale=0.43]{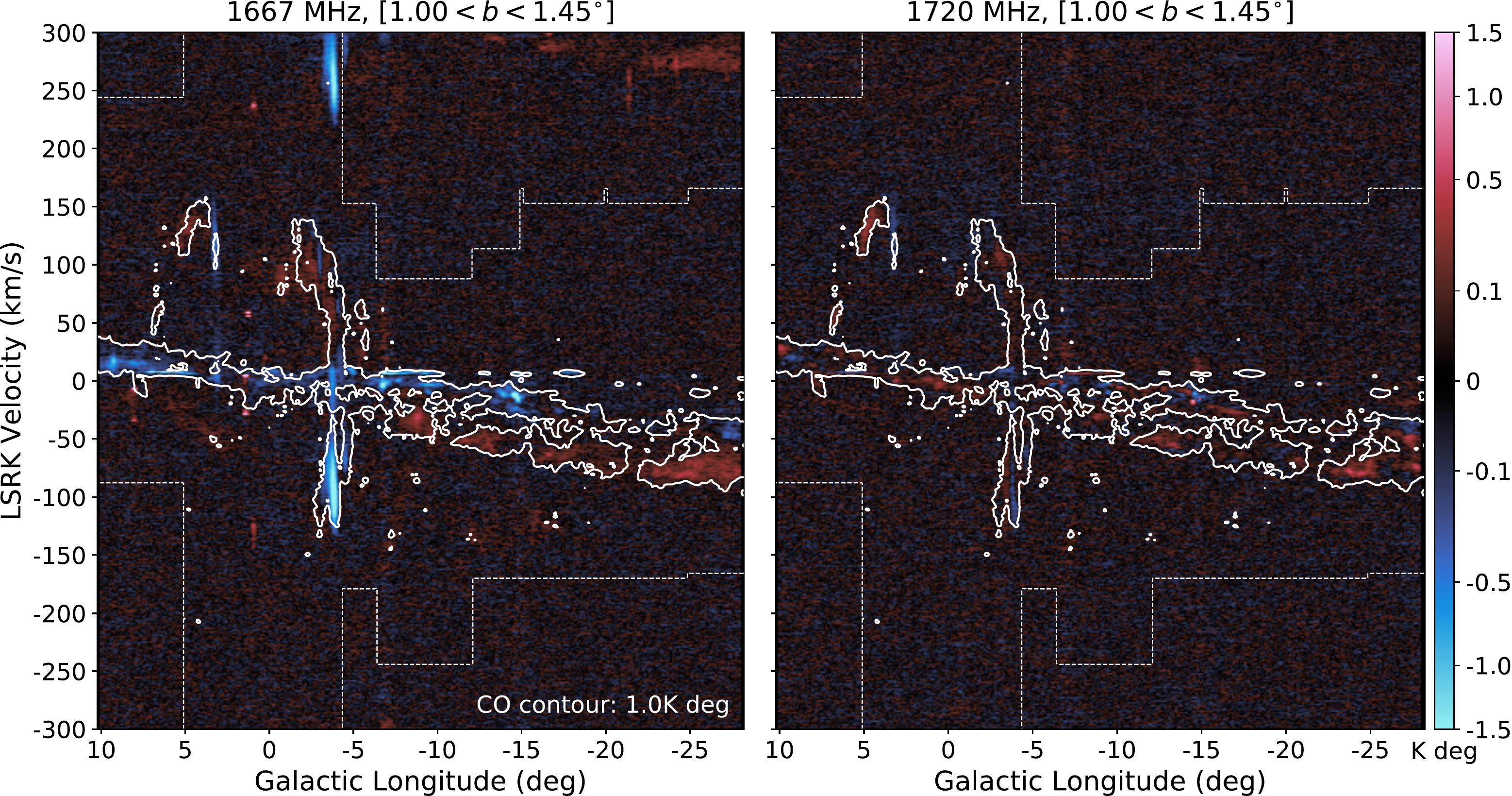}
\end{subfigure}\vspace*{0.25cm}

\begin{subfigure}{1.0\textwidth}
\centering
\includegraphics[scale=0.43]{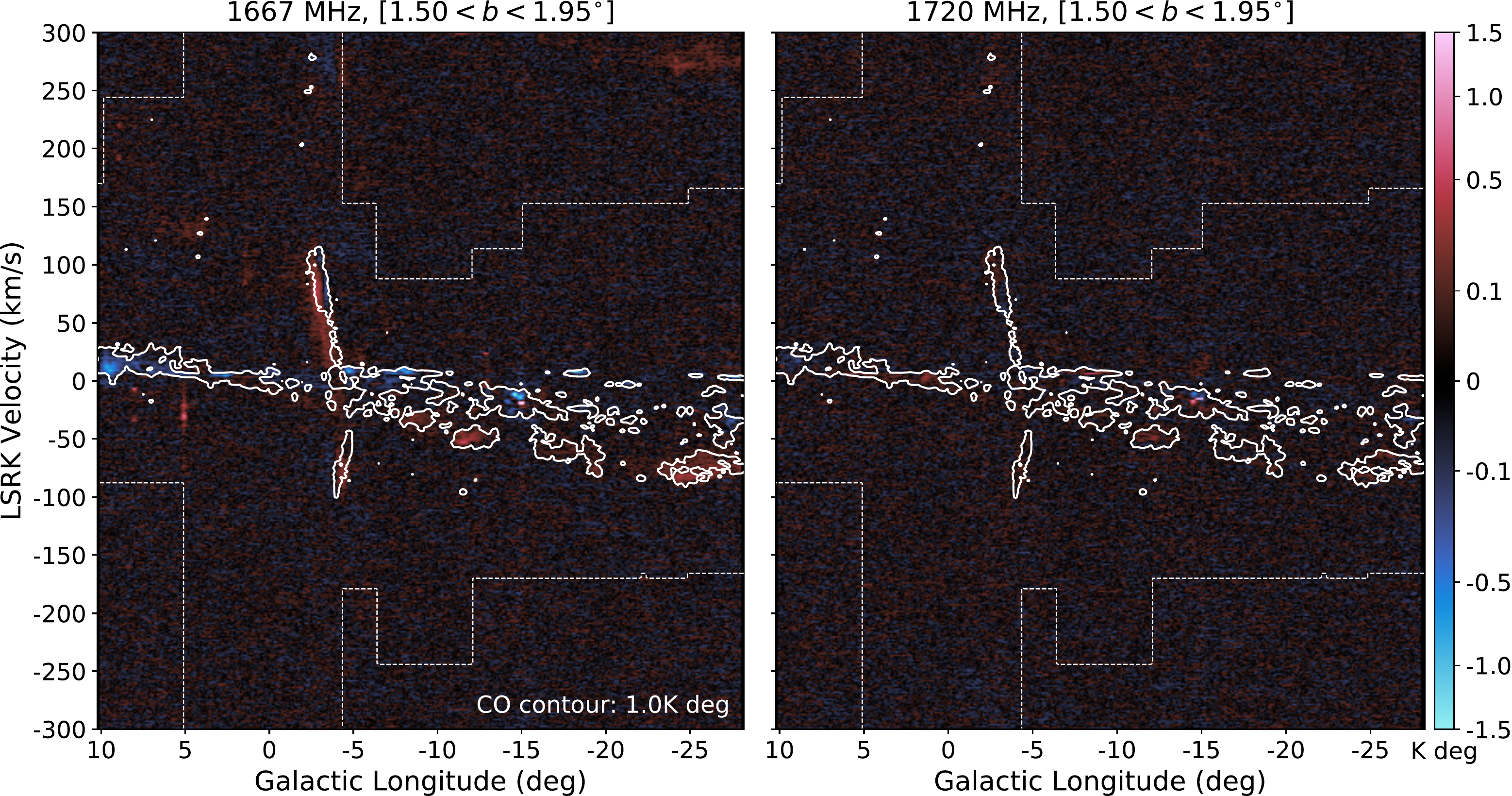}
\end{subfigure}
\caption{(cont.)}
\end{figure*}


\begin{figure*}
\begin{subfigure}{0.32\textwidth}
\includegraphics[width=\linewidth]{flip00.pdf}
\end{subfigure}\hspace*{\fill}
\begin{subfigure}{0.32\textwidth}
\includegraphics[width=\linewidth]{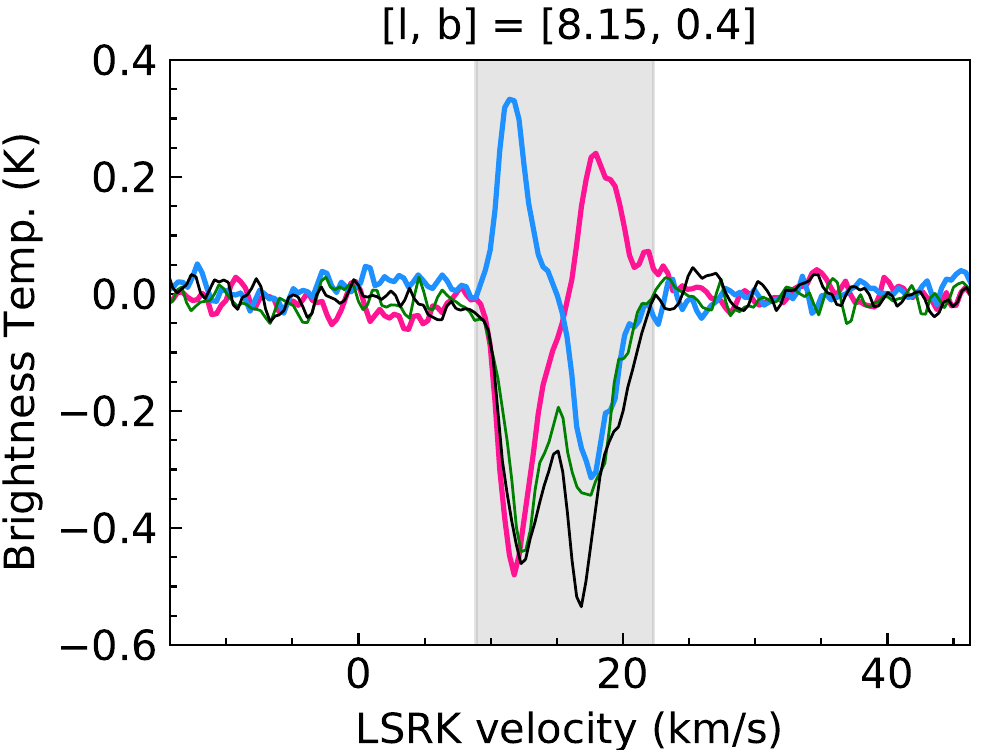}
\end{subfigure}\hspace*{\fill}
\begin{subfigure}{0.32\textwidth}
\includegraphics[width=\linewidth]{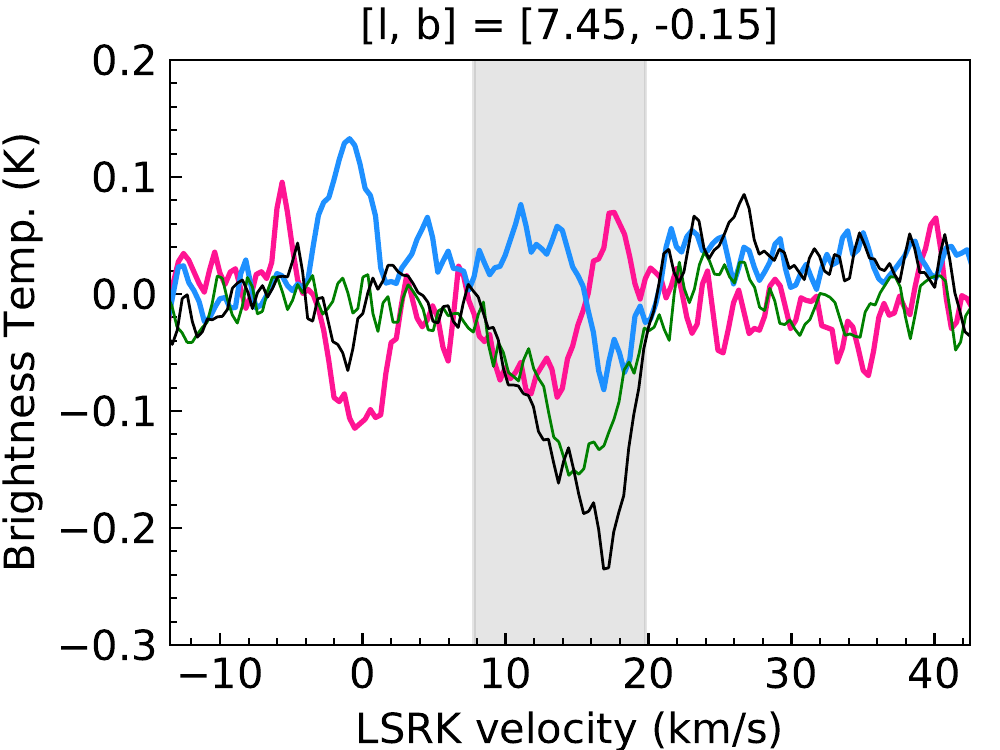}
\end{subfigure}

\medskip
\begin{subfigure}{0.32\textwidth}
\includegraphics[width=\linewidth]{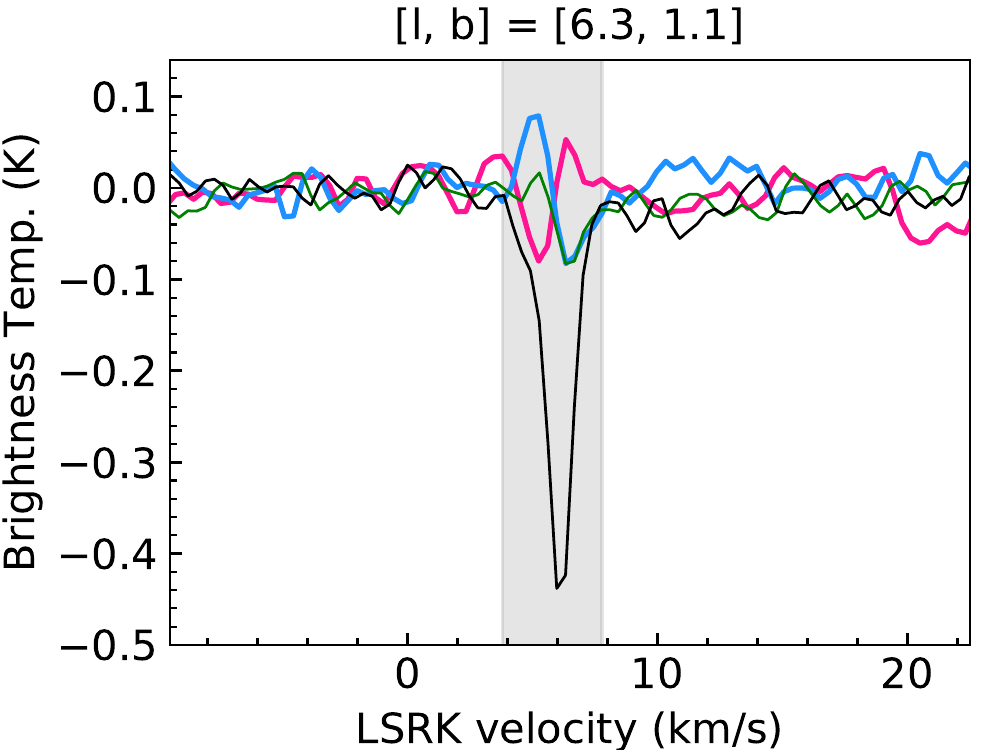}
\end{subfigure}\hspace*{\fill}
\begin{subfigure}{0.32\textwidth}
\includegraphics[width=\linewidth]{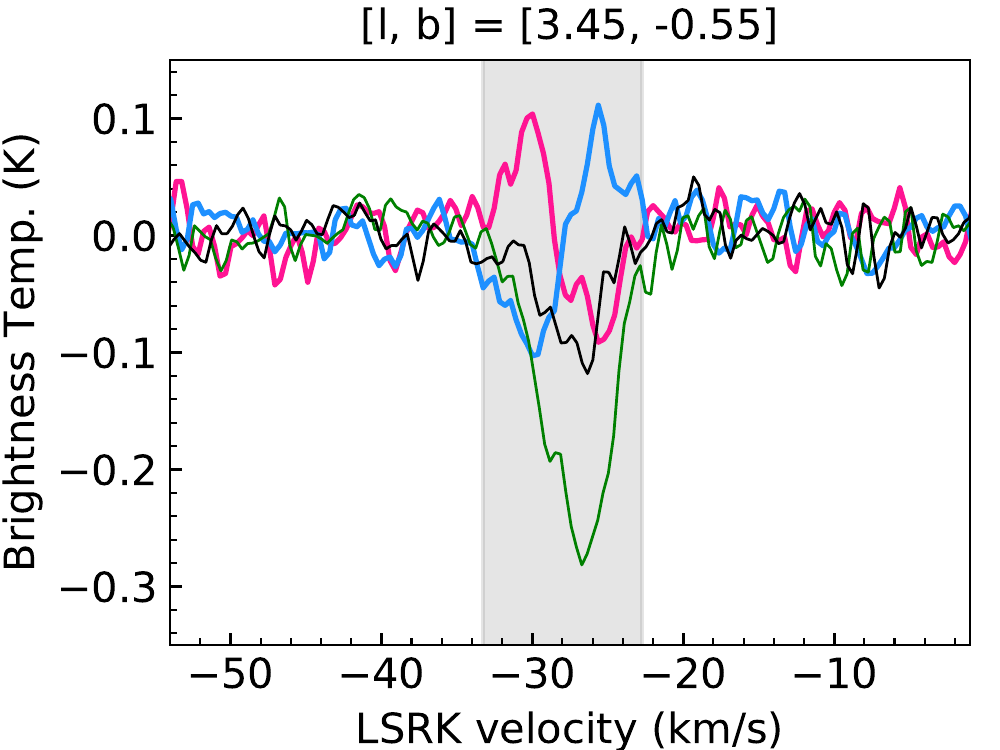}
\end{subfigure}\hspace*{\fill}
\begin{subfigure}{0.32\textwidth}
\includegraphics[width=\linewidth]{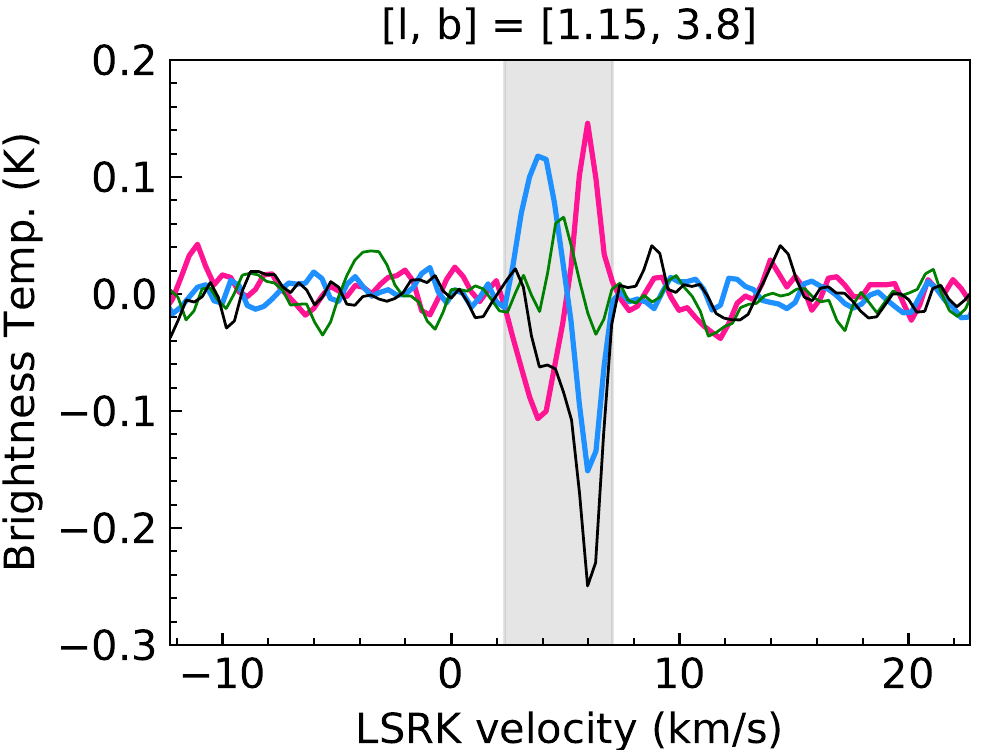}
\end{subfigure}

\medskip
\begin{subfigure}{0.32\textwidth}
\includegraphics[width=\linewidth]{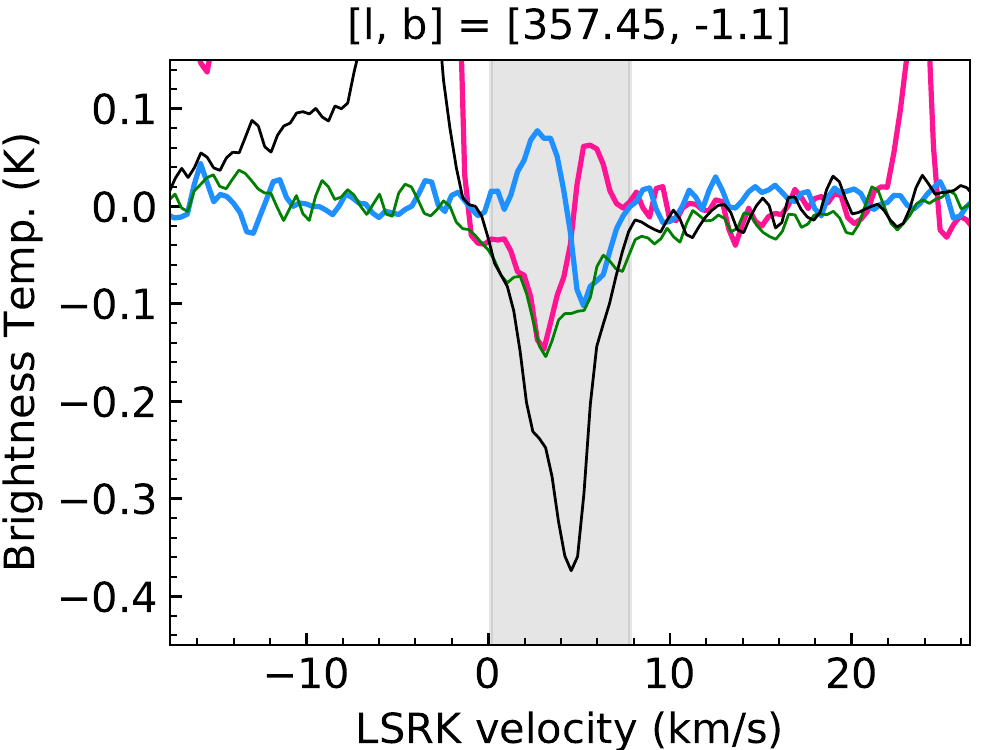}
\end{subfigure}\hspace*{\fill}
\begin{subfigure}{0.32\textwidth}
\includegraphics[width=\linewidth]{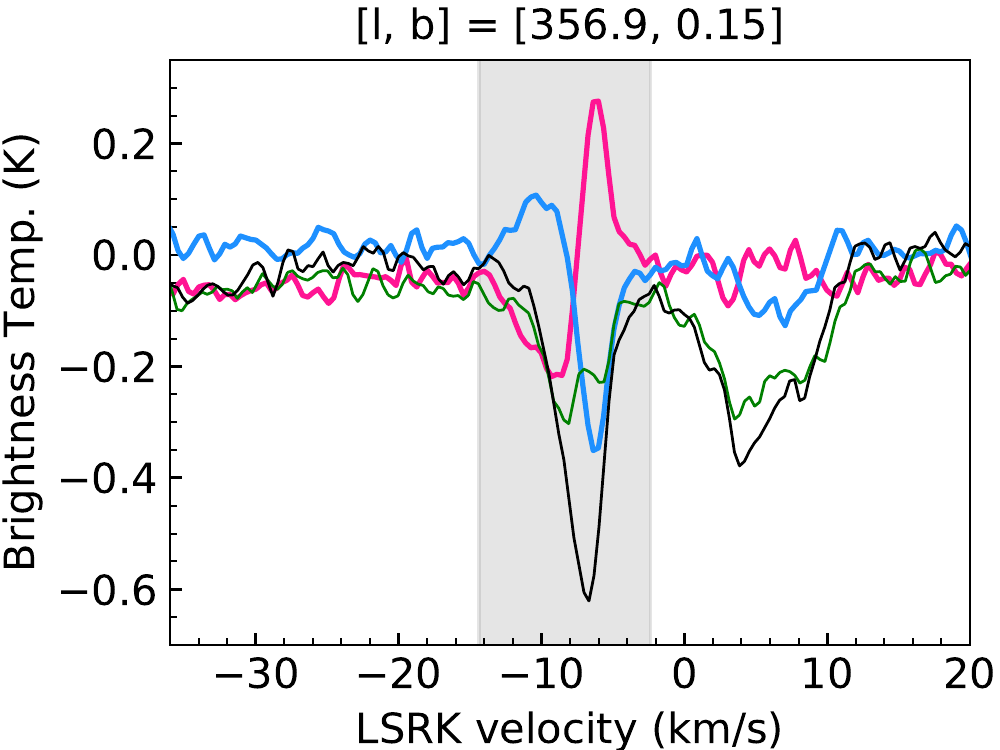}
\end{subfigure}\hspace*{\fill}
\begin{subfigure}{0.32\textwidth}
\includegraphics[width=\linewidth]{flip08.pdf}
\end{subfigure}

\medskip
\begin{subfigure}{0.32\textwidth}
\includegraphics[width=\linewidth]{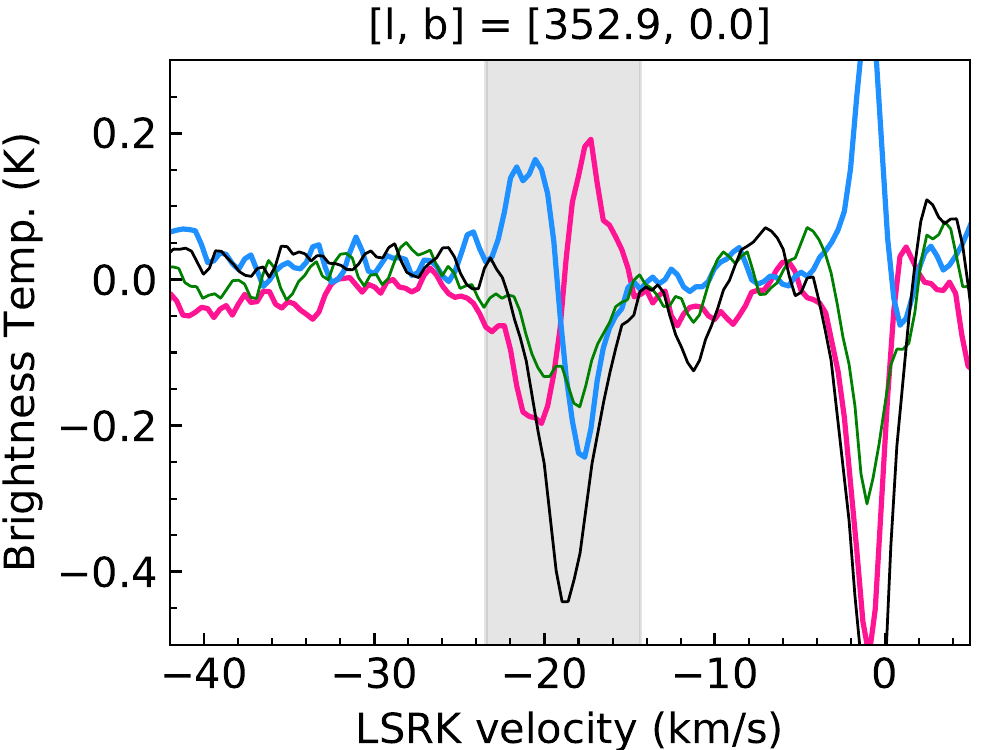}
\end{subfigure}\hspace*{\fill}
\begin{subfigure}{0.32\textwidth}
\includegraphics[width=\linewidth]{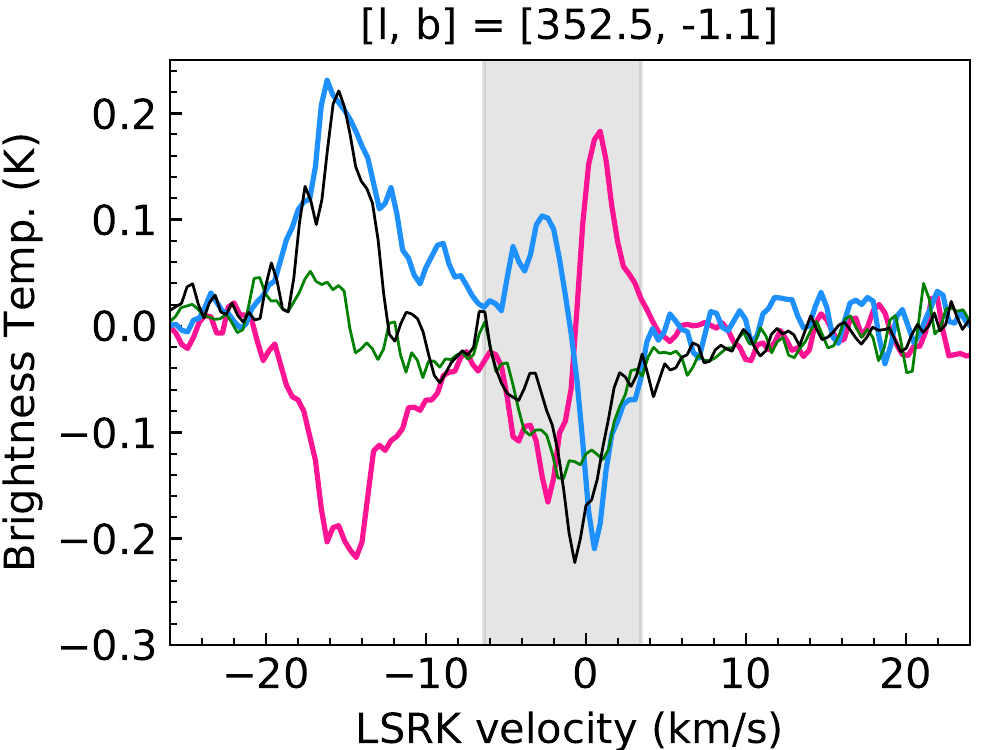}
\end{subfigure}\hspace*{\fill}
\begin{subfigure}{0.32\textwidth}
\includegraphics[width=\linewidth]{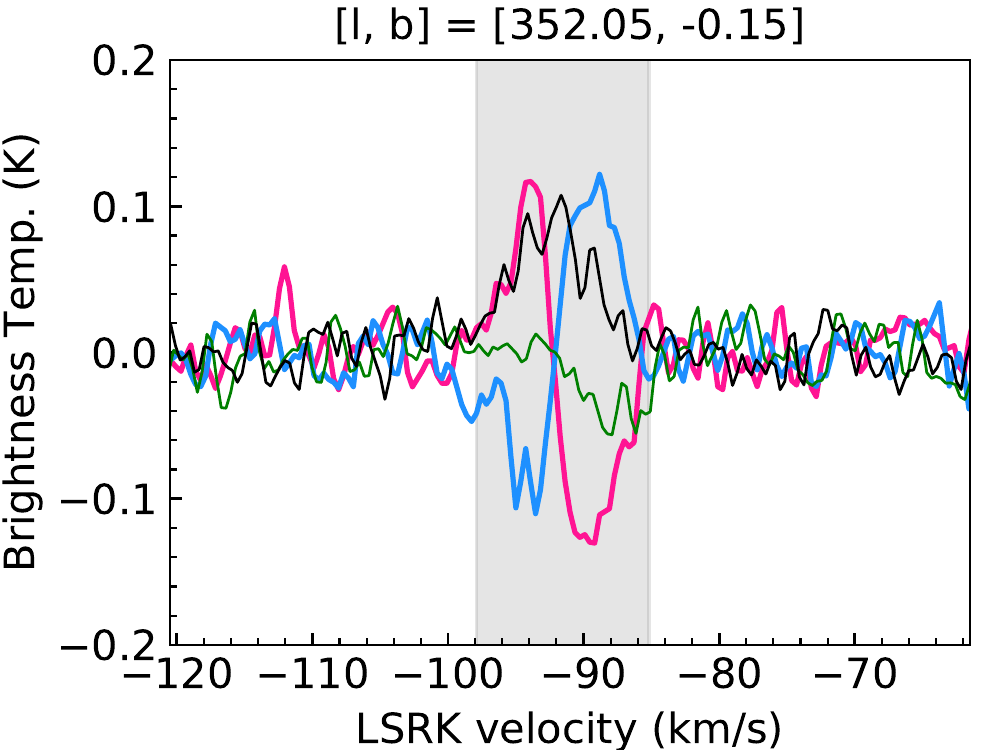}
\end{subfigure}

\medskip
\begin{subfigure}{0.32\textwidth}
\includegraphics[width=\linewidth]{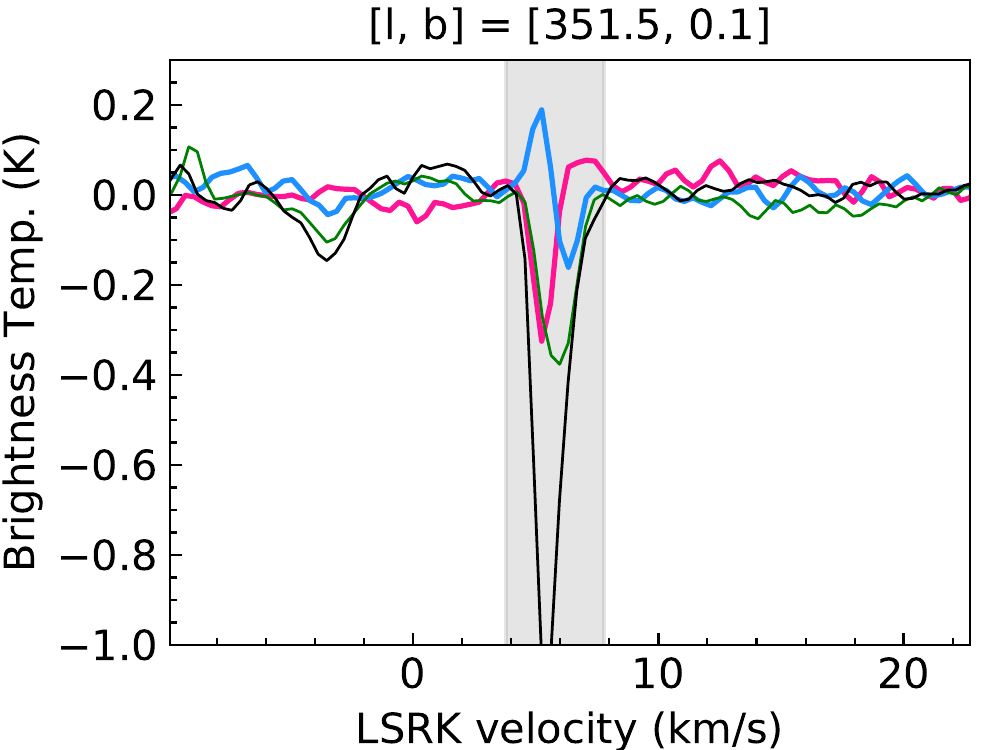}
\end{subfigure}\hspace*{\fill}
\begin{subfigure}{0.32\textwidth}
\includegraphics[width=\linewidth]{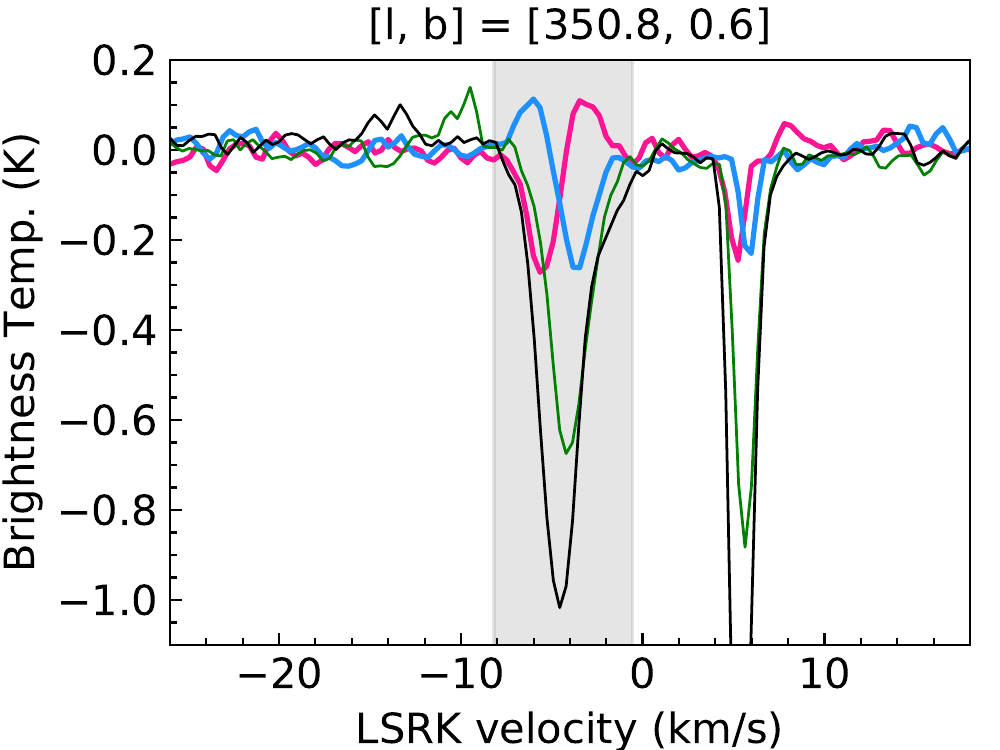}
\end{subfigure}\hspace*{\fill}
\begin{subfigure}{0.32\textwidth}
\includegraphics[width=\linewidth]{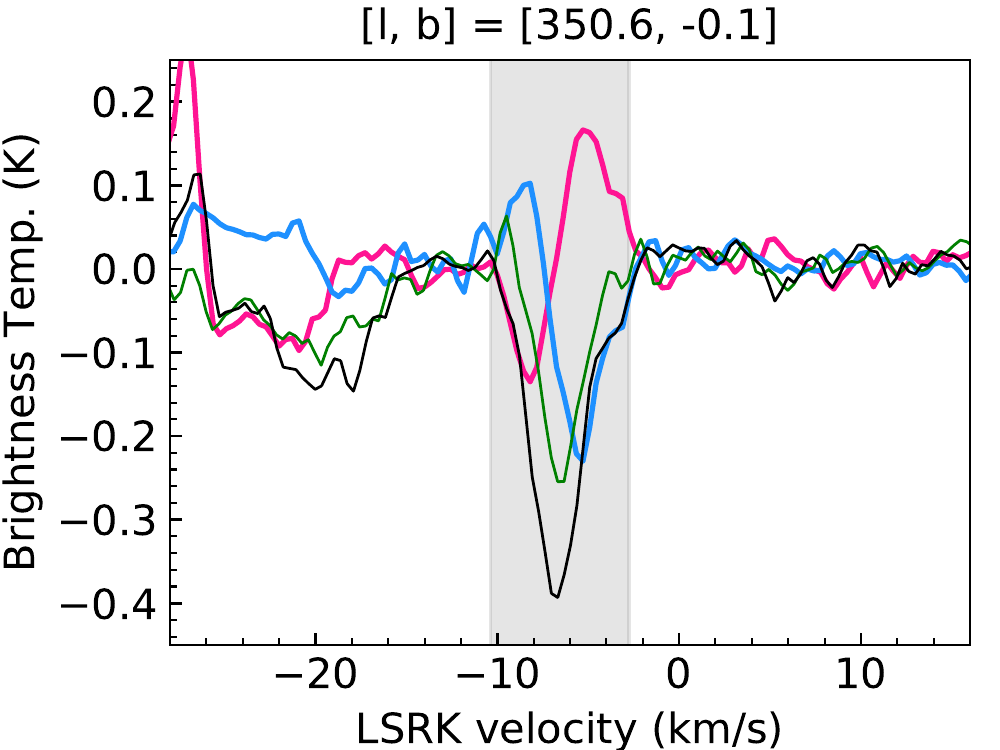}
\end{subfigure}
\caption{Satellite line `flips' in the SPLASH survey region, defined as profiles in which the sense of the conjugate 1612 and 1720\,MHz lines reverses within a closely-blended main line feature. Each panel shows a single example sightline towards a unique instance of the flip in the 3D cubes. Each feature typically extends over multiple beams on the sky.}
\end{figure*}

\addtocounter{figure}{-1}
\begin{figure*}
\begin{subfigure}{0.32\textwidth}
\includegraphics[width=\linewidth]{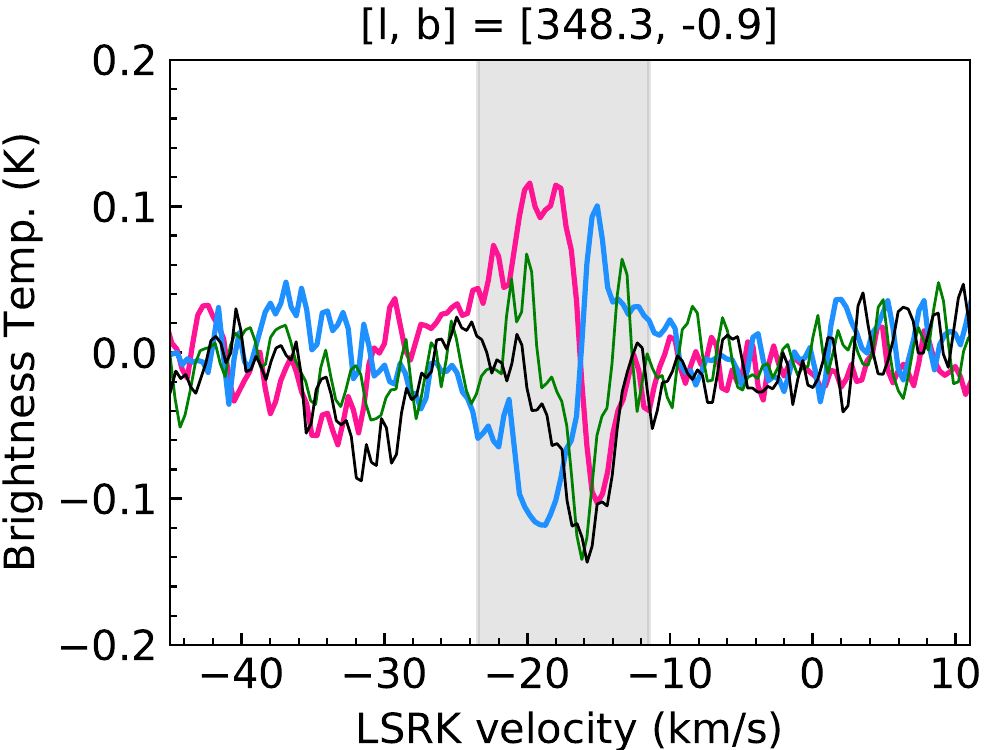}
\end{subfigure}\hspace*{\fill}
\begin{subfigure}{0.32\textwidth}
\includegraphics[width=\linewidth]{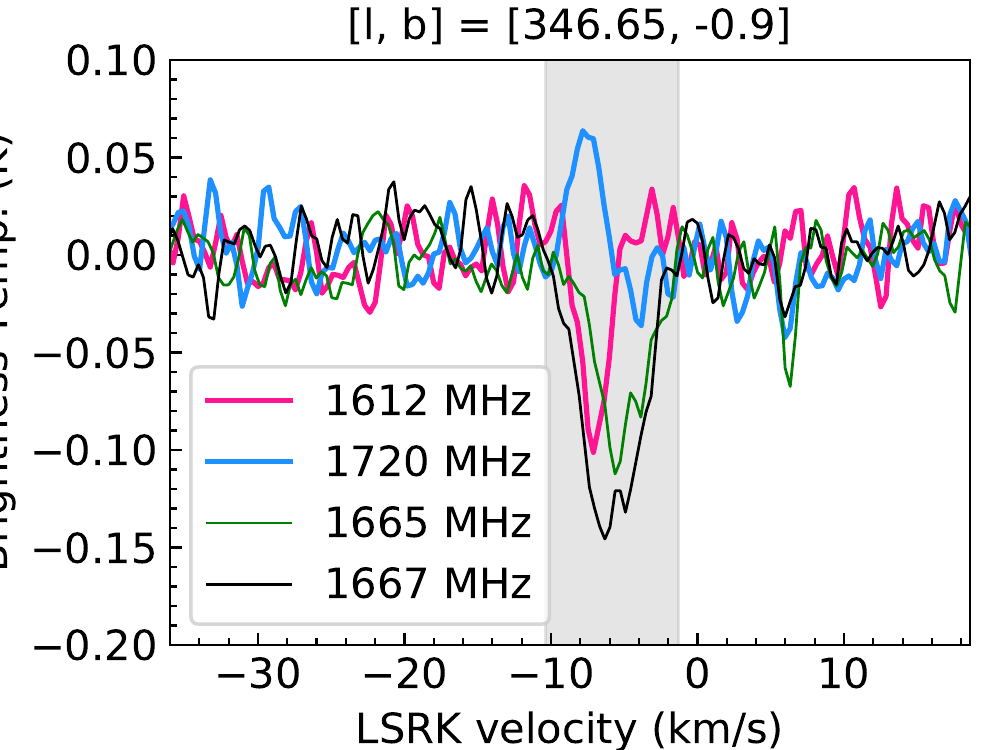}
\end{subfigure}\hspace*{\fill}
\begin{subfigure}{0.32\textwidth}
\includegraphics[width=\linewidth]{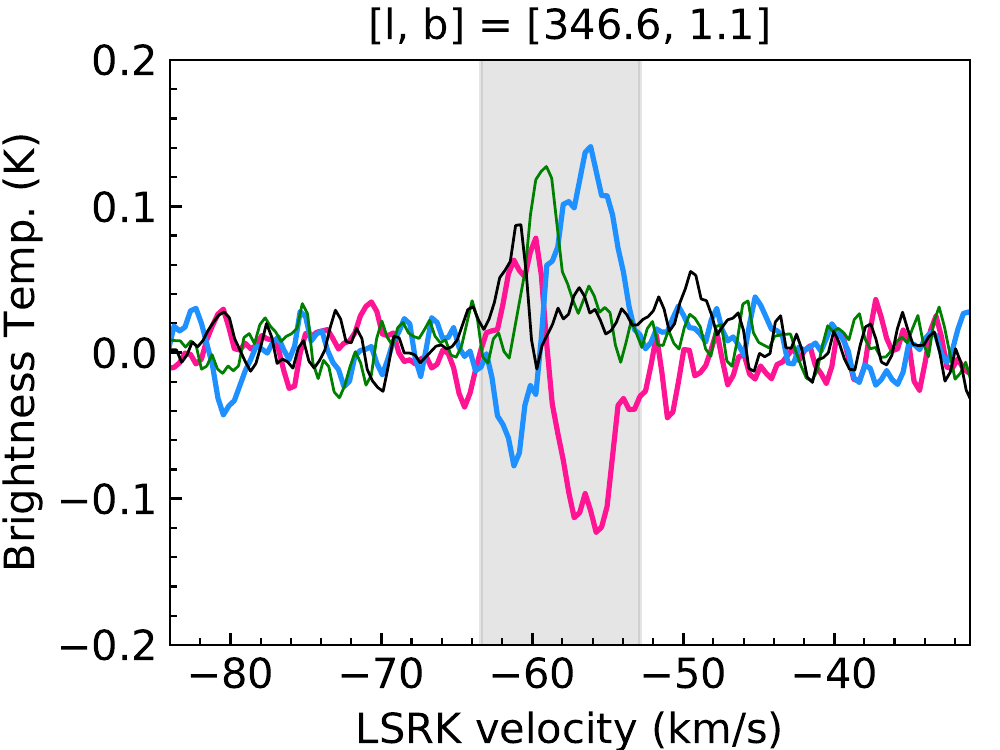}
\end{subfigure}

\medskip
\begin{subfigure}{0.32\textwidth}
\includegraphics[width=\linewidth]{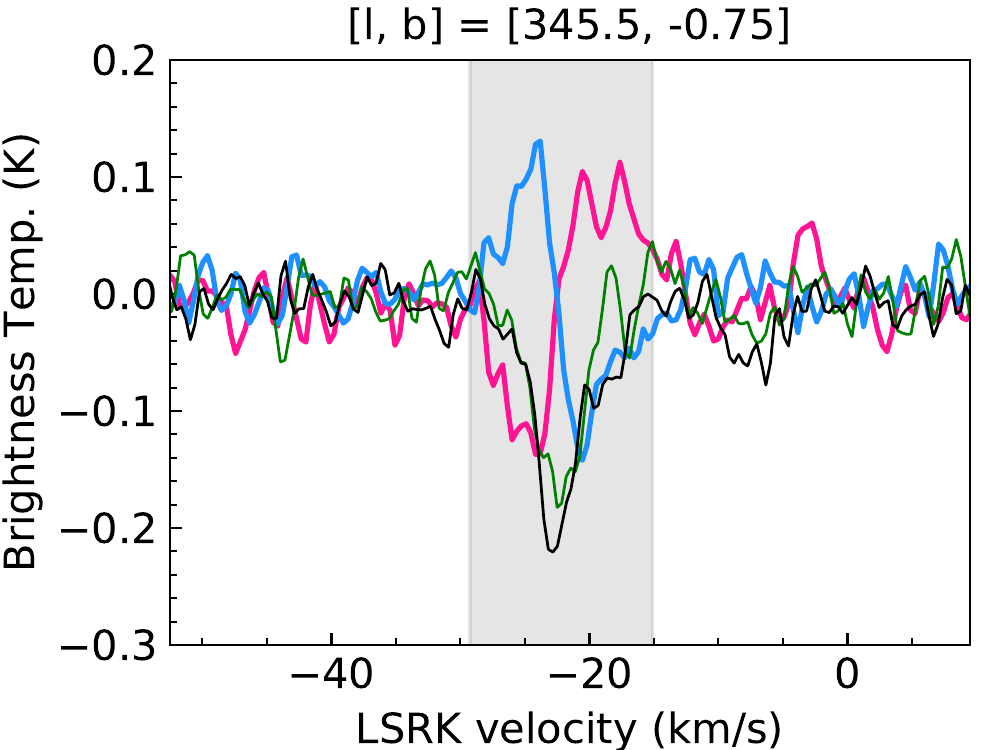}
\end{subfigure}\hspace*{\fill}
\begin{subfigure}{0.32\textwidth}
\includegraphics[width=\linewidth]{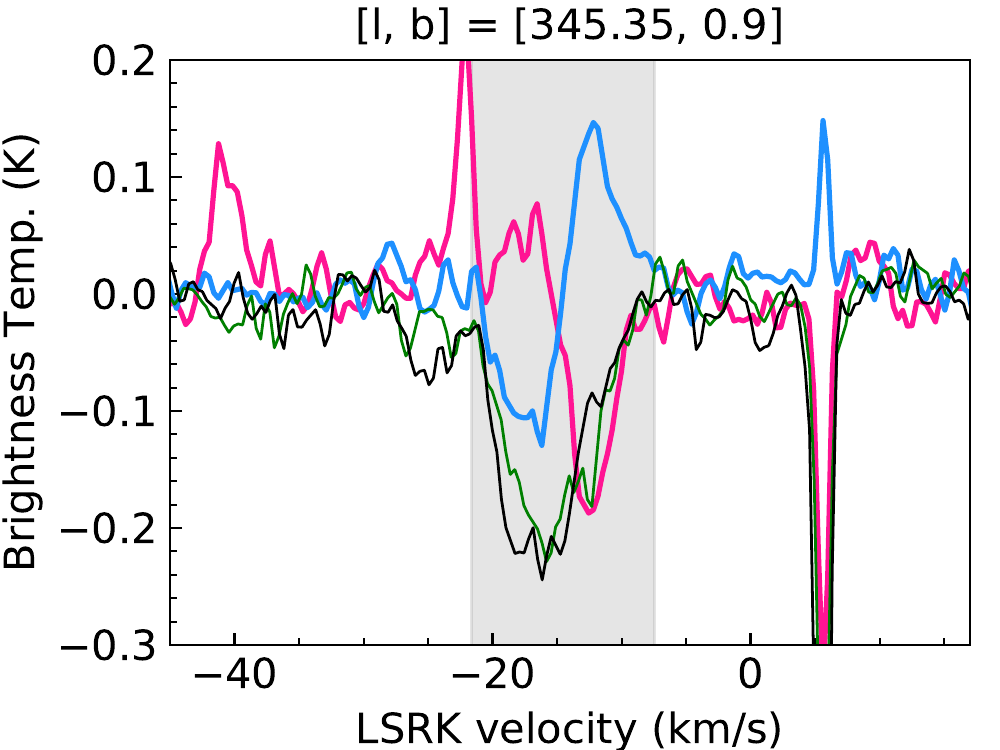}
\end{subfigure}\hspace*{\fill}
\begin{subfigure}{0.32\textwidth}
\includegraphics[width=\linewidth]{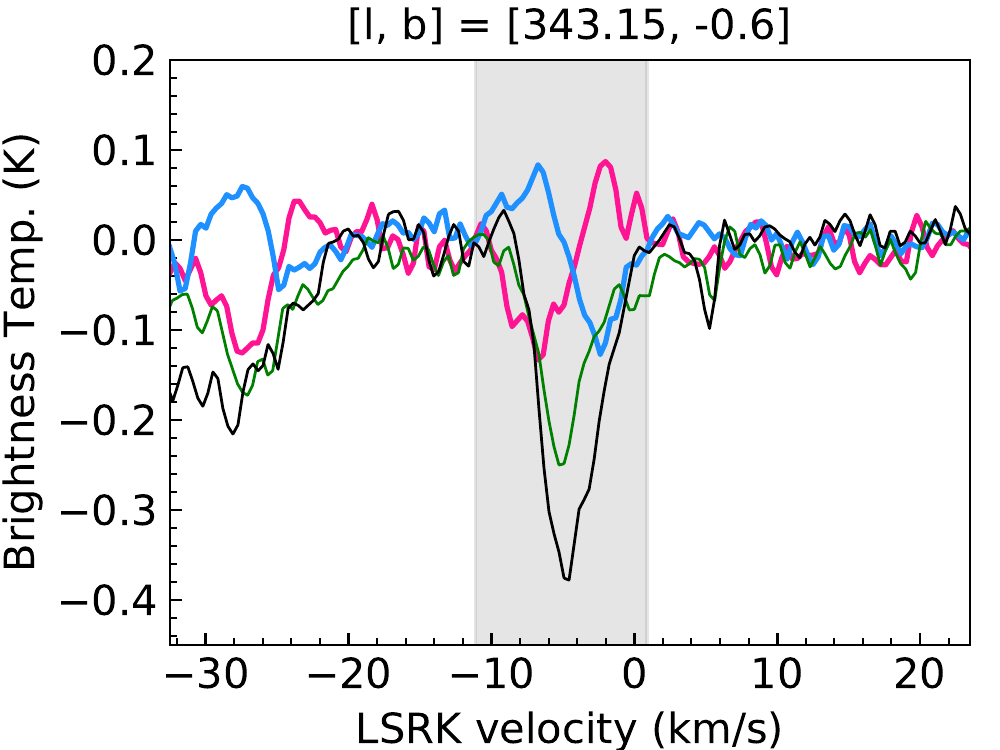}
\end{subfigure}

\medskip
\begin{subfigure}{0.32\textwidth}
\includegraphics[width=\linewidth]{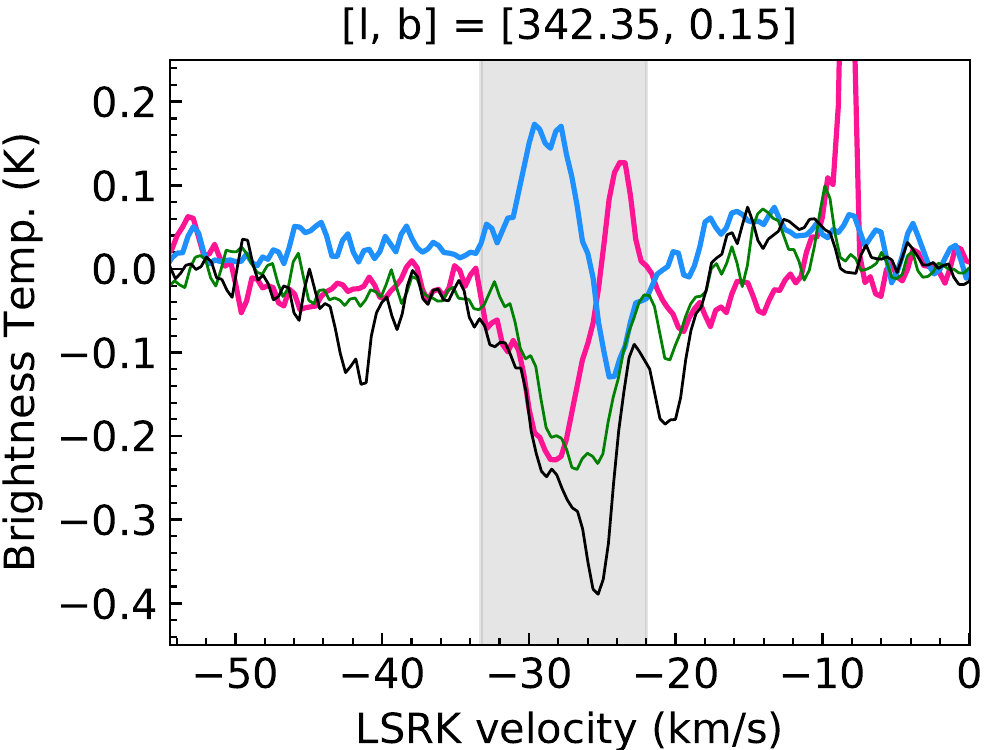}
\end{subfigure}\hspace*{\fill}
\begin{subfigure}{0.32\textwidth}
\includegraphics[width=\linewidth]{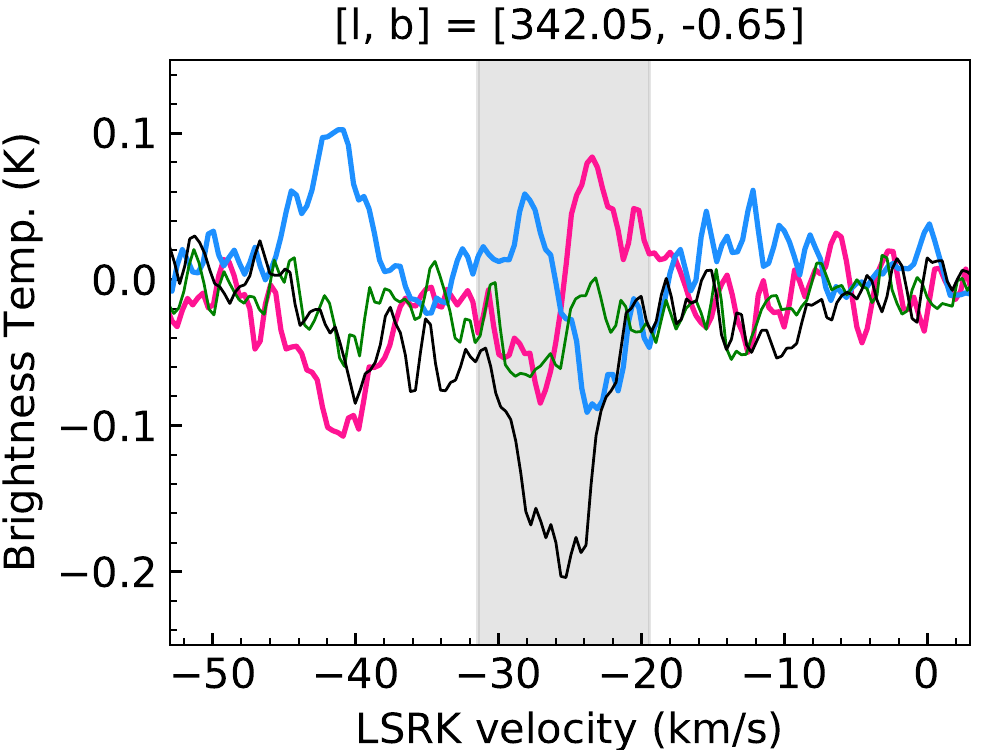}
\end{subfigure}\hspace*{\fill}
\begin{subfigure}{0.32\textwidth}
\includegraphics[width=\linewidth]{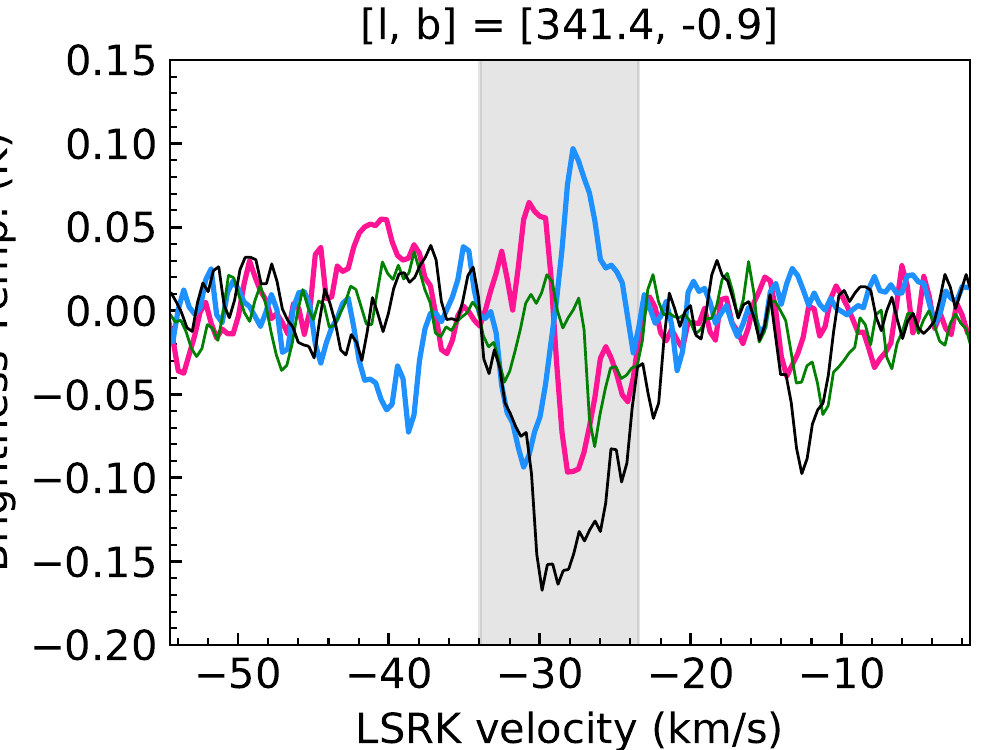}
\end{subfigure}

\medskip
\begin{subfigure}{0.32\textwidth}
\includegraphics[width=\linewidth]{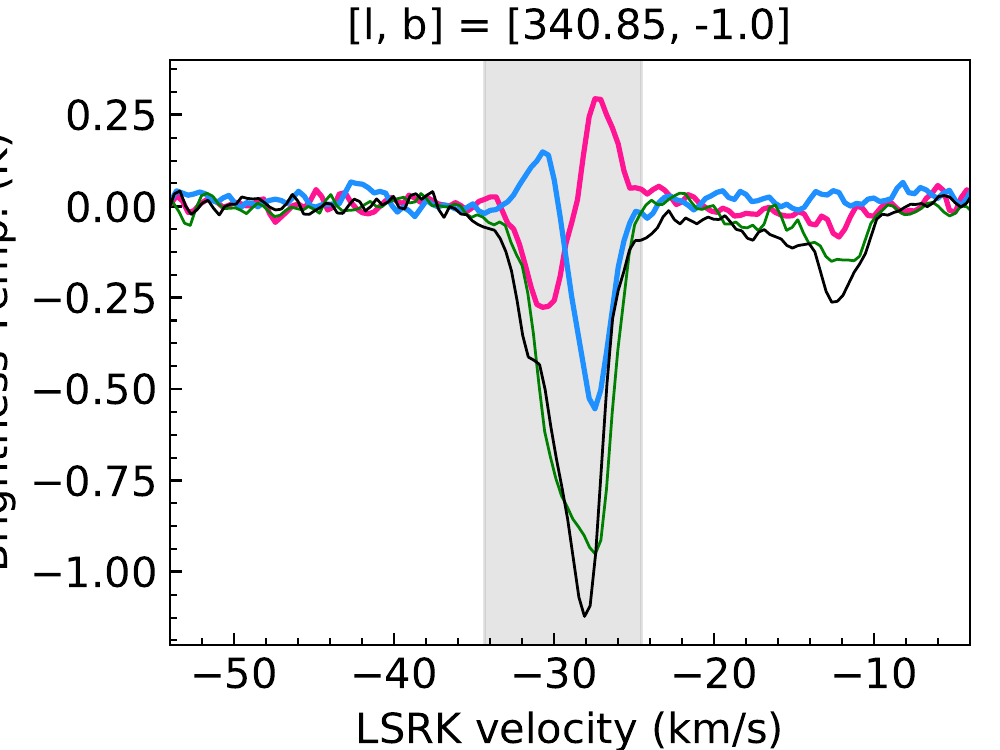}
\end{subfigure}\hspace*{\fill}
\begin{subfigure}{0.32\textwidth}
\includegraphics[width=\linewidth]{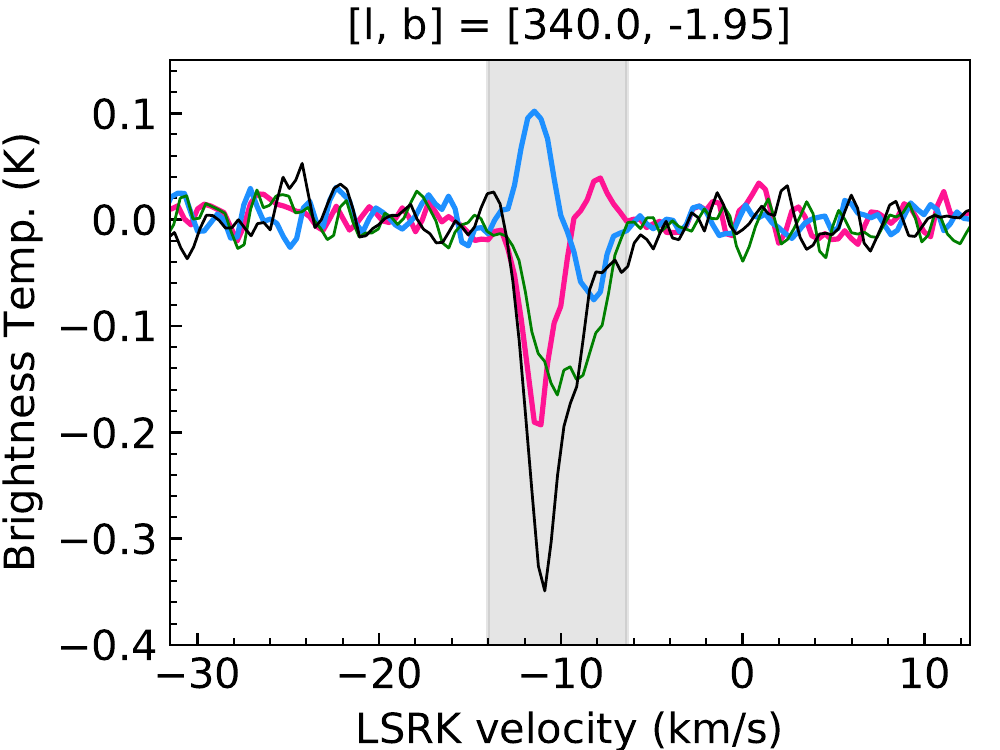}
\end{subfigure}\hspace*{\fill}
\begin{subfigure}{0.32\textwidth}
\includegraphics[width=\linewidth]{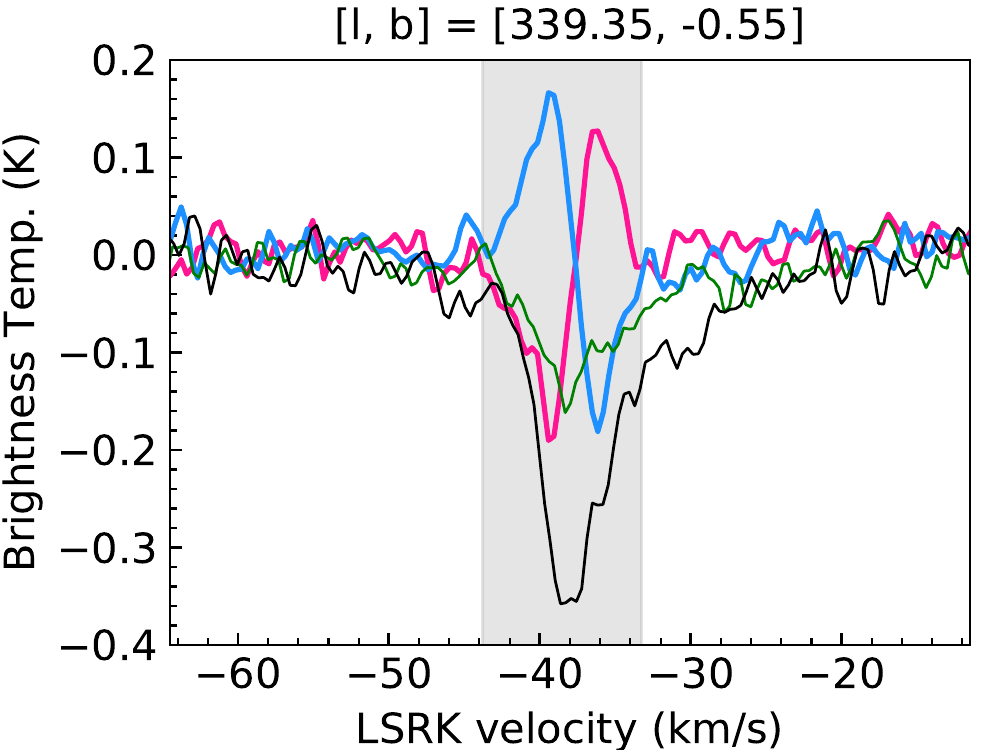}
\end{subfigure}

\medskip
\begin{subfigure}{0.32\textwidth}
\includegraphics[width=\linewidth]{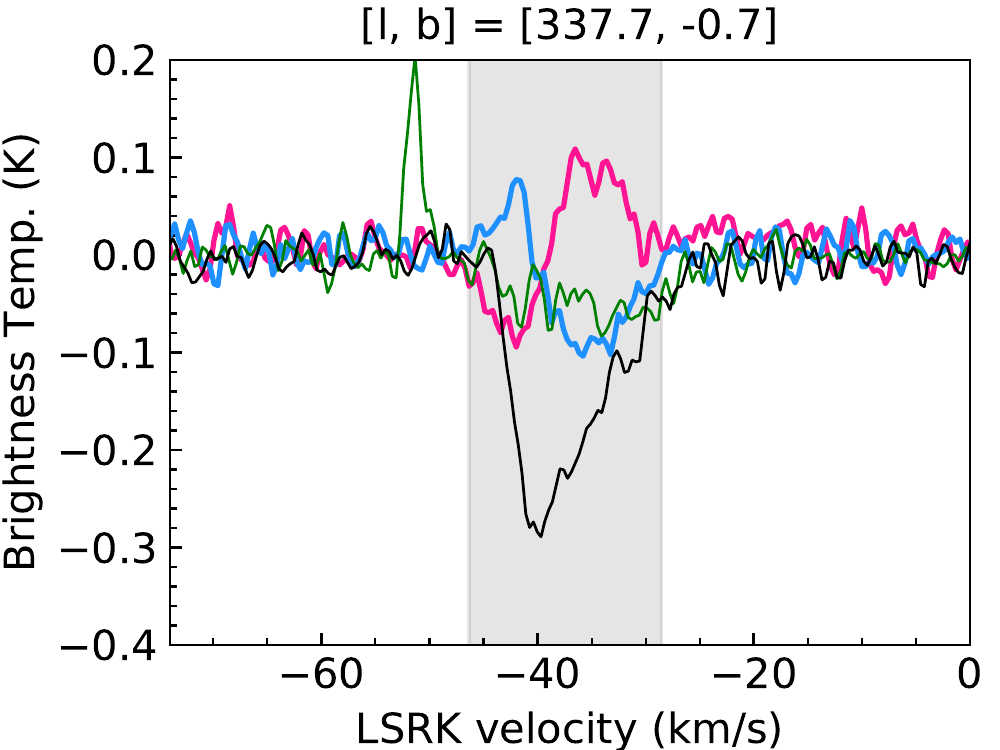}
\end{subfigure}\hspace*{\fill}
\begin{subfigure}{0.32\textwidth}
\includegraphics[width=\linewidth]{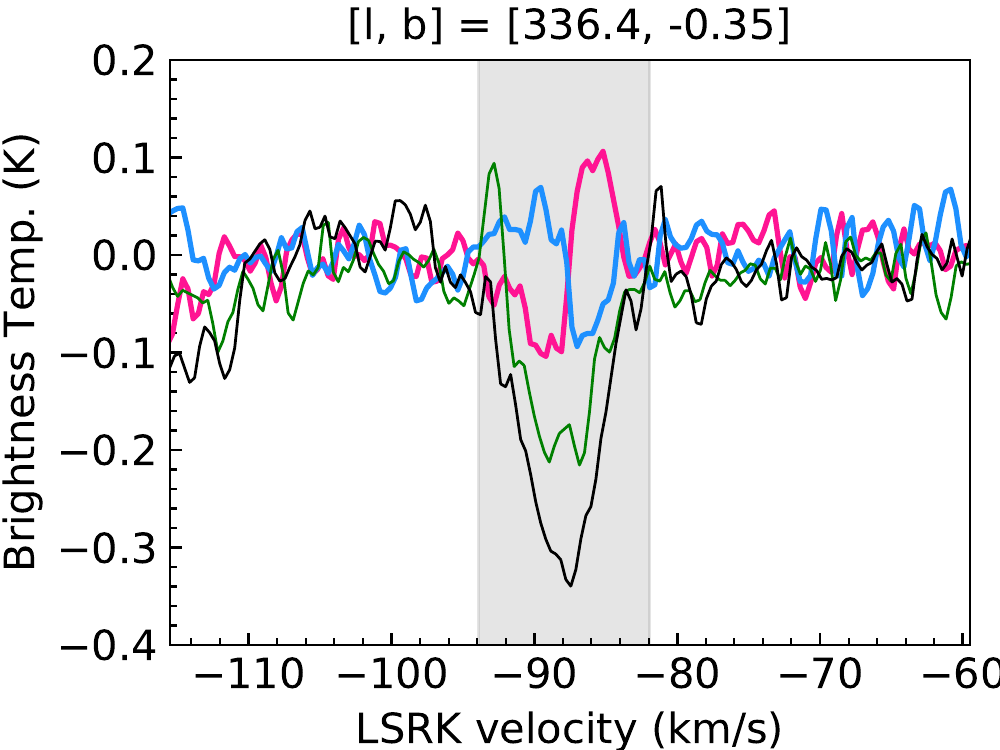}
\end{subfigure}\hspace*{\fill}
\begin{subfigure}{0.32\textwidth}
\includegraphics[width=\linewidth]{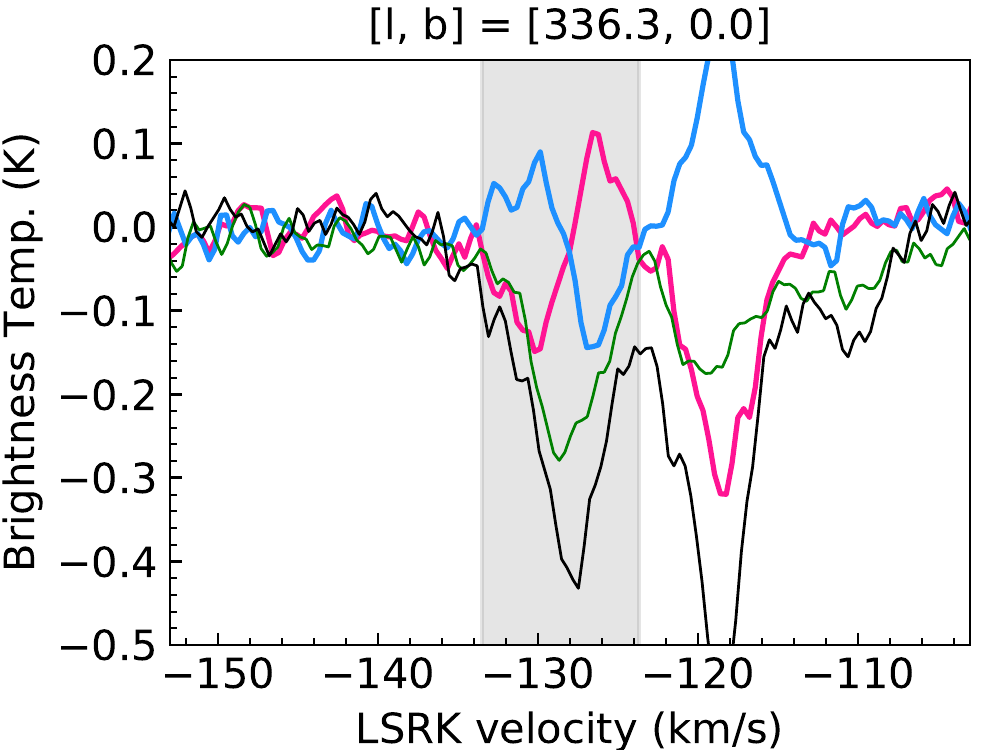}
\end{subfigure}
\caption{(cont.)}
\end{figure*}

\addtocounter{figure}{-1}
\begin{figure*}
\begin{subfigure}{0.32\textwidth}
\includegraphics[width=\linewidth]{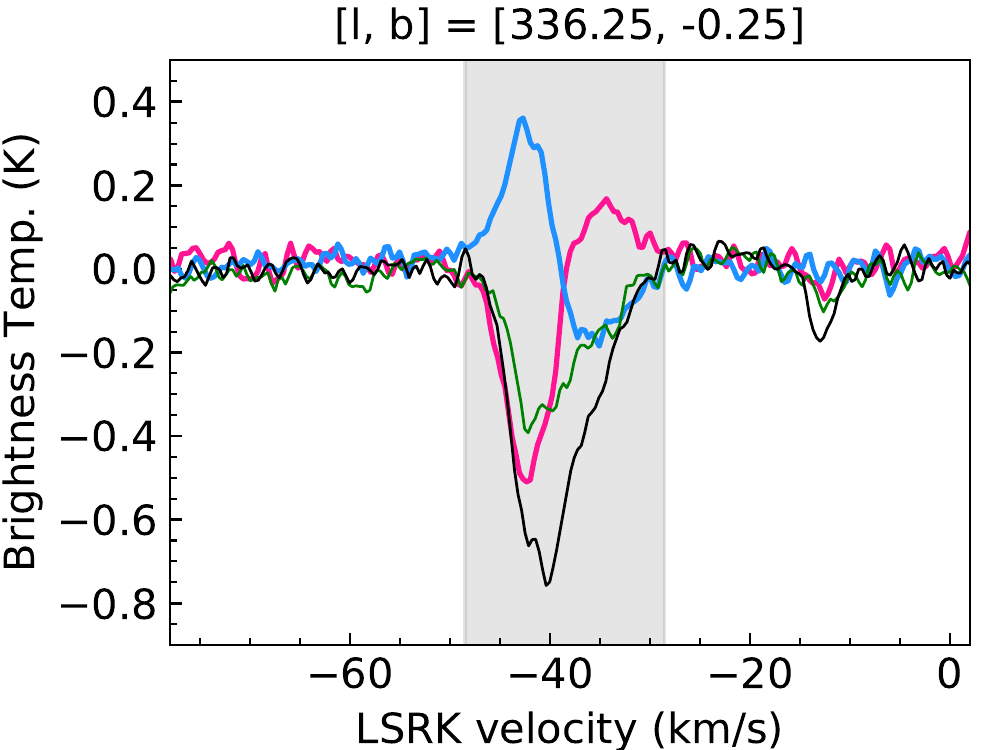}
\end{subfigure}\hspace*{\fill}
\begin{subfigure}{0.32\textwidth}
\includegraphics[width=\linewidth]{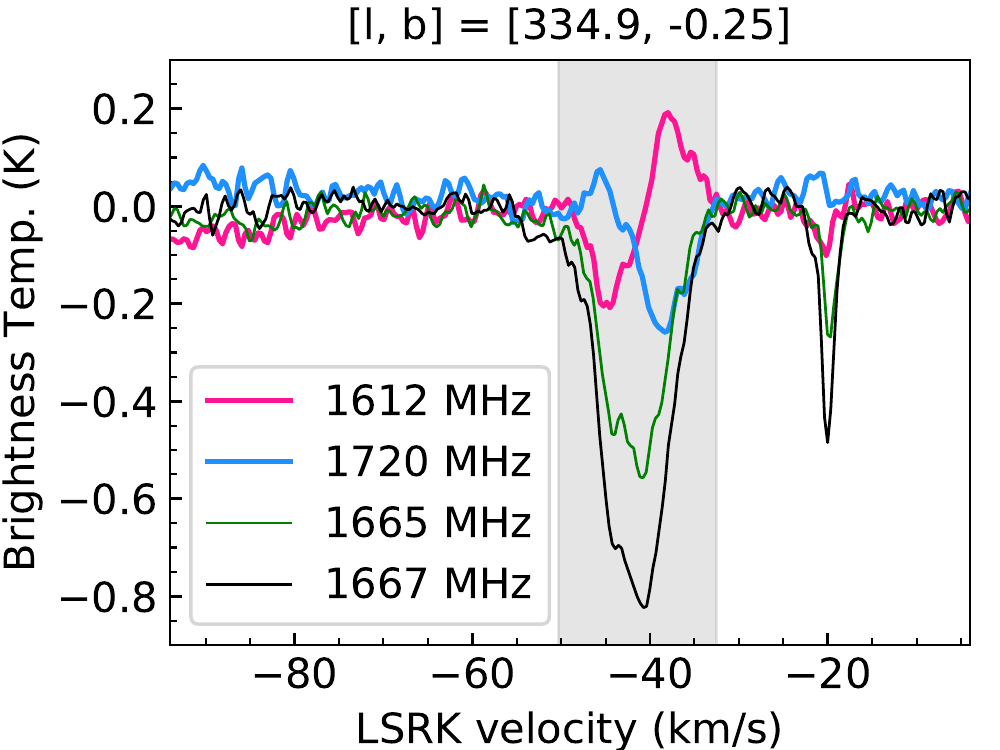}
\end{subfigure}\hspace*{\fill}
\begin{subfigure}{0.32\textwidth}
\includegraphics[width=\linewidth]{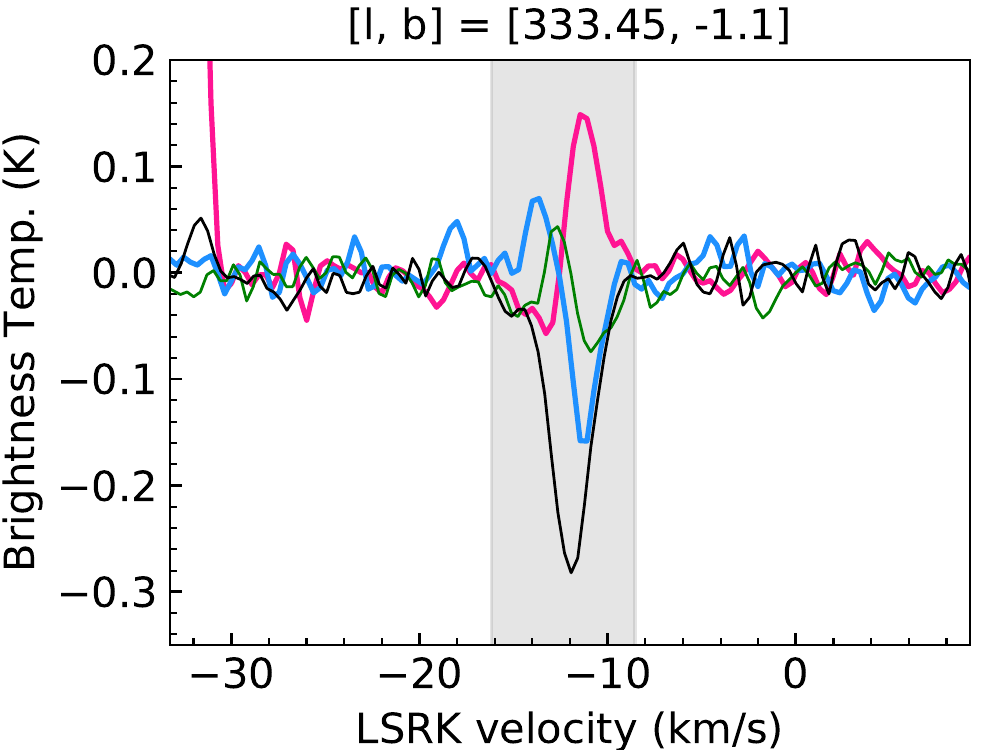}
\end{subfigure}

\medskip
\begin{subfigure}{0.32\textwidth}
\includegraphics[width=\linewidth]{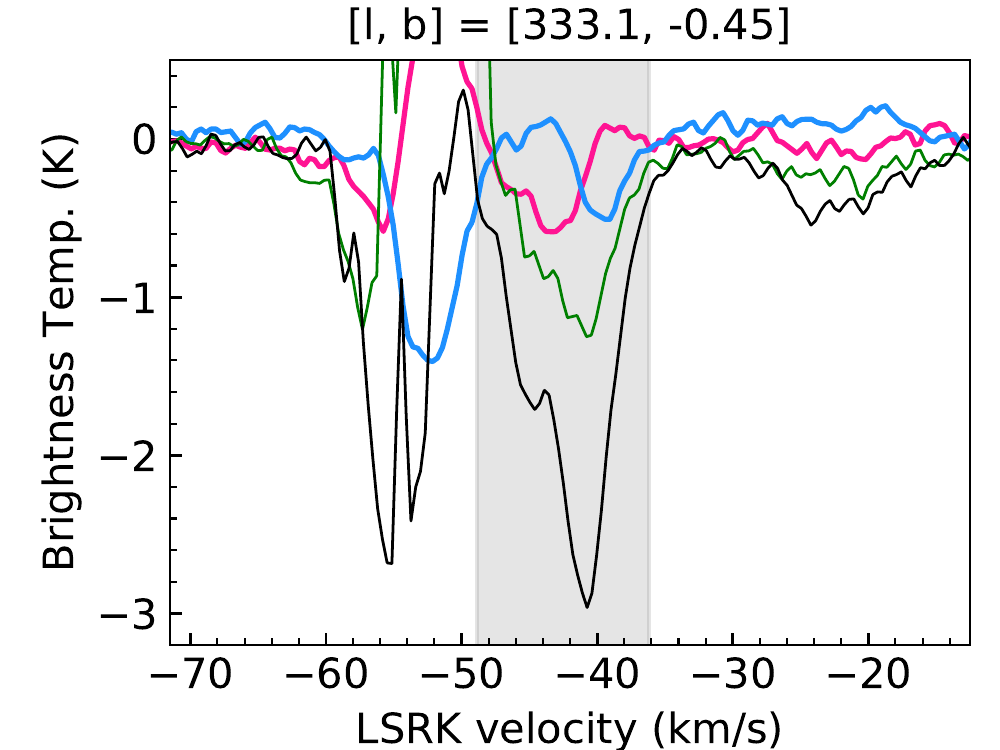}
\end{subfigure}\hspace*{\fill}
\begin{subfigure}{0.32\textwidth}
\includegraphics[width=\linewidth]{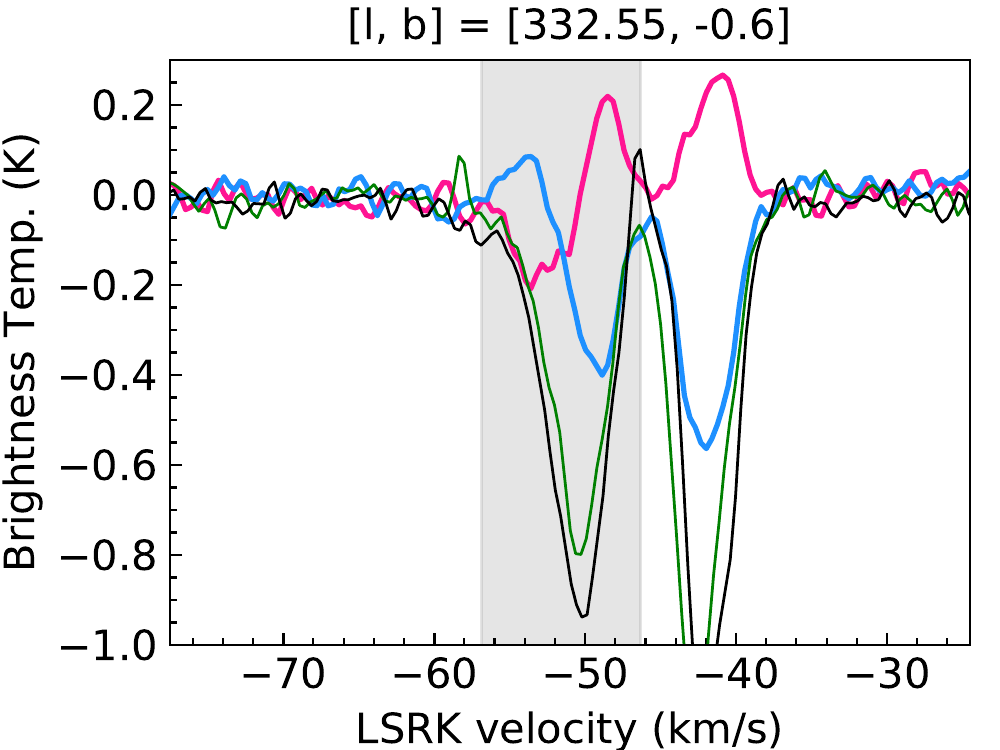}
\end{subfigure}\hspace*{\fill}
\begin{subfigure}{0.32\textwidth}
\includegraphics[width=\linewidth]{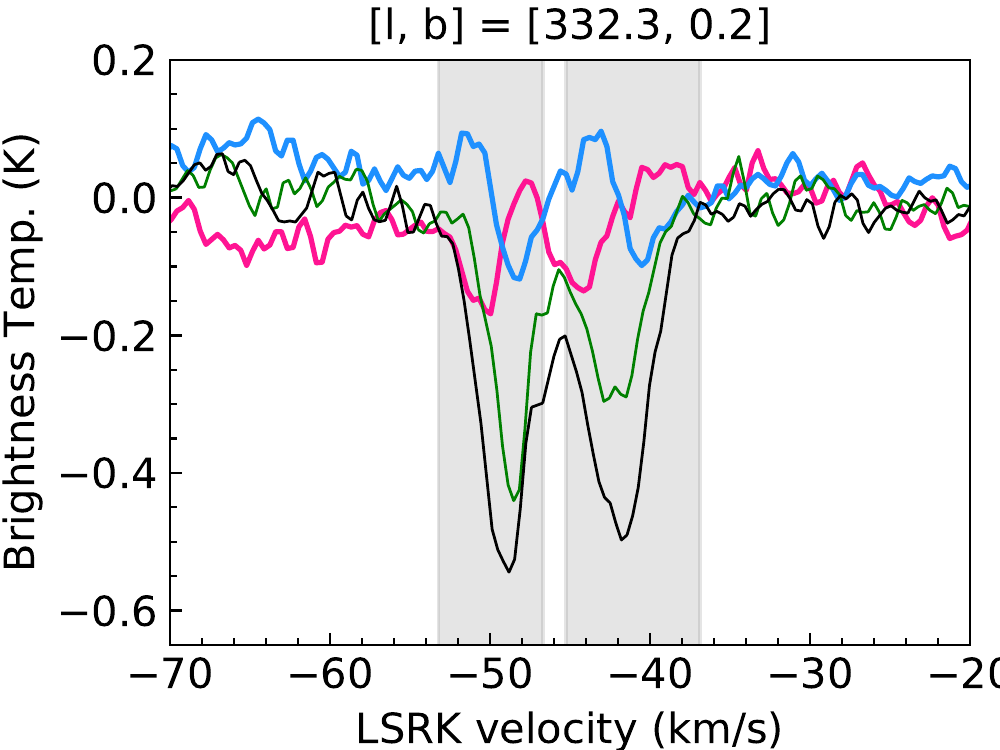}
\end{subfigure}

\medskip
\begin{subfigure}{0.32\textwidth}
\includegraphics[width=\linewidth]{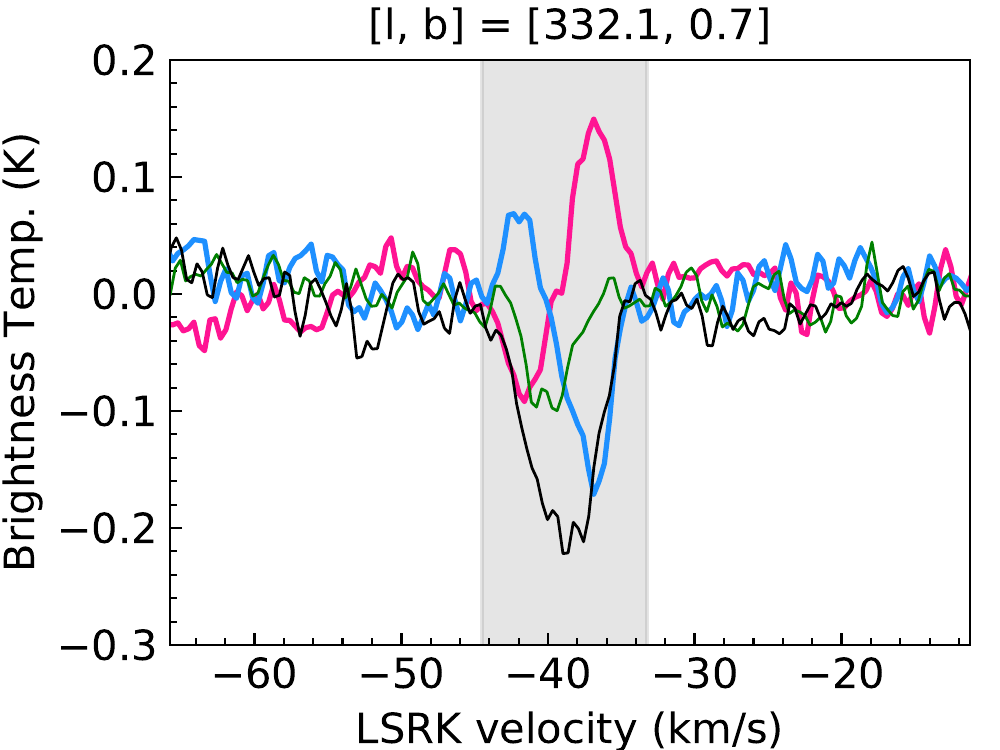}
\end{subfigure}\hspace*{\fill}

\caption{(cont.)}
\end{figure*}

\end{document}